\documentclass[a4paper,11pt]{article}
\pdfoutput=1 

\usepackage{jheppub} 
\usepackage[T1]{fontenc} 
\usepackage{graphicx}
\usepackage{slashed}
\usepackage{epstopdf}

\title{Non-resonant Collider Signatures of a Singlet-Driven Electroweak Phase Transition}

\author[a,b]{Chien-Yi Chen,}
\affiliation[a]{Department of Physics and Astronomy, University of Victoria, Victoria, BC V8P 5C2, Canada}
\affiliation[b]{Perimeter Institute for Theoretical Physics, Waterloo, ON N2L 2Y5, Canada}
\emailAdd{cchen@perimeterinstitute.ca}

\author[c,d]{Jonathan Kozaczuk}
\affiliation[c]{TRIUMF, 4004 Wesbrook Mall, BC V6T 2A3,  Canada }
\affiliation[d]{Amherst Center for Fundamental Interactions, Department of Physics,
 University of Massachusetts, Amherst, MA 01003, USA}
\emailAdd{kozaczuk@umass.edu}

\author[e,f]{ and Ian M. Lewis}
\affiliation[e]{SLAC National Accelerator Laboratory, 2575 Sand Hill Rd, Menlo Park, CA 94025, USA}
\affiliation[f]{Department of Physics and Astronomy, University of Kansas, Lawrence, KS 66045, USA}
\emailAdd{ian.lewis@ku.edu}

\preprint{ACFI-T17-07 \\ \begin{flushright} \vspace{-.75cm} SLAC-PUB-16951 \end{flushright} }

\abstract{We analyze the collider signatures of the real singlet extension of the Standard Model in regions consistent with a strong first-order electroweak phase transition and a singlet-like scalar heavier than the Standard Model-like Higgs. A definitive correlation exists between the strength of the phase transition and the trilinear coupling of the Higgs to two singlet-like scalars, and hence between the phase transition and non-resonant scalar pair production involving the singlet at colliders. We study the prospects for observing these processes at the LHC and a future 100 TeV $pp$ collider, focusing particularly on double singlet production. We also discuss correlations between the strength of the electroweak phase transition and other observables at hadron and future lepton colliders. Searches for non-resonant singlet-like scalar pair production at 100 TeV would provide a sensitive probe of the electroweak phase transition in this model, complementing resonant di-Higgs searches and precision measurements.  Our study illustrates a strategy for systematically exploring the phenomenologically viable parameter space of this model, which we hope will be useful for future work.}

\begin{document}  
\maketitle
\flushbottom

\setcounter{page}{2}

\section{Introduction}

Gauge singlet scalar fields appear in many well-motivated extensions of the Standard Model (SM). An attractive feature of such scenarios is that the singlet can give rise to a strong first-order electroweak phase transition (EWPT), as required for the mechanism of electroweak baryogenesis (EWB)~\cite{Trodden:1998ym,Cline:2006ts,Morrissey:2012db}, without large deviations in the predicted Standard Model-like Higgs properties. This is in contrast with scenarios like minimal supersymmetry, in which a strong first-order electroweak phase transition is excluded by a combination of Higgs measurements and direct searches for light scalar top quarks~\cite{Curtin:2012aa, Cohen:2012zza, Krizka:2012ah, Delgado:2012eu, Katz:2015uja}.

While many electroweak baryogenesis scenarios feature gauge singlet scalar fields along with additional field content (responsible for $CP$-violation, for example), often a singlet scalar is the primary field responsible for strengthening the phase transition. In these cases, the physics associated with a strong first-order electroweak phase transition can be illuminated by simplified models involving only a singlet scalar coupled to the SM through the Higgs field. With this in mind, we will focus on the real singlet extension of the Standard Model~\cite{Barger:2007im, Profumo:2007wc} and attempt to understand how and to what extent strong first-order electroweak phase transitions in this model can be tested by present and future experiments.

There has been much focus in the literature on using resonant Higgs pair production as a probe of the EWPT in the real singlet extension of the Standard Model (see e.g.~Refs.~\cite{  No:2013wsa,Assamagan:2016azc,Kotwal:2016tex,Huang:2017jws,Contino:2016spe} and references therein). This is an attractive channel because of its potentially large cross-section, especially at a 100 TeV collider. A complementary search strategy involves observing the corresponding effects of the singlet on (non-resonant)  Higgs pair production at hadron colliders~\cite{Profumo:2007wc, Noble:2007kk} and/or the couplings of the Higgs to Standard Model states~\cite{Profumo:2014opa}. A particularly powerful probe will be measurements of the $Zh$ production cross-section at lepton colliders, which can deviate from its SM predicted value due to mixing effects and the wavefunction renormalization of the Higgs-like scalar $h$~\cite{Craig:2013xia, McCullough:2013rea}. A combination of these approaches, in addition to gravitational wave experiments such as LISA~\cite{Caprini:2015zlo}, show promise in probing the EWPT in singlet models~\cite{Profumo:2007wc, Noble:2007kk, Huang:2016cjm, No:2013wsa, Assamagan:2016azc, Kotwal:2016tex, Caprini:2015zlo, Huang:2017jws, Contino:2016spe, Profumo:2014opa,  Curtin:2014jma, Chala:2016ykx, Beniwal:2017eik}.

There are two primary observations motivating the present study. For one, there exist several cases in which the strategies mentioned above are unlikely to probe the parameter space associated with a singlet-driven first-order electroweak phase transition. For example, if the singlet-like state is lighter than twice the SM-like Higgs mass, resonant di-Higgs production will be absent. Even if the singlet-like state is heavier than twice the Higgs mass, if the mixing angle between the two scalar fields is small, the coupling of the singlet-like state to the SM particles is suppressed, rendering resonant di-Higgs production practically unobservable.  The effects of the singlet will also be difficult to detect in precision Higgs observations and measurements of the Higgs cubic self-coupling if the Higgs-singlet mixing angle is not very large. Despite being difficult to probe, the parameter space below the di-Higgs threshold, as well as that with small mixing angles\footnote{Note that, as mentioned in Ref.~\cite{Huang:2016cjm}, the zero-mixing limit of the model without a $\mathbb{Z}_2$ symmetry is not technically natural and can require some amount of tuning to realize.}, is known to support a strong first-order EWPT, and so it is important to consider ways to access these regions experimentally.

Secondly, in portions of the parameter space known to be testable at present and future experiments it is still crucial to consider all possible independent probes of the electroweak phase transition. If a discovery is made in one experiment, several additional and independent observations will likely be required to definitively determine whether or not the discovery is consistent with a first-order electroweak phase transition in the early Universe. Furthermore, many signatures often considered in the literature, such as alterations of the Standard Model Higgs properties, are \emph{indirect}, and do not provide access to the new state(s) responsible for the deviations. It is thus worthwhile to consider additional direct experimental signatures of strong, singlet-driven first-order phase transitions.

Motivated by the observations above, in this study we address the possibility of directly probing the electroweak phase transition in singlet models via \emph{non-resonant} scalar pair production involving the singlet-like state at hadron colliders. We consider the processes
\begin{equation}
p p \rightarrow s s, \,sh \qquad s \rightarrow {\rm visible}
\end{equation} 
where $s$ is a singlet-like scalar. We will do so in the general real singlet extension of the Standard Model, without an accompanying $\mathbb{Z}_2$ symmetry, such that $s$ decays to visible Standard Model states. Searching for evidence of these processes at colliders can complement Higgs self-coupling and other precision measurements in their coverage of the parameter space, especially for small Higgs-singlet mixing angles.  We demonstrate this by comparing the various leading-order scalar pair production cross-sections across the parameter space of the model, and by studying the prospects for observing non-resonant $ss$ production at the LHC and a future 100 TeV collider. 

The production cross-section for $p p \rightarrow ss$ is highly correlated with the strength of the EWPT in this scenario. Furthermore, it is not suppressed in the small-mixing limit, unlike direct production, resonant di-Higgs and non-resonant $hs$ production, allowing it to provide experimental coverage to a significant range of mixing angles not accessible otherwise. This type of non-resonant scalar pair production has been studied in the past in the $\mathbb{Z}_2$-limit of the singlet model in Refs.~\cite{Curtin:2014jma, Craig:2014lda}. In this case, the $s$ is stable, and so can be searched for in final states involving missing energy. Refs.~\cite{Arkani-Hamed:2015vfh, Huang:2016cjm} both briefly discuss some of the prospects away from the $\mathbb{Z}_2$ limit and we build on their observations here.  We proceed much in the spirit of Ref.~\cite{Curtin:2014jma} in asking to what extent strong EWPTs can be probed in the singlet model at present and in future experiments through $ss$ production and other non-resonant processes. 

To this end, a thorough investigation of the parameter space is needed. Requiring compatibility with current experimental results, along with perturbativity, high-energy perturbative unitarity, and weak-scale vacuum stability, we will show how one can, in principle, explore \emph{all} of the parameter space consistent with a strong first-order electroweak phase transition and the aforementioned assumptions for a given mass and mixing angle, up to the scan resolution and uncertainties in the phase transition calculation. This provides a systematic strategy for surveying the parameter space of this model, which we hope will be useful for future work.

The remainder of this study is structured as follows: in Sec.~\ref{sec:model} we briefly review the real singlet extension of the SM along with its current and projected experimental status. Sec.~\ref{sec:EWPT} comprises a discussion of the electroweak phase transition and the trilinear $hss$ coupling as a diagnostic of the EWPT in this model. In Sec.~\ref{sec:nonres} we compare the leading-order cross-sections for the various non-resonant scalar pair production processes at colliders, showing that they provide sensitivity to complementary regions of the singlet model parameter space.  We then proceed to analyze one such process, $ss$ production, in Sec.~\ref{sec:trilepton}, focusing on the trilepton final state $2j 2\ell ^{\pm} \ell ^{\prime \mp} 3\nu$ with $\ell^{\prime}\neq \ell$. The prospects for accessing regions of the model supporting a strong first-order EWPT at the LHC and a future 100 TeV collider in this channel are presented in Secs.~\ref{sec:LHC} and~\ref{sec:100TeV}, respectively, along with a comparison to the sensitivity expected from $hh$ and $Zh$ observations at the LHC and future colliders. We conclude in Sec.~\ref{sec:concl}. Additional information regarding our renormalization scheme, the non-resonant scalar pair production cross-sections, the kinematic distributions relevant for our trilepton study, and our calculation of higher order effects on the effective $ZZh$ coupling is included in Appendices~\ref{sec:app_renorm},~\ref{sec:app_signal},~\ref{sec:app_kin}, and~\ref{sec:app_Zh}, respectively.

\section{The Model}\label{sec:model}

The real singlet extension of the Standard Model augments the SM by including a real scalar field $S$ that transforms as a singlet under $SU(3)_c \times SU(2)_L\times U(1)_Y$. The most general gauge-invariant renormalizable scalar potential involving the new field is 
\begin{equation}
\begin{aligned}
V_0(H,S)=&-\mu^2 \left|H\right|^2 + \lambda \left|H\right|^4 +\frac{1}{2} a_1 \left|H\right|^2 S + \frac{1}{2} a_2 \left|H\right|^2 S^2  \\
&+ b_1 S+ \frac{1}{2}b_2 S^2 +\frac{1}{3} b_3 S^3+\frac{1}{4}b_4 S^4
\end{aligned}
\end{equation}
where $H$ is the $SU(2)_L$ Higgs doublet of the Standard model. Without making any field redefinitions, the singlet will generically obtain a non-zero vacuum expectation value (VEV) at zero temperature. We can then expand
\begin{equation}
H = \frac{1}{\sqrt{2}}\left( \begin{array}{c}
\sqrt{2} \varphi^+\\
\phi_h+h+i \varphi^0
\end{array}
\right),
\hspace{0.3 cm} S = \frac{1}{\sqrt{2}}\left( \phi_s+ s \right)
\end{equation}
where $\varphi^{0,\pm}$ are the Goldstone fields, $\phi_{h,s}$ are the Higgs and singlet background fields, and at zero temperature, $\phi_h = v = 246$ GeV, $\phi_s = v_s$ in the electroweak vacuum. The two neutral $CP$-even gauge eigenstates will generally mix. The mass eigenstates can be ordered in mass and parametrized as
\begin{equation}
\begin{aligned}
&h_1 = h \cos \theta + s \sin\theta \\
&h_2 = -h \sin \theta + s \cos\theta \label{eq:mix1}
\end{aligned}
\end{equation}
In the rest of our study, we will use the parametrization of Ref.~\cite{Chen:2014ask} in which the $T=0$ singlet VEV is taken to be zero by appropriately shifting the singlet field (see also Ref.~\cite{Espinosa:2011ax}). We will also assume that $h_2$ is the mostly singlet-like state, with $h_1$ the Standard Model-like Higgs with $m_1=125$ GeV $<m_2$. We anticipate revisiting the case of a lighter singlet-like state in future work.

One-loop radiative corrections to the spectrum (at zero  external momentum) are encoded in the Coleman-Weinberg potential, $\Delta V_1$. The Coleman-Weinberg potential is
\begin{equation}\label{eq:V1loopT0}
V_{\rm eff}^1(\phi_h, \phi_s, T=0) = V_0(\phi_h, \phi_s) - i \sum_j \frac{\pm n_j}{2} \int \frac{d^4k}{(2\pi)^4} \log \left[-k^2+m^2_j(\phi_h, \phi_s) - i \epsilon \right] + \Delta V_{\rm ct}
\end{equation}
where the sum is over all species coupling to $\phi_{h,s}$ with $n_j$ degrees of freedom and $m^2_j(\phi_h,\phi_s)$ the corresponding field-dependent mass squared. The upper (lower) sign applies to bosons (fermions). $\Delta V_{\rm ct}$ is a renormalization scheme-dependent counterterm contribution required to renormalize the effects of the divergent momentum integral in Eq.~\ref{eq:V1loopT0}.  Cutting off the integral at $\Lambda$ yields~\cite{Anderson:1991zb}
\begin{equation}\label{eq:V1loopT0_2}
\Delta V_1 = \sum_j \frac{\pm n_j}{32 \pi^2} \left\{ \frac{1}{2} m_j^4(\phi_h,\phi_s) \left[ \log\left(\frac{m^2_j(\phi_h,\phi_s)}{\Lambda^2}\right)-\frac{1}{2} \right] + m^2_j(\phi_h, \phi_s) \Lambda^2 \right\}.
\end{equation}
Similarly to Refs.~\cite{Espinosa:2011ax, Curtin:2014jma}, we choose to renormalize the 1-loop effective potential in a pseudo--on-shell scheme which minimizes the one-loop contributions to the scalar trilinear and quartic couplings at zero temperature. This is detailed in Appendix~\ref{sec:app_renorm}. The resulting effective potential is independent of the cutoff $\Lambda$ at one loop.
This scheme also leaves the location of the tree-level electroweak minimum and the scalar mass matrix unaltered, and so the tree-level mass spectrum is retained. 

Throughout our study, we will be interested in how the strength of the electroweak phase transition is correlated with processes observable at colliders. While the EWPT is governed by the effective potential, the various couplings in $V(H,S)$ are not directly observable. As emphasized in e.g.~Refs.~\cite{Profumo:2007wc, Noble:2007kk}, they do, however, enter into the various multi-linear scalar interactions after electroweak symmetry breaking. We will therefore investigate processes that depend on these couplings, in particular those that are cubic in $h_1$ and $h_2$. These couplings can be obtained directly by rewriting the potential at cubic order in the mass basis:
\begin{equation}
V_{\rm cubic} = \frac{1}{6}\lambda_{111} h_1^3 + \frac{1}{2}\lambda_{211}h_2 h_1^2 + \frac{1}{2}\lambda_{221}h_2^2 h_1 + \frac{1}{6}\lambda_{222} h_2^3,
\end{equation}
where 
\begin{equation}
\lambda_{ijk} \equiv \frac{\partial^3V(h_1,h_2)}{\partial h_i \partial h_j \partial h_k}
\end{equation}
and $h_{1,2}$ are understood as the corresponding background fields. Up to small finite-momentum effects, these $\lambda_{ijk}$ are those that then enter the expressions for the various multi-scalar production cross-sections at hadron colliders. Detailed tree-level expressions relating the mass eigenstate couplings to those of the gauge eigenstate basis can be found in Ref.~\cite{Chen:2014ask}. Note that, in our renormalization scheme, $\lambda_{221}^{\rm 1-loop} \simeq \lambda_{221}^{\rm tree}$ and $\lambda_{222}^{\rm 1-loop} \simeq \lambda_{222}^{\rm tree}$. In our computation of the various di-scalar production cross-sections, we will typically use the tree-level values (neglecting finite-momentum effects) to maintain a consistent leading-order collider treatment, and we will take $\lambda_{ijk}$ to denote the corresponding tree-level couplings derived from the scalar potential, unless otherwise specified.

\subsection{Current Constraints and Projected Sensitivities}

Before proceeding, let us briefly comment on the current and projected experimental sensitivity to the parameter space of the singlet-extended SM.  A summary of the current constraints on this model can be found in various places in the literature (see e.g.~\cite{Chen:2014ask, Profumo:2014opa, Robens:2015gla, Buttazzo:2015bka, Chalons:2016jeu}). For our purposes, the most important conclusions from these studies are that currently all values of $\left|\sin \theta\right| \lesssim 0.2$ are allowed for $m_2 < 2 m_1$, while for $m_2 \gtrsim 2 m_1$ resonant di-Higgs production places an additional constraint on the parameter space and provides another discovery channel for this model.

A number of future and planned experiments are also expected to impact the parameter space of real singlet extension of the SM:
\begin{itemize}
\item Higgs coupling measurements at the LHC are expected to probe mixing angles as small as $\sin\theta \sim 0.2$ with 300 fb$^{-1}$ at 14 TeV, independently of $m_2$~\cite{Dawson:2013bba, Profumo:2014opa, Buttazzo:2015bka}.
\item Direct searches for $h_2$ production at the high-luminosity LHC will likely be able to reach $\sin\theta \sim 0.1$ for $m_2 \gtrsim 2 m_{W}$ with 3000 fb$^{-1}$ at 14 TeV~\cite{Buttazzo:2015bka}.
\item Future lepton colliders, such as the ILC, FCC-ee, CEPC, and CLIC, would likely be able to probe values of $\sin\theta \gtrsim 0.05$ via precision Higgs coupling measurements~\cite{Dawson:2013bba, Profumo:2014opa, Buttazzo:2015bka}.
\item As mentioned in the Introduction, resonant di-Higgs production at the LHC~\cite{No:2013wsa, Chen:2014ask,  Huang:2017jws} and a future 100 TeV hadron collider~\cite{Chen:2014xra, Assamagan:2016azc, Kotwal:2016tex, Contino:2016spe} would be expected to probe portions of the parameter space with $m_2>2m_1$. Ref.~\cite{Kotwal:2016tex} found that a 100 TeV collider could have sensitivity to portions of the parameter space down to $\sin\theta \sim 0.03$ for particular values of the $\lambda_{211}$ coupling. However, as $\sin\theta$ decreases, the $h_2$ production cross-section falls as $\sin^2 \theta$, and so for small enough mixing angle, this channel will be unlikely to provide sensitivity to the model, even for $m_2>2m_1$.
\end{itemize} 

Given the above considerations, we will focus on $|\sin \theta| \lesssim 0.2$ in this work, paying particular attention to small mixing angles to demonstrate the usefulness of non-resonant scalar pair production in probing this difficult region. We will take positive values for $\sin \theta$; negative values yield qualitatively very similar results.

\subsection{A Comprehensive Analysis of the Parameter Space} \label{sec:param_space}

To assess the degree to which experiments can conclusively probe the nature of the electroweak phase transition in this model, we would like to investigate the corresponding parameter space as comprehensively as possible.

In addition to a 125 GeV SM-like Higgs boson, in most of what follows we will impose the following requirements on the model:
\begin{itemize}
\item \textbf{Absolute weak-scale vacuum stability - } no vacuum exists at $T=0$ that is deeper than the electroweak vacuum with $v=246$ GeV, $v_s=0$ GeV. To enforce this condition, we minimize the one-loop effective potential at $T=0$ using the \texttt{Minuit} routine~\cite{James:1975dr}. We also require that there are no runaway directions in the tree-level scalar potential. We do not check whether or not a deeper vacuum exists at very large field values, a problem already present in the Standard Model~\cite{Degrassi:2012ry}.
\item \textbf{Perturbativity - } We require all dimensionless couplings to be less than $4\pi$ at the electroweak scale. We also require $|b_3|/v<4\pi$. Note that we do not impose any perturbativity requirements on the theory above the electroweak scale or check for the existence of low-lying Landau poles. These considerations would only reduce the parameter space available for a strongly first-order EWPT, and so not affect our conclusions. See e.g.~Refs.~\cite{Gonderinger:2009jp,  Gonderinger:2012rd} for analyses including these constraints in singlet models.
\item \textbf{Perturbative Unitarity - } We exclude points that violate perturbative unitarity at high energies. The strongest resulting constraint is on the singlet quartic coupling $b_4$, and results in the requirement $b_4 < 8\pi/3$\footnote{There is another constraint on the quartic coupling $\lambda$, which requires $\lambda < 4\pi/3$. However, the constraint is always trivially satisfied in the small angle region as indicated by eq.~\ref{relation}.}. See also Refs.~\cite{Chen:2014ask, Robens:2015gla, Chalons:2016jeu} for similar considerations in singlet models.
\end{itemize}

To systematically explore the parameter space consistent with the above assumptions, we will make use of the following strategy: for a given value of  $m_2$ and $\sin\theta$, choose $\lambda$, $\mu^2$, $b_1$, $a_1$ and $b_2$ accordingly and such that $m_1=125$ GeV, $v=246$ GeV, $v_s=0$. This corresponds to setting
\begin{equation}
\begin{aligned}
\label{relation}
a_1 = \frac{1}{v}\left(m_1^2-m_2^2\right)\sin 2\theta&, \quad b_1 = -\frac{1}{4}v^2 a_1 ,\quad \mu^2 = \lambda v^2 \\
b_2 = m_1^2 \sin^2\theta + m_2^2 \cos^2 \theta - \frac{a_2}{2} v^2&, \quad \lambda = \frac{1}{2v^2} \left( m_1^2 \cos^2\theta + m_2^2 \sin^2 \theta \right).
\end{aligned}
\end{equation}
Three free parameters remain: $a_2$, $b_3$, and $b_4$. We can then continuously vary these parameters in the range 
\begin{equation}
|a_2|,\, |b_3|/v < 4\pi, \quad b_4 < 8\pi/3
\end{equation}
while imposing the vacuum stability requirements discussed above. This allows us, in principle, to scan over the complete parameter space of the model for a given $m_2$, $\sin \theta$, given our assumptions (and the finite resolution of the scan). Since, in our conventions, all of the experimental observables of interest are independent of $b_4$, we can then project onto the $a_2-b_3$ plane without losing any relevant information.

\begin{figure}[!t]
\centering
\includegraphics[width=.45\textwidth]{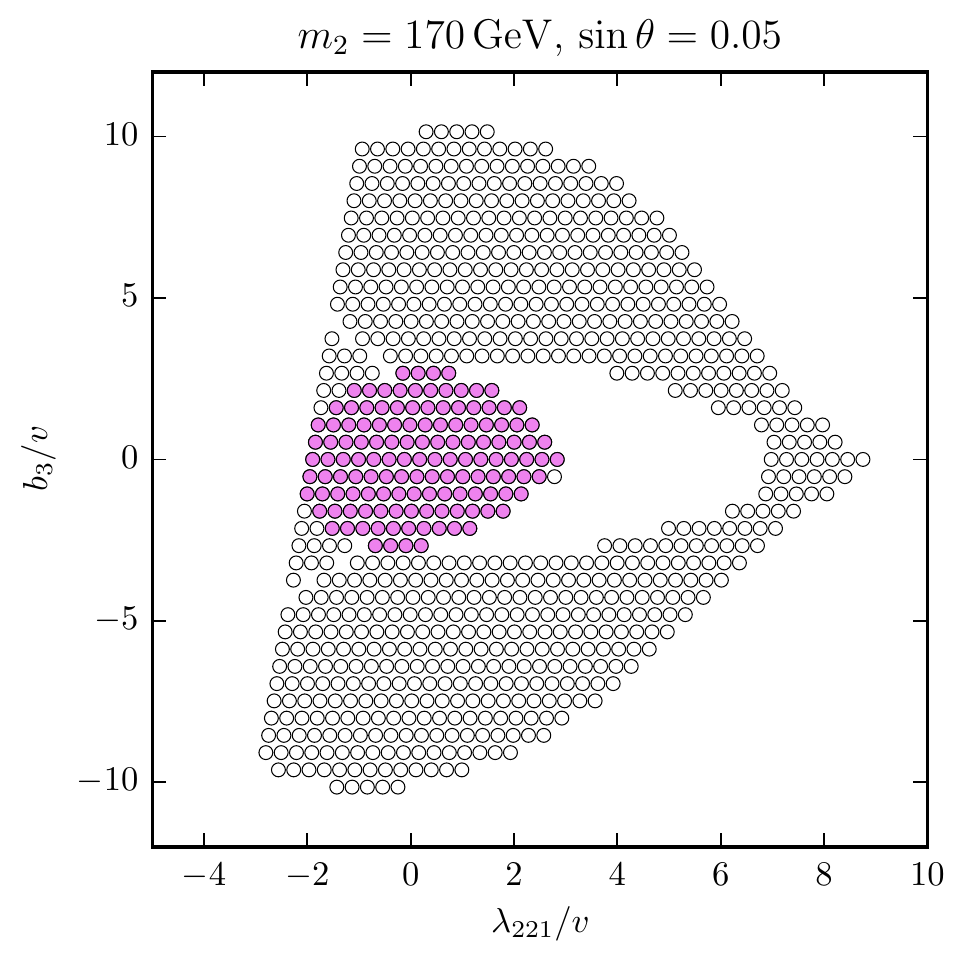}\, \includegraphics[width=.45\textwidth]{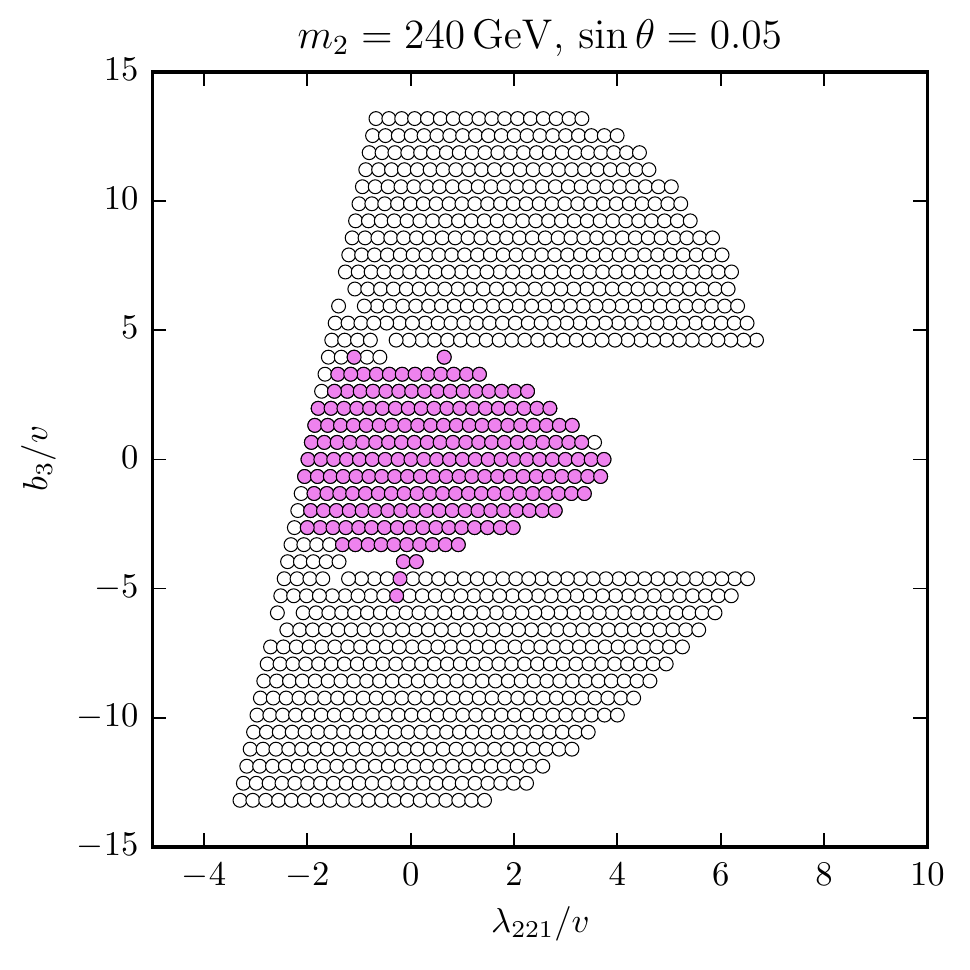}

\caption{\label{fig:initial_scan} The parameter space of interest for $m_2=170$ GeV (left) and $m_2=240$ GeV (right) with $\sin\theta=0.05$ consistent with our requirements of perturbativity, vacuum stability, and perturbative unitarity. The parameter $b_4$ has been marginalized over, such that the points shown are found to have some value of $b_4 < 8\pi/3$ such that these requirements hold (we scan down to $b_4 = 0.01$). These points were obtained by a grid scan over $a_2$, $b_3$ and $b_4$. The darker shaded points satisfy the above requirements at both tree- and one-loop level, while the lighter points satisfy these requirements at one-loop but not tree-level. The white regions (without points) are disallowed by our requirements at 1-loop for all values of $b_4$ considered.}
\end{figure}

The results of such a scan for $m_2 = 170$, 240 GeV (the particular masses we will focus on in our collider study below) and $\sin \theta = 0.05$ (below the current and projected sensitivity of precision Higgs measurements) are shown in Fig.~\ref{fig:initial_scan}. The results for larger mixing angles  look qualitatively similar for $|\sin \theta | \lesssim 0.2$, as we will show below. We display the results in terms of $\lambda_{221}$ instead of $a_2$, since this coupling will be important in our phase transition analysis. In these figures, we have marginalized over $b_4$. Points indicate that, for the corresponding values of $a_2$ and $b_3$, some value of $b_4 < 8\pi/3$ is found such that all of the above requirements are satisfied, with $b_4>0.01$ (the lower cutoff for our scan). The white regions with no points are disallowed by our requirements for all values of $b_4$ considered. Note that, as mentioned above, $\lambda_{221}$ is independent of $b_4$ in our conventions.

In Fig.~\ref{fig:initial_scan}, we also show points that satisfy the above requirements at both 1-loop and tree-level; the corresponding points are shaded purple. These plots make clear the regions where radiative corrections become important; as expected, this occurs for large values of the various couplings. In these regions , the one-loop contributions can uplift the non-electroweak tree-level vacua and stabilize the potential as in the well-known Coleman-Weinberg scenario. For example, for $m_2=170$ GeV, this occurs at large $|b_3|$ and $a_2$ values, enclosing the central void region. Note that the corresponding region would also be enclosed for $m_2=240$ GeV, however this requires larger couplings than are allowed by our perturbativity requirements.

Other features of the viable parameter space are also straightforward to understand. The leftmost boundaries in Fig.~\ref{fig:initial_scan} feature values of $a_2$ that are sufficiently negative to produce a run-away direction in the tree-level potential. The rightmost region does not feature any points due to our absolute vacuum stability requirements for perturbative values of the couplings. The upper and lower boundaries for $m_2=170$ GeV also arise from vacuum stability requirements, while for $m_2=240$ GeV, some points are also cut off by our perturbativity requirement on $b_3$. Note that, if the upper limit on $b_4$ were lowered, the parameter space shown would shrink.

While points with large couplings are technically allowed by our scan, we caution the reader that our one-loop perturbative treatment will likely be insufficient to capture the physics of these regions. Also, additional requirements such as perturbativity up to scales above the electroweak scale or the non-existence of low-lying Landau poles are likely to further reduce the parameter space at large $|b_3|$ and $a_2$, along the lines of Refs.~\cite{Gonderinger:2009jp, Gonderinger:2012rd}. Given these considerations, our results are most reliable in the purple regions, where the perturbative expansion is clearly under control. It will turn out that this region is also where our phase transition predictions are most robust and the most difficult region to probe experimentally, thus providing a compelling target for new search strategies. 

With these features in mind, we will now turn to analyzing the electroweak phase transition across this parameter space.

\section{The Electroweak Phase Transition in Singlet Models}\label{sec:EWPT}

First-order cosmological phase transitions can occur for a given set of parameters in a theory if two or more distinct vacua coexist for some range of temperatures. A scalar background field trapped in a metastable phase can then thermally fluctuate (or quantum mechanically tunnel) to an energetically favorable ``truer'' vacuum. In perturbation theory, such transitions can be studied semi-classically using the finite-temperature effective potential.

\subsection{The finite temperature effective potential}

Assuming a homogeneous background field configuration, the various vacua of the theory correspond to the minima of the effective potential, $V_{\rm eff}$. At zero temperature, $V_{\rm eff}$ is given by Eq.~\ref{eq:V1loopT0}. At finite temperature there are additional contributions to the effective potential, given by
 \begin{align}\label{eq:VT}
 \Delta V_1^T(\phi_h, \phi_s,T)=\frac{T^4}{2\pi^2}\left[\sum_i \pm n_i J_{\pm}\left(\frac{m^2_i(\phi_h,\phi_s)}{T^2}\right)\right],
 \end{align}
 where
  \begin{equation}
  J_{\pm}(x)=\int_0^\infty dy\, y^2\log\left[1\mp \exp(-\sqrt{x^2+y^2})\right].
 \end{equation}
 There are several technical challenges and outstanding problems related to obtaining reliable predictions from the finite-temperature effective potential (see e.g.~Refs.~\cite{  Dolan:1973qd, Nielsen:1975fs, Patel:2011th,  Garny:2012cg, Martin:2014bca, Andreassen:2014eha, Plascencia:2015pga, Espinosa:2016uaw, Curtin:2016urg}). To ensure that our results are as robust as possible, we will employ two different strategies for computing $\Delta V_1^T$.
 
  In the first approach, we consider the full $T=0$ 1-loop Coleman-Weinberg potential in Landau gauge (neglecting the Goldstone boson contributions) and evaluate the finite temperature functions $J_{\pm}(x)$ numerically. This is the historically conventional approach in the literature. It is well-known that the thermal contribution above suffers from an IR problem: infrared bosonic loops of zero Matsubara frequency spoil the perturbative expansion for small field-dependent masses. This effect can be mitigated by resumming the so-called ``daisy diagram'' contributions, giving rise to a self-energy shift in the zero mode finite-temperature bosonic propagators. Effectively, this amounts to adding a term to the finite-temperature effective potential of the form
 \begin{equation} \label{eq:daisy}
 \Delta V_{\rm ring}^T(\phi_h,\phi_s, T) = \sum_{j} \frac{n_j T}{12 \pi} \left[m_j^3(\phi_h,\phi_s) - m_j^3(\phi_h, \phi_s, T)\right]
 \end{equation}
In doing so, we use the high-$T$ approximation for the thermal self-energies in $m_j^3(\phi_h, \phi_s, T)$. The numerical accuracy of this approximation and methods for improving it are discussed in Ref.~\cite{Curtin:2016urg}. The sum in the above expression is over longitudinal gauge bosons and scalars (the transverse contribution vanishes in the high-$T$ approximation we use). Expressions for the field-dependent zero-temperature and effective thermal masses are found in Appendix~\ref{sec:app_renorm}.

Unfortunately, the strategy described above is known to yield gauge-dependent results for the critical temperature and order parameter of the phase transition~\cite{  Dolan:1973qd, Nielsen:1975fs, Patel:2011th}. This gauge dependence arises from  the gauge fixing, gauge and Goldstone boson contributions to the Coleman-Weinberg potential and to the finite-$T$ cubic term (see e.g.~Ref.~\cite{Patel:2011th} for a comprehensive discussion). In singlet models, a strong first-order electroweak phase transition is typically catalyzed by the singlet contributions to the potential, which are gauge invariant. This roughly suggests that the results obtained by the method outlined above in Landau gauge should be quite insensitive to small variations of the gauge fixing parameter $\xi$. Nevertheless, it may be morally dissatisfying that there is still residual dependence of our results on an unphysical parameter.

To obtain an explicitly gauge-invariant result\footnote{Another method for obtaining a gauge-invariant result is the so-called ``$\hbar$-expansion'' described in Ref.~\cite{Patel:2011th}. While we do not utilize it here, it would be interesting to compare our results with those obtained from the $\hbar$-expansion in the future.}, we will also analyze the finite-$T$ behavior of the model by retaining only the tree-level potential at $T=0$, performing a high-$T$ expansion of the thermal functions, whereby
\begin{equation}
\label{eq:highT}
\begin{aligned}
 T^4 J_+\left(\frac{m^2}{T^2}\right) =&-\frac{\pi^4 T^4}{45}+\frac{\pi^2m^2 T^2}{12}-\frac{T\pi (m^2)^{3/2}}{6}-\frac{(m^4)}{32}\log\frac{m^2}{a_b T^2}, \\
 T^4 J_-\left(\frac{m^2}{T^2}\right)= &\frac{7\pi^4 T^4}{360}-\frac{\pi^2m^2 T^2}{24}-\frac{(m^4)}{32}\log\frac{m^2}{a_f T^2}, 
\end{aligned}
\end{equation}
and dropping all terms except those proportional to $T^2$, which are explicitly gauge-invariant (see e.g.~Ref.~\cite{Quiros:1999jp} for definitions of $a_f$, $a_b$, and a pedagogical discussion of this approximation). We will refer to this strategy as the ``high-$T$ approximation''. Of course this method will neglect terms that can be numerically important, especially for large tree-level couplings (we have already seen that loop corrections can have important implications for vacuum stability, for example). However, the regions of parameter space predicting a strong first-order EWPT in both approaches are a particularly compelling target for experimental searches, since the agreement of both methods suggests a robust prediction for the PT.
 
\subsection{Searching for Strong First-Order Electroweak Phase Transitions}

The presence of additional singlet scalars with electroweak scale masses can give rise to a strong first order phase transition through a combination of different mechanisms~\cite{Profumo:2007wc, Espinosa:2011ax}. A barrier at finite temperature between an electroweak-symmetric and -broken phase can be produced by new tree-level cubic terms in the scalar potential, by zero-temperature loop effects, or through significant thermal contributions. In general one expects a combination of these mechanisms at work. Previous studies of the EWPT in the $\mathbb{Z}_2$-symmetric singlet extension of the SM have made use of simple analytic criteria for determining whether or not a first-order phase transition is possible at one loop~\cite{Curtin:2014jma, Craig:2014lda, Beniwal:2017eik}. In the more general case without the discrete symmetry, the additional terms in the scalar potential make a simple analytic treatment more complicated. This is due to the appearance of various additional minima at zero and finite temperature (see e.g.~Refs.~\cite{Profumo:2007wc, Espinosa:2011ax, Xiao:2015tja} for detailed discussions of the various possibilities in the high-temperature approximation). We thus proceed numerically, as described below.

At high temperatures, electroweak symmetry is typically unbroken\footnote{We assume that the reheat temperature after inflation is above the electroweak scale.} and the true vacuum of the theory\footnote{Throughout our analysis we ignore minima with Higgs and singlet field values greater than 1 and 10 TeV, respectively, since our one-loop perturbative analysis begins to break down for large field values. To consider such vacua, an RG-improved effective potential should be used. Since the tunneling rate to such far vacua is typically very slow, including such minima in our analysis should not affect our conclusions.} features $\phi_h= 0$. As the temperature of the Universe drops, electroweak symmetry breaking can occur once it becomes energetically favorable for the $SU(2)_L$ background field $\phi_h$ to take on a non-zero value. If at this temperature there is a barrier separating  the two phases, the field can then transition out of the metastable $\phi_h = 0$ vacuum to one with $\phi_h \neq 0$ via a first order electroweak phase transition. The temperature at which the two vacua become degenerate is known as the critical temperature, $T_c$.  Such a phase transition is said to be ``strongly first order'' if 
\begin{equation}\label{eq:Tc}
\frac{\phi_h(T_c)}{T_c} \gtrsim 1.
\end{equation}
There are several uncertainties and assumptions implicit in the above criterion~\cite{Patel:2011th}, but overall it is known to provide a reliable guide to finding points compatible with electroweak baryogenesis (see e.g.~Ref.~\cite{Fuyuto:2014yia} for a discussion of this criterion in the singlet model).

Hypothetically, it is possible for electroweak symmetry to be broken, then restored, then broken again, or for electroweak symmetry breaking to proceed via a multi-step transition~\cite{Espinosa:2011ax, Patel:2012pi, Patel:2013zla, Blinov:2015sna, Inoue:2015pza}. In all cases, the relevant transition for electroweak baryogenesis is the one with the lowest critical temperature such that the metastable phase features $\phi_h=0$, since this is the transition that shuts off the sphalerons for the last time. Thus, to find viable points with a strong first-order electroweak phase transition we employ the following strategy: \emph{starting from $T=0$, scan up in temperature until a vacuum with $\phi_h\neq 0$ is no longer the global minimum of the potential~\footnote{A $\phi_h\neq 0$ vacuum must be the global minimum of the potential at low temperatures by our assumption of vacuum stability.}. Denote the temperature at which the $\phi_h=0$ vacuum becomes the global minimum as $T_*$. If a first order electroweak phase transition is possible, this $\phi_h=0$ must be degenerate with the $\phi_h \neq 0$ minimum at some temperature $T_c\approx T_*$. If this is the case, identify the field value in the broken phase as $\phi_h(T_c)$. If $\phi_h(T_c)/T_c \geq 1$, we consider this point as having a strongly first-order phase transition.}

An implicit assumption of this method is that the field efficiently tunnels whenever it is energetically favorable to do so. In parts of the parameter space with sizable tree-level barriers between the vacua, it is likely that the phase transition to the physical vacuum will not complete (this is a concern whenever one uses the criterion in Eq.~\ref{eq:Tc} to determine the viability of electroweak baryogenesis). We include these points in our analysis anyway, since we do not compute the tunneling rate and we would like to retain as much of the potentially viable parameter space as possible. It could instead be the case that the field never reaches the initial $\phi_h=0$ vacuum (identified as the metastable phase for the first electroweak-symmetry breaking transition) due to a small tunneling rate out of another phase with $\phi_h=0$. However, this would mean that the true electroweak symmetry--breaking transition occurs at a higher temperature, and hence very likely with reduced strength relative to that predicted by our method. Regardless of the pattern of symmetry breaking in the early universe, we therefore expect our treatment to effectively capture all points compatible with a strong first-order electroweak phase transition at one loop, given our assumptions about vacuum stability and perturbative unitarity (as well as the resolution of our temperature scan and our methods for computing the finite-$T$ effective potential).

To ensure that we find all the minima of the potential at a given temperature, we use the \texttt{Minuit} routine~\cite{James:1975dr} for gradient-based global minimization. At each temperature, we feed the algorithm all tree-level extrema of the potential and allow it to flow to the nearest minimum. This is similar to the strategy used by the software package \texttt{VEVacious}~\cite{Camargo-Molina:2013qva} to find the minima of the one-loop $T=0$ potential.  This procedure is not necessarily guaranteed to find all minima, however for parameter space points such that the one-loop corrections to the scalar potential are under perturbative control, it is quite reliable. Nevertheless, at each temperature we feed additional starting points to the algorithm to safeguard against missing minima that may appear far away from tree level minima, maxima, and saddle points.

Applying the above strategy to the parameter space consistent with the requirements laid out in Sec.~\ref{sec:param_space} yields the results shown in Fig.~\ref{fig:PT} for $m_2= 170$, $240$ GeV and $\sin\theta = 0.05$, 0.2. The results are again projected onto the $\lambda_{221}-b_3$ plane, to show the maximal extent of the corresponding parameter space. The blue colored points feature a strong first order electroweak phase transition for some value of $b_4>0.01$ in our full (gauge-dependent) approach. Purple points feature a strong first-order EWPT in both the full approach and gauge-invariant high-$T$ approximation. Since the latter method drops the 1-loop Coleman-Weinberg piece, it is only applied to regions of the parameter space with tree-level vacuum stability (e.g.~points shaded purple in Fig.~\ref{fig:initial_scan}). We once again stress that, for a given mass and mixing angle, these figures should show the full extent of the parameter space consistent with a strong first-order EWPT, given our assumptions, requirements, scan resolution, and numerical accuracy. Points that are not shaded blue or purple do not feature a strongly first-order EWPT detected by our scans for any value of $b_4$  considered. Other more sophisticated methods for computing the phase transition properties could be applied to the same parameter space in the future and would provide an interesting comparison. Our strategy for systematically surveying the parameter space makes it straightforward to definitively analyze the correlation of various observables with the  strength of the phase transition.

\subsection{A Strong Electroweak Phase Transition and the Triscalar Couplings}

\begin{figure}[!t]
\centering
\includegraphics[width=.45\textwidth]{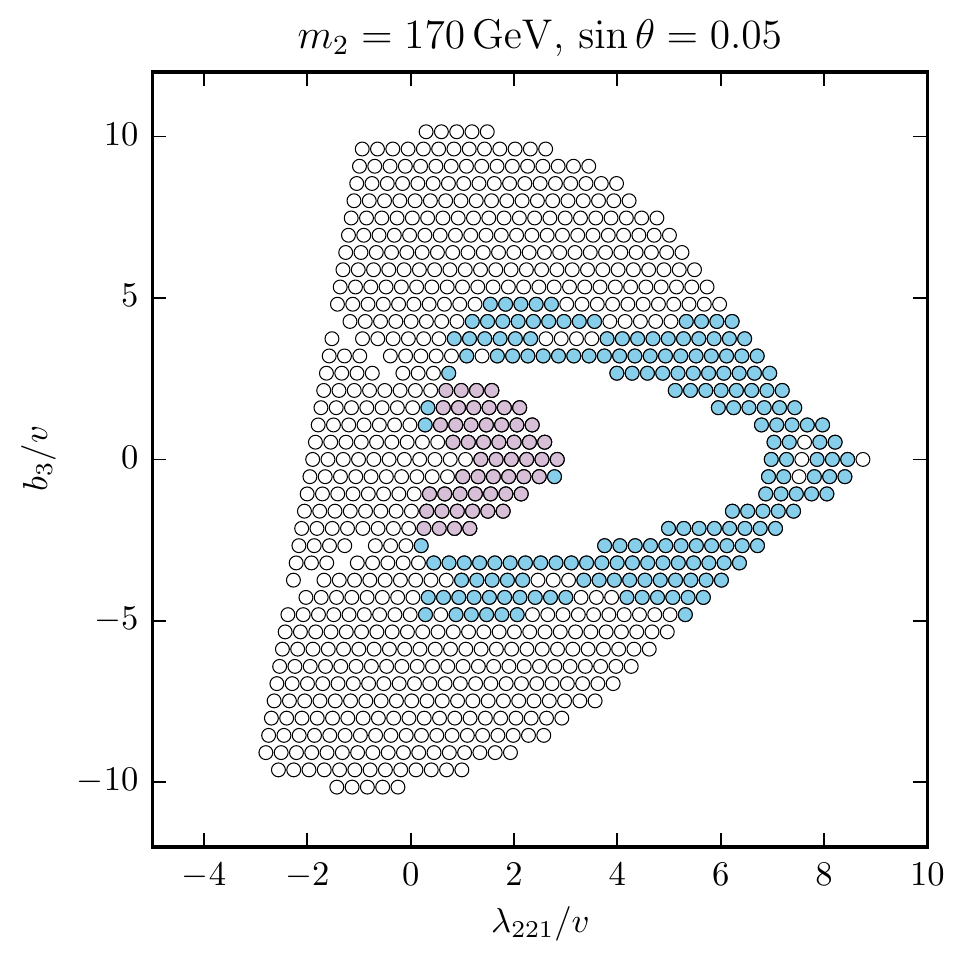}\, \includegraphics[width=.45\textwidth]{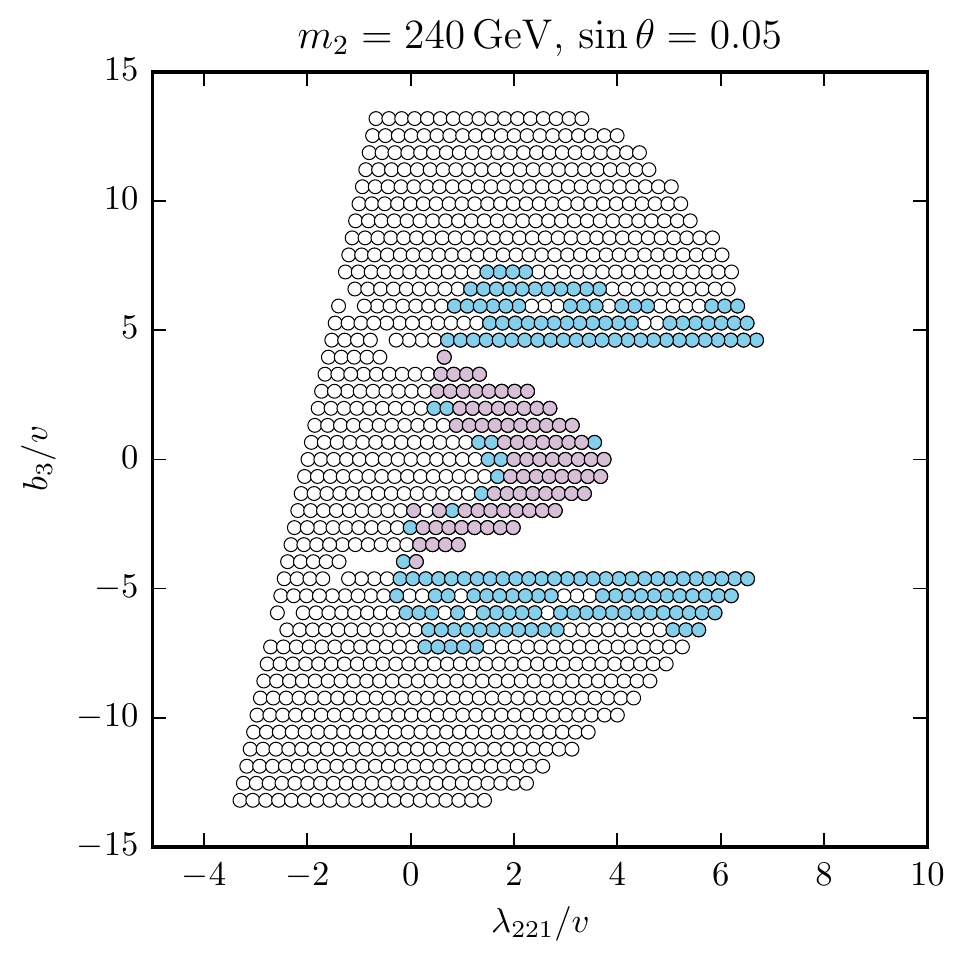}
\includegraphics[width=.45\textwidth]{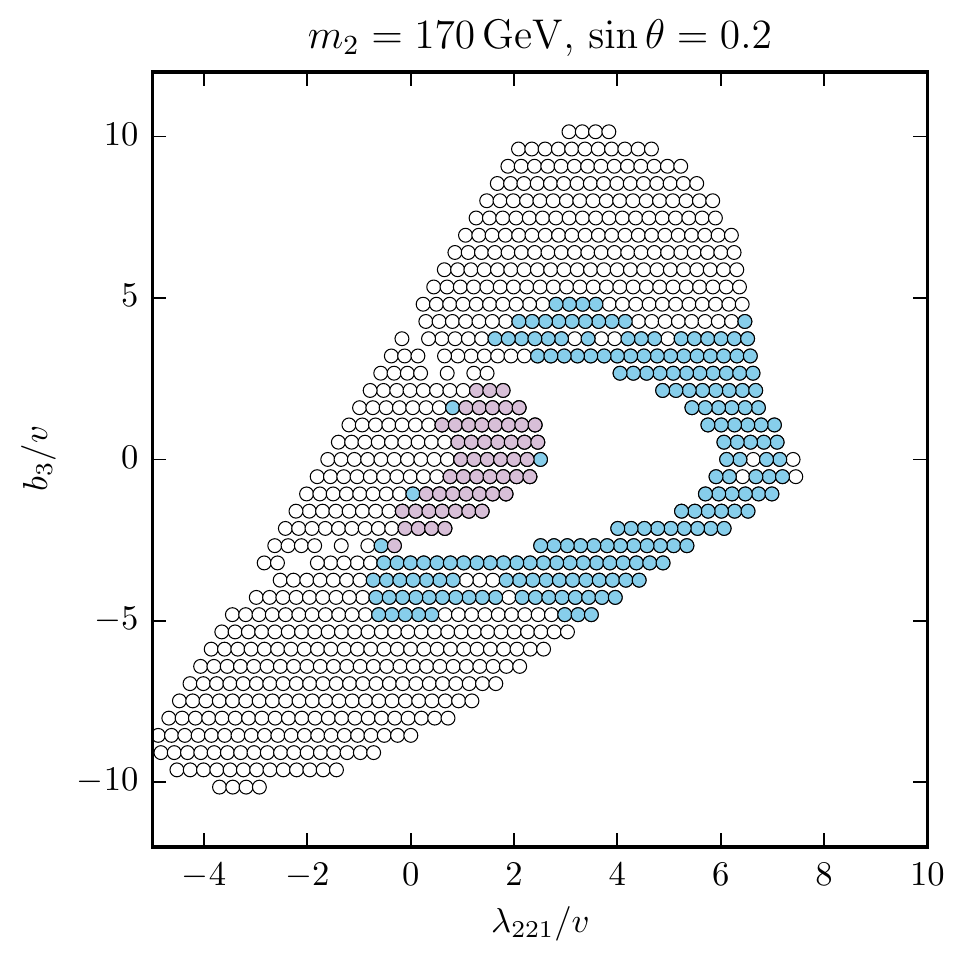}\, \includegraphics[width=.45\textwidth]{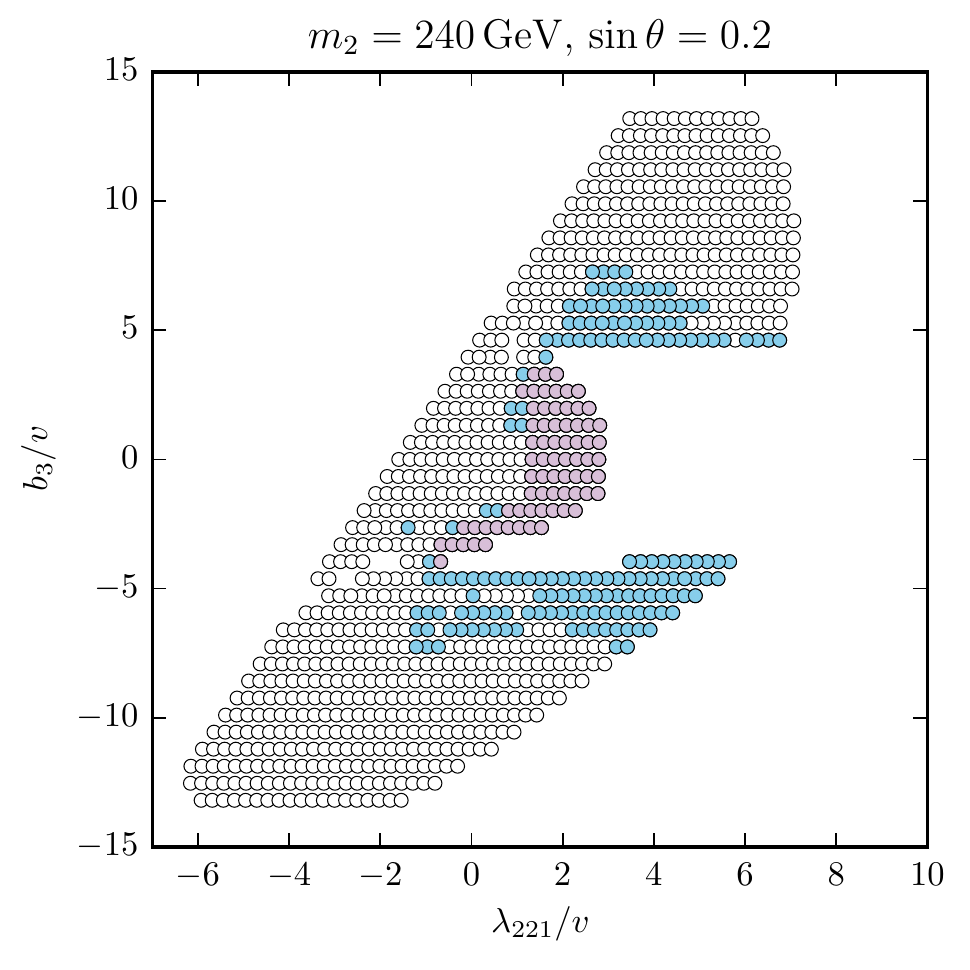}

\caption{\label{fig:PT} The parameter space of the model consistent with our requirements for $m_2=170$, 240 GeV and $\sin \theta = 0 .05$, 0.2 , now showing regions with a strong first-order electroweak phase transition. Results for both $\sin \theta = 0 .05$ and 0.2 are shown. Blue points feature an EWPT with $\phi_h(T_c)/T_c \geq 1$ for some value of $b_4 > 0.01$ in our approach utilizing the one-loop daisy-resummed thermal effective potential. Purple points additionally feature a strong first-order electroweak phase transition as predicted by the gauge-invariant high-$T$ approximation (which drops the Coleman-Weinberg potential and is thus only applied to regions with tree-level vacuum stability). Strong electroweak phase transitions are typically correlated with sizable values of $\lambda_{221}$.}
\end{figure}

How might one probe the regions compatible with a strong first-order electroweak phase transition, such as those shown in Fig.~\ref{fig:PT}, experimentally? If regions with a strong first-order EWPT robustly predict that a particular process should be observable at colliders, its experimental observation would hint at a strong EWPT (a hint that would be made more concrete by other independent observations), while its absence would, in principle, conclusively rule out a strong EWPT in this model. To this end, Fig.~\ref{fig:PT} suggests to focus on processes that are sensitive to the coupling $\lambda_{221}$ at leading order. 

It is straightforward to see why $\lambda_{221}$ should be correlated with the strength of the phase transition (for singlet-like $h_2$). Higgs coupling measurements already restrict $\sin \theta$ to be small. In the small-$\theta$ limit, $h_1 \sim h$ and $h_2 \sim s$. If the singlet is to have any impact on the EWPT, it must do so via its couplings to the $h$. This singles out $\lambda_{211}$ and $\lambda_{221}$ at tree-level. However, 
\begin{equation}
\lambda_{211} \propto \sin \theta, \, \, \lambda_{221}\propto \cos\theta \, \, \, {\rm for} \, \sin\theta \ll 1,
\end{equation}
thus, in the small mixing angle limit, $\lambda_{221}$ must be non-negligible for $s$ to have an impact on the EWPT at tree-level. The singlet can also induce substantial radiative corrections to $\lambda_{111}$ in regions with a strong first-order EWPT, however these effects are typically subdominant to those of the tree-level couplings (i.e.~of $\lambda_{221}$). We will show this explicitly below, when we consider the impact of Higgs self-coupling measurements on the viable parameter space with a strong first-order EWPT.

One can also phrase this explanation in terms of the $\mathbb{Z}_2$-symmetric limit of the theory, considered in e.g.~Refs.~\cite{Curtin:2014jma, Craig:2014lda, Beniwal:2017eik}. In our parametrization, this corresponds to the limit $\sin\theta$, $b_3 \rightarrow 0$, and is thus a particular case of the model we are considering. In the exact $\mathbb{Z}_2$ limit, the only term coupling $s$ to $h$ in the scalar potential is
\begin{equation}
\frac{1}{2}a_2 \left|H\right|^2 S^2.
\end{equation}
Thus, if $s$ is to affect the strength of the EWPT, $a_2$ must be non-negligible. Since in this limit $\lambda_{221} = a_2 v/2$, and since the $\mathbb{Z}_2$ limit lies within the parameter space of the general singlet model at small mixing angle, we again conclude that $\lambda_{221}$ should be correlated with the strength of the EWPT at small $\sin \theta$.

These simple analytic arguments are confirmed by the results shown in Fig.~\ref{fig:PT}. While our reasoning is only formally correct in the limit $\sin\theta \ll 1$, Fig.~\ref{fig:PT} shows that this correlation persists for larger $|\sin\theta|$ as well. This motivates us to consider non-resonant pair production processes involving the singlet-like scalar in the final state.

\section{Comparison of Scalar Pair Production Modes at Colliders} \label{sec:nonres}

\begin{figure}[tb]
\begin{center}
\includegraphics[width=0.3\textwidth,clip]{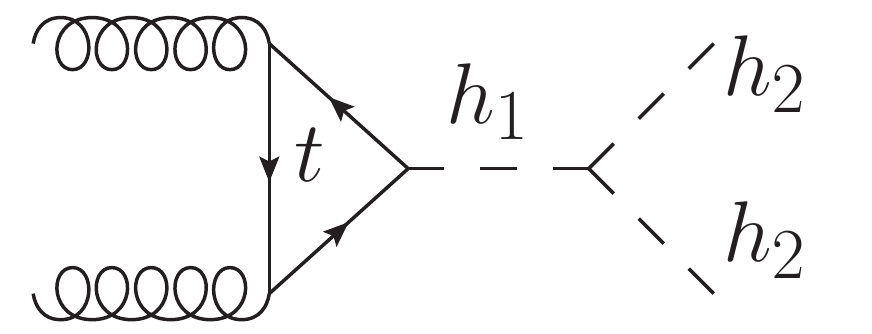}
\includegraphics[width=0.3\textwidth,clip]{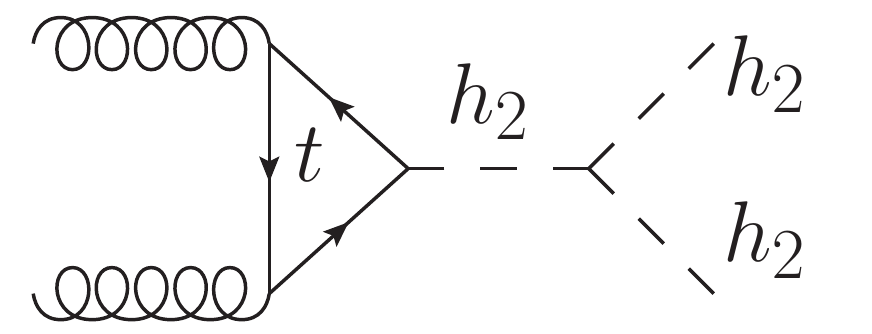}
\includegraphics[width=0.3\textwidth,clip]{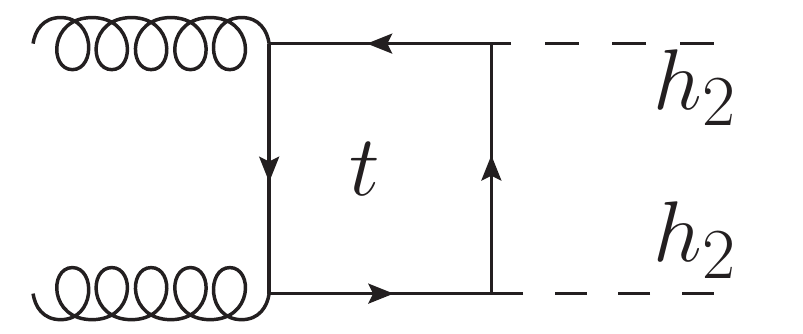}
\end{center}
\caption{Representative diagrams for $h_2h_2$ production via gluon fusion through top quark loops: (left) $s$-channel $h_1$, (center) $s$-channel $h_2$, and (right) box diagram.}
\label{fig:h2h2prod}
\end{figure}

The coupling $\lambda_{221}$ enters at leading order into the processes $pp \rightarrow h_1 h_2, \, h_2 h_2$. For example, the diagrams contributing to $h_2 h_2$ production are shown in Fig.~\ref{fig:h2h2prod}; the leftmost diagram contributes a term to the amplitude proportional to $\lambda_{221}$. Because of the different parametric dependence of the various triscalar couplings, $h_1 h_2$ and $h_2 h_2$ production can provide sensitivity to regions of the parameter space not covered by processes primarily dependent on the Higgs ($h_1$) self-coupling $\lambda_{111}$ or $\lambda_{211}$ alone. In this section, we make this observation more precise, putting aside for the moment the correlation with the EWPT.  We stress that, throughout this section, the trilinear scalar couplings are calculated at leading order. In some regions of parameter space higher order effects can be significant, as seen in Fig.~\ref{fig:initial_scan} and discussed further below.

\begin{figure}[tb]
\begin{center}
\includegraphics[width=0.3\textwidth,clip]{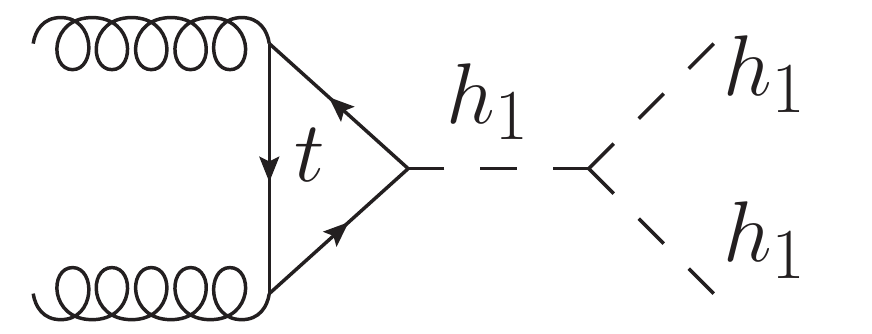}
\includegraphics[width=0.3\textwidth,clip]{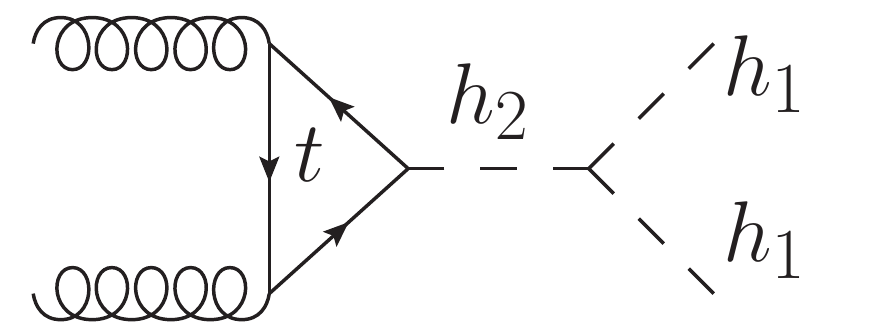}
\includegraphics[width=0.3\textwidth,clip]{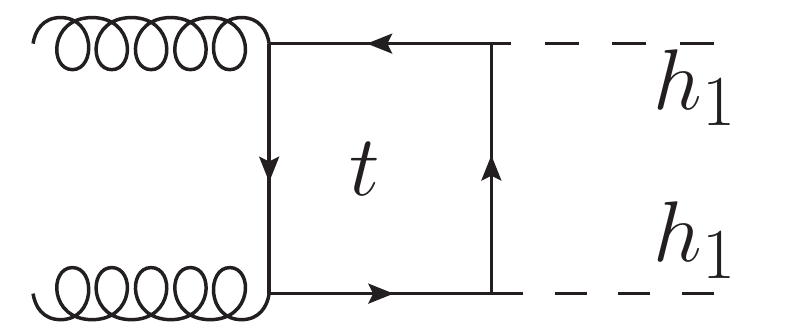}
\end{center}
\caption{Representative diagrams for $h_1h_1$ production via gluon fusion through top quark loops: (left) $s$-channel $h_1$, (center) $s$-channel $h_2$, and (right) box diagram.}
\label{fig:h1h1prod}
\end{figure} 
We consider the various non-resonant production cross-sections across the parameter space, scanning over all parameters of the model. We demand only that the potential be bounded from below at tree-level. Constraints such as requiring a strong first order phase transition and that the electroweak symmetry breaking minimum be the global minimum can be found by comparing to Fig.~\ref{fig:PT}. At each point we can compute the $h_1h_1$, $h_2 h_1$, and $h_2 h_2$ production cross-sections. The cross sections are generated by implementing our model into $\texttt{FeynArts}$~\cite{Hahn:2000kx} via $\texttt{FeynRules}$~\cite{Christensen:2008py,Alloul:2013bka} and using $\texttt{FormCalc}$~\cite{Hahn:1998yk}. We use the NNPDF2.3QED leading order~\cite{Ball:2013hta} parton distribution functions (pdfs) with $\alpha_s(M_Z)=0.119$.  These are implemented via LHAPDF~\cite{Buckley:2014ana}.  The factorization and renormalization scales, $\mu_f,\mu_r$, are both set to be the diboson invariant mass.  Our results are cross checked using $\texttt{HPAIR}$~\cite{Dawson:1998py}.   All cross sections are calculated at leading order at 14 TeV.  The results of this section are nearly identical for a 100 TeV proton proton collider.

\begin{figure}[tb]
\begin{center}
\includegraphics[width=0.45\textwidth,clip]{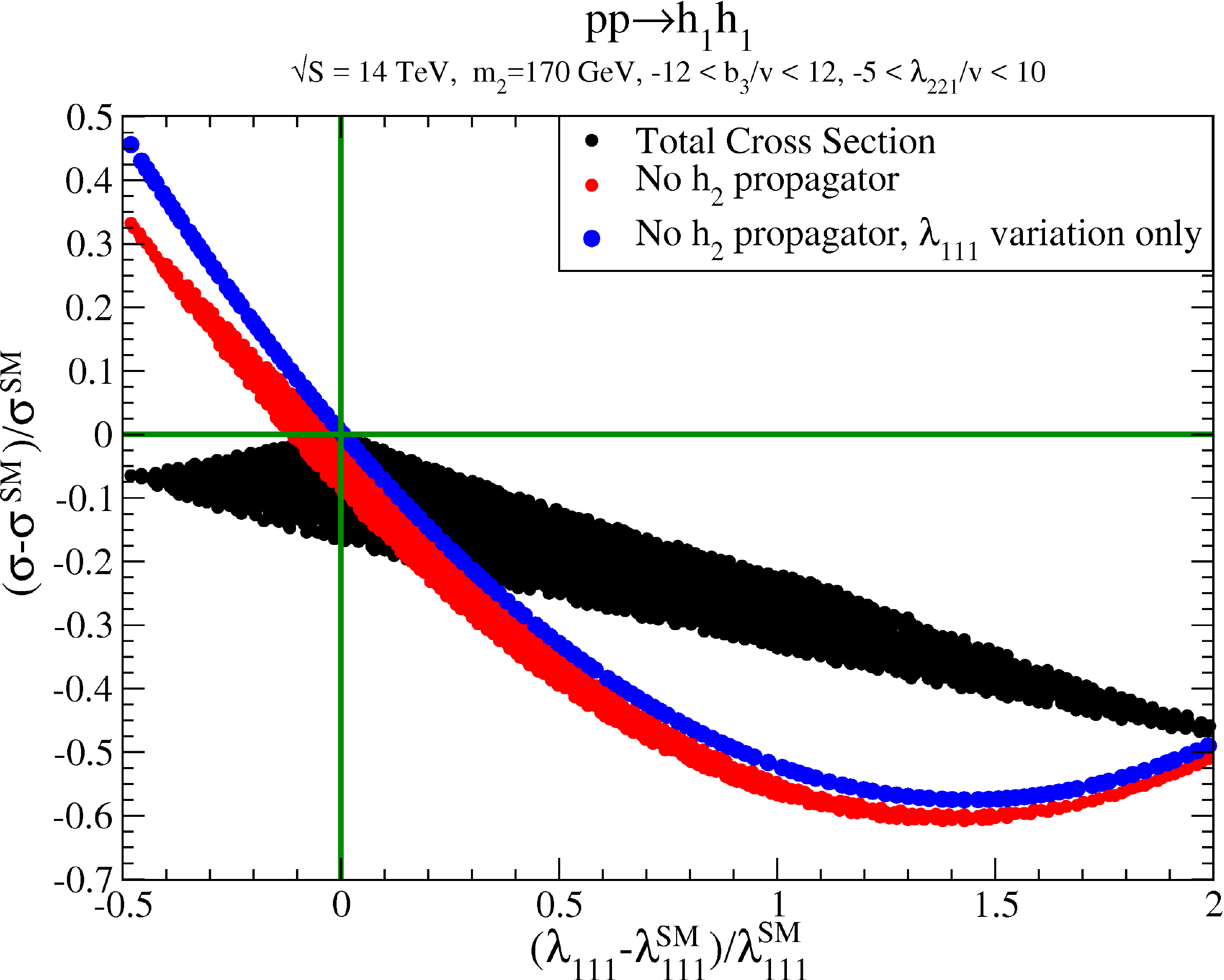}
\includegraphics[width=0.45\textwidth,clip]{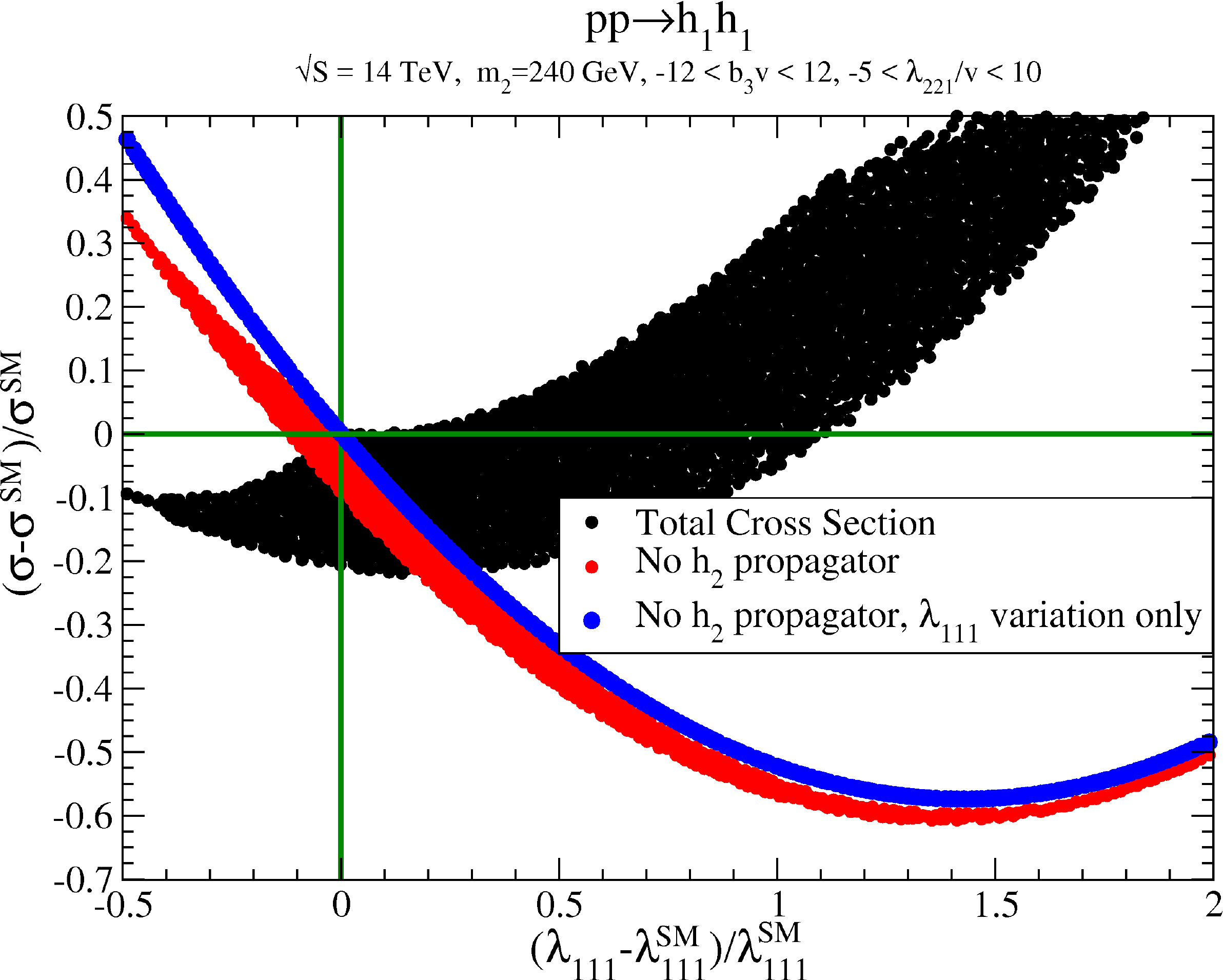}
\end{center}
\caption{\label{fig:h1h1} Fractional variation of $h_1h_1$ production cross section $\sigma$ and $\lambda_{111}$ away from the SM values denoted with superscript $SM$.  Total cross section considering all relevant diagrams (black dots), cross sections computed with $s$-channel $h_2$ propagators removed (blue dots), and cross sections considering only $\lambda_{111}$ variation with the top quark Yukawa fixed at the SM value and $s$-channel $h_2$ propagators removed (blue dots) are shown.  Two masses (left) $m_2=170$~GeV and (right) $m_2=240$~GeV are shown. The parameter region relevant of the strong first order EWPT [see Fig.~\ref{fig:PT}] is considered: $|\sin\theta|\le 0.35$, $-5<\lambda_{221}/v<10$ and $-12<b_3/v<12$.}
\end{figure}

\begin{figure}[tb]
\begin{center}
\includegraphics[width=0.45\textwidth,clip]{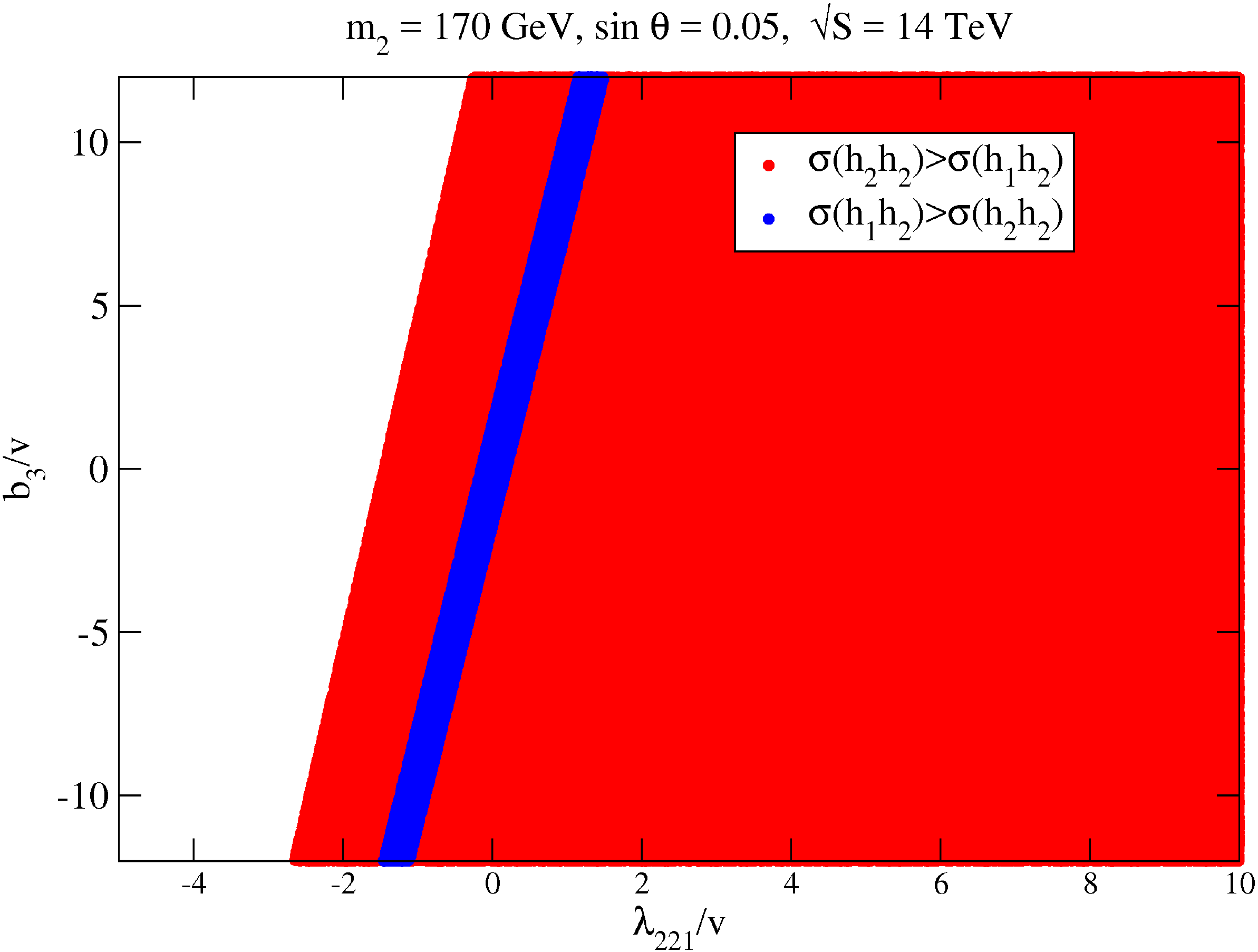}
\includegraphics[width=0.45\textwidth,clip]{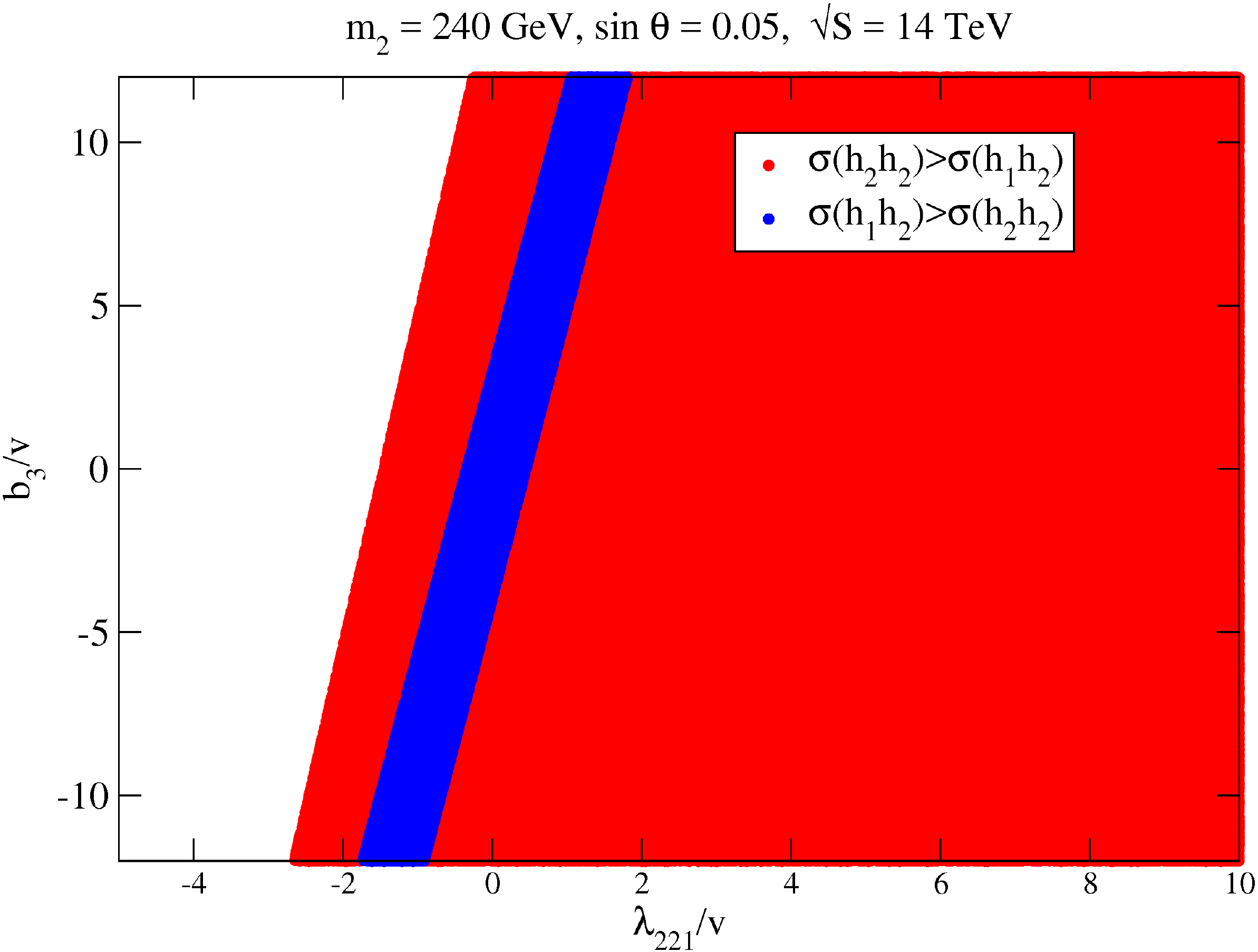}
\includegraphics[width=0.45\textwidth,clip]{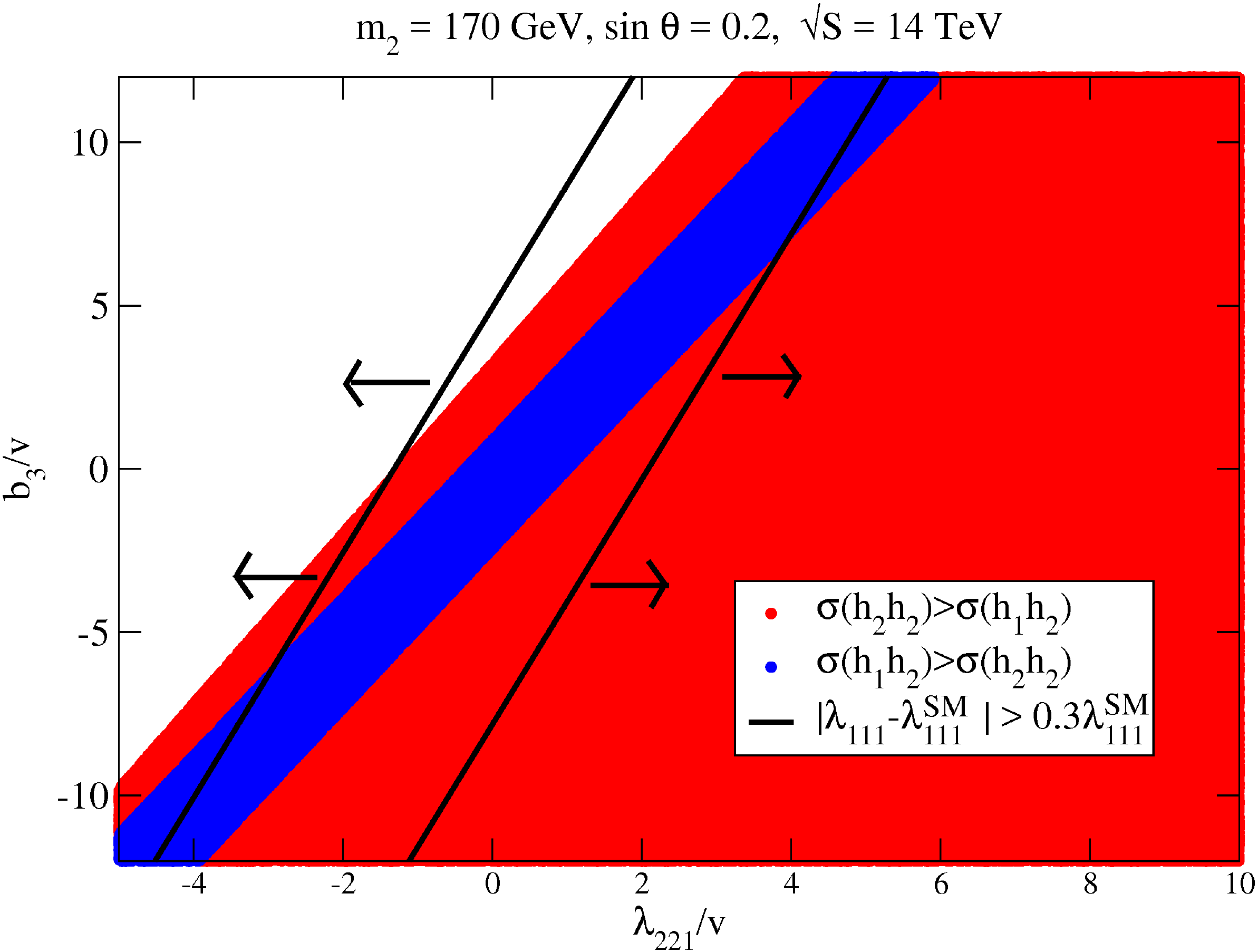}
\includegraphics[width=0.45\textwidth,clip]{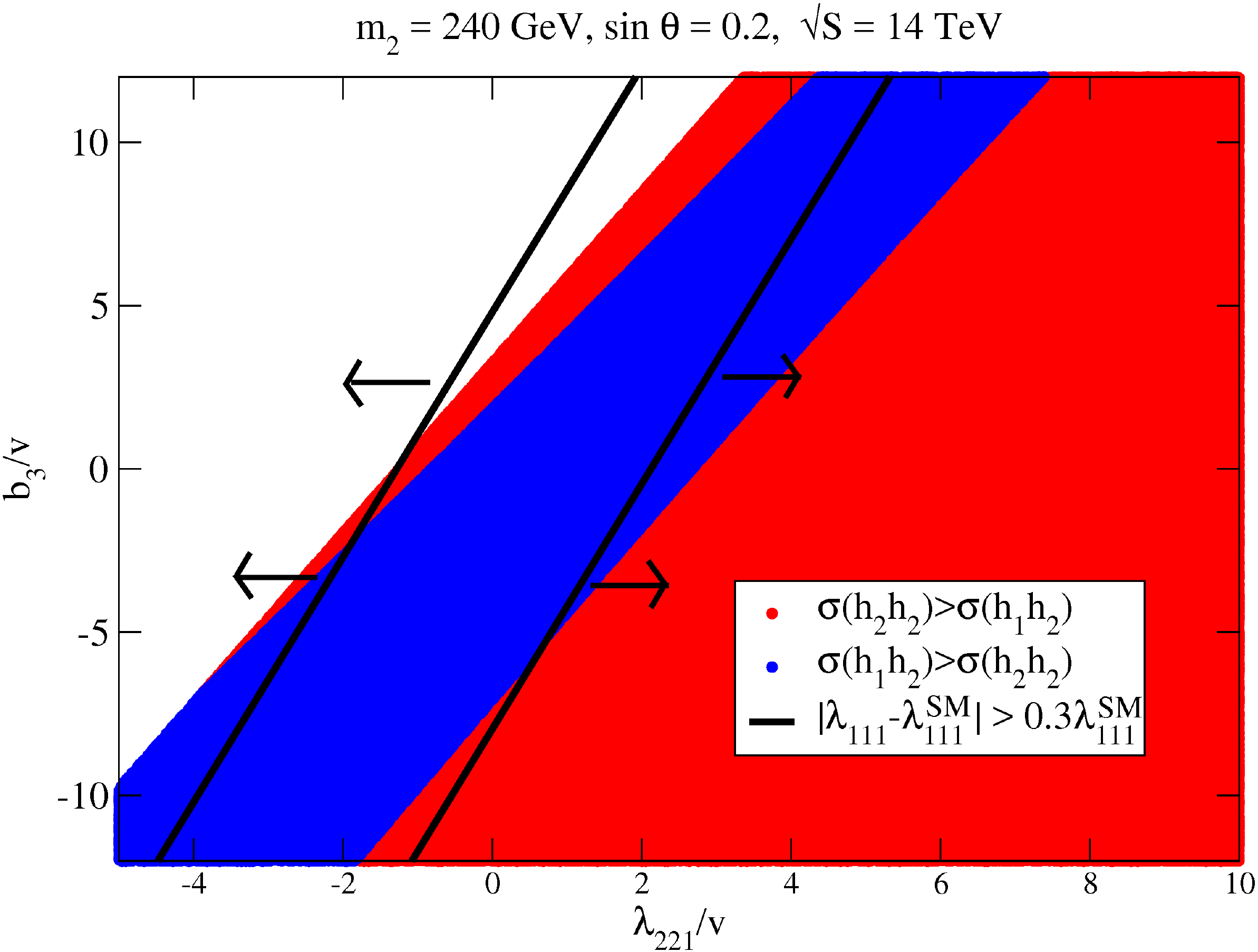}
\end{center} 
\caption{\label{fig:b3vslam221prod} Regions where (red) $\sigma(h_2h_2)>\sigma(h_1h_2)$, (blue) $\sigma(h_1h_2)>\sigma(h_2h_2)$, and (black) where $\lambda_{111}$ (computed at tree-level) is more than $30\%$ different from the SM prediction for (top) $\sin\theta=$0.05 and (bottom) $\sin\theta=$0.2 with (left) $m_2=170$~GeV and (right) $m_2=240$~GeV.  The black arrows indicate the regions for which $|\lambda_{111}-\lambda_{111}^{\rm SM}|>0.3\lambda_{111}^{\rm SM}$.}
\end{figure}

There are two main regions of interest: $m_2>2m_1$ where resonant $h_1h_1$ production is possible and $m_2<2m_1$ where only non-resonant production of $h_1h_1$ is allowed.  The purpose here is to determine in which regions of parameter space the different production modes are relevant. Equations for the partonic level cross section for diboson final states can be found in Appendix~\ref{sec:app_signal}, along with numerical formulas for the non-resonant hadronic cross sections.  These various final states have also been studied in~\cite{Dolan:2012ac}, with a different emphasis than ours.

 We first focus on the non-resonant $m_2<2m_1$ region.  Since we are interested in detecting new physics, we estimate the effect of measuring $h_1h_1$ production by using the fact that the LHC is expected to limit $\lambda_{111}$ to within $30-50\%$ of the SM value~\cite{Curtin:2014jma,Kling:2016lay}.  Although this may be optimistic~\cite{ATLAS,CMS:2015nat}, many other theory studies have found similar results~\cite{Baglio:2012np,Barger:2013jfa,Barr:2013tda,Lu:2015jza,Huang:2015tdv}.  Importantly, these studies consider only variations of the trilinear $\lambda_{111}$ coupling, while in the singlet model the scalar-top quark Yukawa couplings are suppressed by the scalar mixing angle and there is an additional $h_2$ propagator contributing to $h_1h_1$ production.  Representative Feynman diagrams for $h_1h_1$ production are shown in Fig.~\ref{fig:h1h1prod}. 

To investigate the importance of the various contributions to $h_1h_1$ production, in Fig.~\ref{fig:h1h1} we show the fractional deviation of the $h_1h_1$ cross section and $\lambda_{111}$ from the SM predictions considering (black dots) all relevant diagrams, (red dots) removing the $s$-channel $h_2$ diagrams, and (blue dots) considering only $\lambda_{111}$ variation with the $s$-channel $h_2$ diagrams removed and the top quark Yukawa coupling fixed to its SM value.  
 As can be seen, if only $\lambda_{111}$ variation is considered there is a direct correspondence between a limit on the $h_1h_1$ cross section and a limit on $\lambda_{111}$.  Now, if the top quark Yukawa coupling is allowed to change with the scalar mixing angle (red dots), this direct correspondence begins to break down.  Finally, if the total rate is calculated correctly with the $h_2$ propagator, the relationship between the cross section and $\lambda_{111}$ almost completely breaks down.  In fact, as can be clearly seen, requiring the $h_1h_1$ production rate to be within 50\% of the SM value is considerably less constraining when the cross sections are calculated correctly as opposed to only considering $\lambda_{111}$ variation.  Note that with a new propagator at a different mass than the SM Higgs boson, kinematic distributions may very well be more sensitive than total rate measurements, as has been shown for the SM case~\cite{Kling:2016lay,Huang:2015tdv}.  Nevertheless, constraining this model using a 30\% uncertainty on $\lambda_{111}$ is optimistic.  Despite this observation, we will assume $\lambda_{111}$ deviations up to 30\% can be measured and show that even in this optimistic scenario other double scalar production modes are still required for colliders to fully explore the relevant parameter region.

In Fig.~\ref{fig:b3vslam221prod} we show the regions where (red) $\sigma(h_2h_2)>\sigma(h_1h_2)$ and (blue) $\sigma(h_1h_2)>\sigma(h_2h_2)$ for (top) $\sin\theta=0.05$ and (bottom) $\sin\theta=0.2$ with (left) $m_2=170$~GeV and (right) $m_2=240$~GeV.  The black arrows indicate regions in which $\lambda_{111}$ deviates by 30\% or more from the SM prediction at tree-level.  The absence of black lines for $\sin\theta=0.05$ indicate that limits on the tree level $\lambda_{111}$ are not constraining in this parameter region (we will extend this discussion to include higher order effects below).  As can be clearly seen, precision measurements of the Higgs trilinear coupling $\lambda_{111}$ are not sufficient to be sensitive to all of the viable parameter space at leading order (see  Fig.~\ref{fig:PT}). Indeed, all three di-scalar final states, $h_1h_1$, $h_1h_2$, and $h_2h_2$, should be probed to fully explore the parameter space of the model.  This conclusion becomes more striking as the scalar mixing angle $|\sin\theta|$ decreases.  Comparing the top and bottom plots of Fig.~\ref{fig:b3vslam221prod}, it can be clearly seen that as $|\sin\theta|$ decreases the parameter regions that the $\lambda_{111}$ measurements are sensitive to shrink.  In particular, for $\sin\theta=0.05$ the $h_2h_2$ production mode is dominant in the majority of the relevant parameter region at leading order. This point will be seen again when comparing the sensitivity of $\lambda_{111}$ precision measurements to our findings regarding $h_2h_2$ production below.  

\begin{figure}[tb]
\begin{center}
\includegraphics[width=0.45\textwidth,clip]{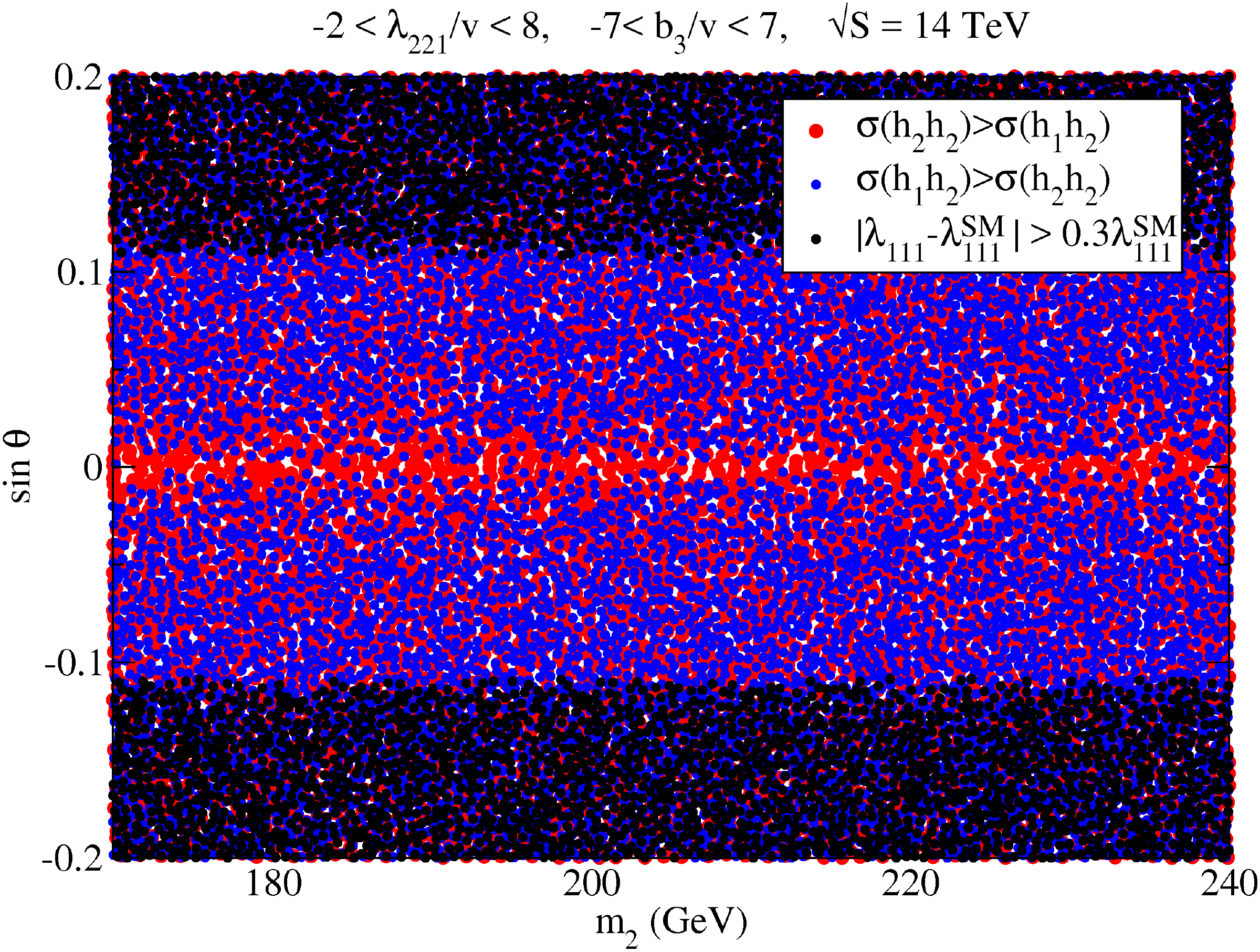}
\includegraphics[width=0.45\textwidth,clip]{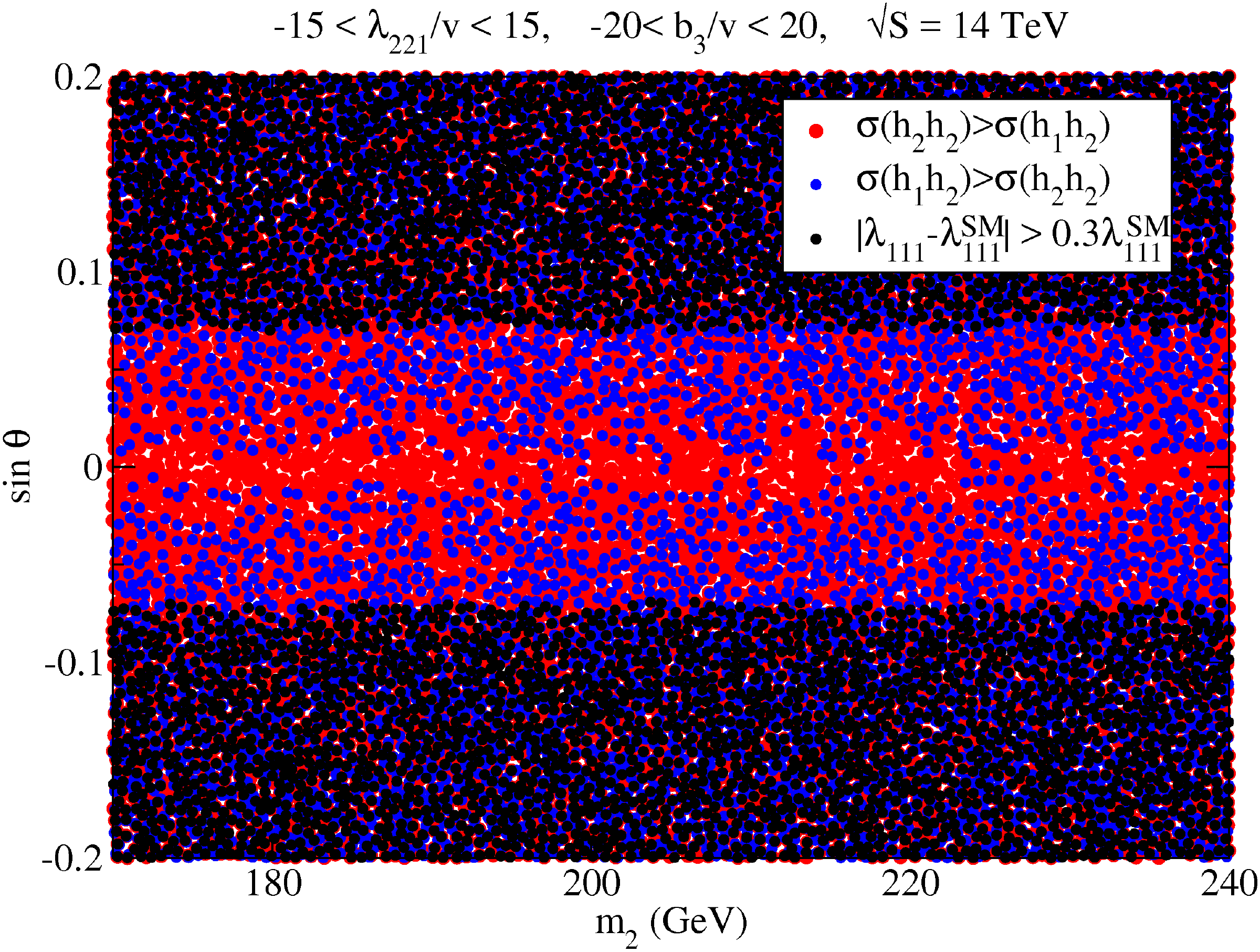}
\end{center}
\caption{\label{fig:sthvsm2prod} Regions where (red) $\sigma(h_2h_2)$, (blue) $\sigma(h_1h_2)$, (black) $\lambda_{111}$ is more than $30\%$ different from the SM prediction for parameter regions (left) $-2<\lambda_{221}/v<8$, $-7<b_3/v<7$ and (right) $-15<\lambda_{221}/v<15$, $-20<b_3/v<20$.  These results are in the $\sin\theta-m_2$ plane in the mass range where $h_1h_1$ production is non-resonant.}
\end{figure}

In Fig.~\ref{fig:sthvsm2prod} we again show results in the nonresonant region $m_2<2m_1$, this time in the $\sin\theta-m_2$ plane considering two parameter regions.  The left plot has $-2<\lambda_{221}/v<8$ and $-7<b_3/v<7$. This region satisfies all constraints from vacuum stability and a strong first order EWPT for perturbative values of $b_4$ as seen in Fig.~\ref{fig:PT}.  The right plot considers a wider region $-15<\lambda_{221}/v<15$ and $-20<b_3/v<20$.  Comparing these two plots allows us to determine how the vacuum stability and EWPT constraints affect the phenomenology of this model.  For (left) the parameter region relevant for EWPT and vacuum stability, the $\lambda_{111}$ measurement is only sensitive to mixing angles $|\sin\theta|\gtrsim0.12$ (at leading order).  For the (right) larger parameter region the $\lambda_{111}$ measurement is sensitive to $|\sin\theta|\gtrsim0.08$.  The requirement of a strong first order EWPT and vacuum stability makes the $\lambda_{111}$ measurement less relevant for small mixing angles. In fact, it has been shown that a strong first order electroweak phase transition is viable in the  $\sin\theta\rightarrow 0$ limit in which $h_2$ is stable~\cite{Curtin:2014jma}.  The implication of Fig.~\ref{fig:sthvsm2prod} is that to fully probe the parameter region consistent with a strong first order EWPT the $\lambda_{111}$ measurement does not appear to be sufficient.  Hence, it will be necessary to search for di-boson final states with $h_2$: $h_1h_2$ and $h_2h_2$.

\begin{figure}[tb]
\begin{center}
\includegraphics[width=0.45\textwidth,clip]{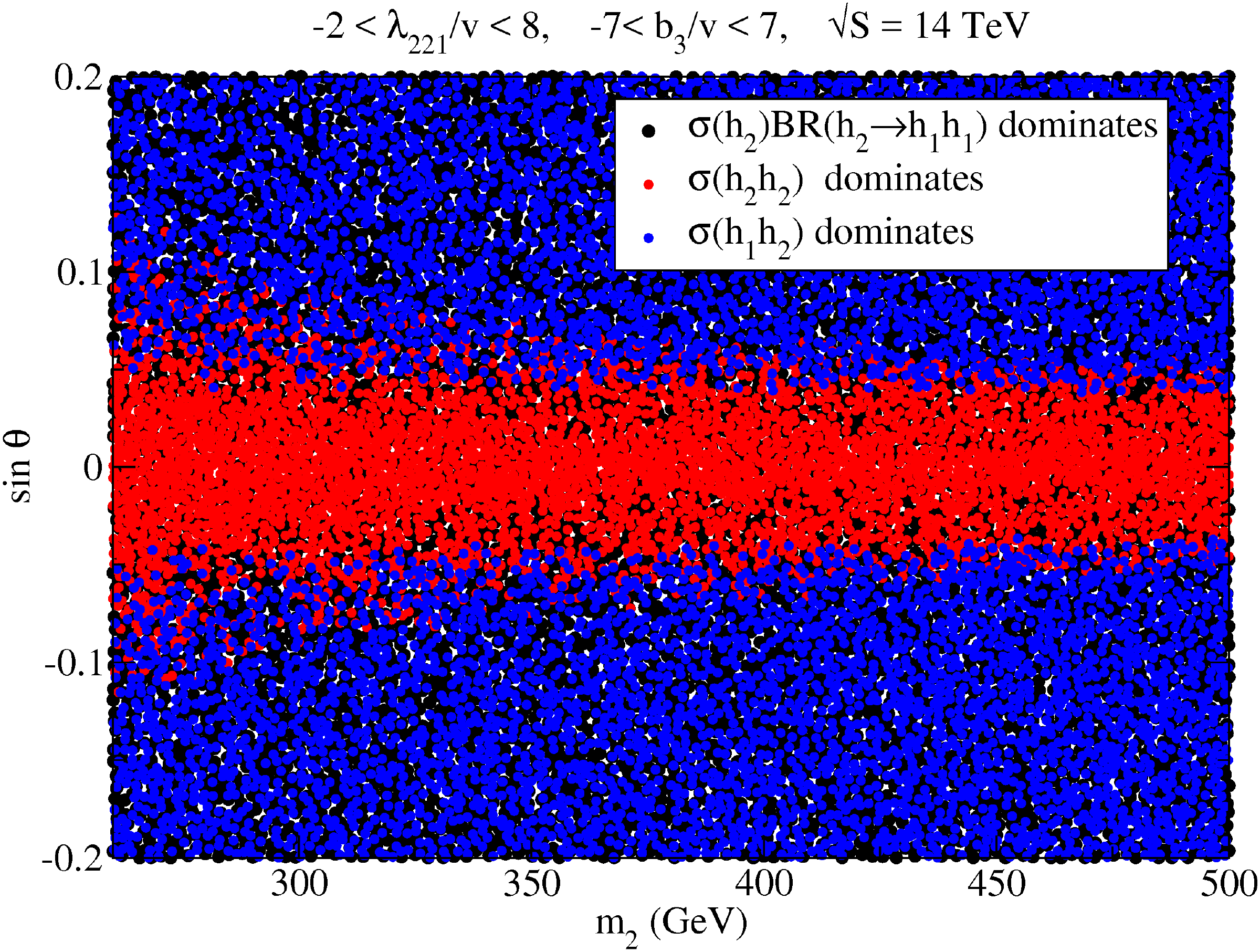}
\end{center}
\caption{\label{sthvsm2prodres} Regions where (red) $\sigma(h_2h_2)>\sigma(h_1h_2)$, (blue) $\sigma(h_1h_2)$, and (black) $\sigma(h_2){\rm BR}(h_2\rightarrow h_1h_1)$ are the largest production cross sections for the parameter region $-2<\lambda_{221}/v<8$, $-7<b_3/v<7$.  These results are in the $\sin\theta-m_2$ plane in the mass range where $h_1h_1$ production is resonant.}
\end{figure}

Finally, we consider the region $m_2>2m_1$, where resonant production of the $h_1h_1$ final state is possible.  In this case we compare the production cross sections of all diboson final states, while using the narrow width approximation for $h_1h_1$ production: $\sigma(h_1h_1)\approx \sigma(h_2)\times{\rm BR}(h_2\rightarrow h_1h_1)$.  In Fig.~\ref{sthvsm2prodres} we show where the cross sections of the various final states dominate: (red) $h_2h_2$, (blue) $h_1h_2$, and (black) $h_1h_1$.  Throughout this parameter space, there are points in which the resonant $h_1h_1$ production dominates.  However, even when resonant production of $h_1h_1$ is possible, there are points at small $\sin\theta$ for which $h_2h_2$ dominates and larger $\sin\theta$ where $h_1h_2$ dominates.  Hence, non-resonant $h_1h_2$ and $h_2h_2$ production can still be important {\it even when resonant $h_1h_1$ production is possible}.  However, a full collider study of each diboson production channel would be needed to determine which mode is most sensitive to the relevant parameter space.  

Summarizing the results of this section, Figs.~\ref{fig:b3vslam221prod}-\ref{sthvsm2prodres} clearly show that the various scalar production modes $h_1 h_1$, $h_1 h_2$, $h_2 h_2$ provide sensitivity to different regions of parameter space and hence are each deserving study. In particular, we see that of all these processes, $h_2 h_2$ production is the most sensitive to small mixing angles at leading order, which are difficult to probe via other means. It is also strongly dependent on $\lambda_{221}$, which is correlated with the strength of the phase transition. For the remainder of this study, we will therefore focus on non-resonant $h_2 h_2$ production. Given the sensitivity of resonant $h_1 h_1$ production for $m_2\geq 2m_1$, we will also restrict our attention to $m_2<2m_1$. We expect to study the other regions and production modes more thoroughly in a dedicated future study. 
\section{Probing Singlet-like Scalar Pair Production with Trileptons}\label{sec:trilepton}

We now investigate to what extent the LHC and a future 100 TeV collider can probe the electroweak phase transition in this model via non-resonant $h_2 h_2$ production. 

 For the purposes of this work we will consider $m_2 > 140$ GeV so that $h_2$ decays primarily to gauge bosons. For lower masses, a separate collider study is required to consider the viability of final states involving $b$'s, $\tau$'s, and photons. 

\subsection{Signal}

To reduce QCD and Drell-Yan backgrounds, we consider final states with leptons of the same charge (``same-sign leptons''). We will focus on the process
\begin{equation}
pp \rightarrow h_2 h_2 \rightarrow 4W \rightarrow 2j 2\ell ^{\pm} \ell ^{\prime \mp} 3\nu, \qquad \ell \neq \ell^{\prime}\label{eq:signal}.
\end{equation}
Similar topologies were considered before the Higgs discovery as a way of measuring the Higgs self-coupling~\cite{Baur:2002rb, Baur:2002qd, Blondel:2002nta}. As pointed out in these studies, as well as Ref.~\cite{Huang:2016cjm}, the $h_2 h_2 \rightarrow 4W \rightarrow 4j 2\ell ^{\pm} 2\nu$ channel can also be promising; it has a larger branching fraction, however the trilepton final state has the advantage of being less susceptible to backgrounds from fake leptons and tends to allow for larger signal-to-background ratios than the dilepton channel~\cite{Baur:2002qd}. 

We perform a Monte Carlo collider study of the trilepton channel for both the LHC and a future 100 TeV collider. To generate a signal event sample, we first implement this model with the top quark integrated out into $\texttt{Madgraph 5}$~\cite{Alwall:2014hca} using the $\texttt{FeynRules}$~\cite{Christensen:2008py,Alloul:2013bka} package.  This is the so-called Higgs effective theory and leads to dimension-5 and dimension-6 effective interactions $h_i G^{A,\mu\nu}G^{A}_{\mu\nu}$ and $h_i h_j G^{A,\mu\nu} G^{A}_{\mu\nu}$, where $G^{A}_{\mu\nu}$ are the gluon field strength tensors.  Here we use the default NNPDF2.3 leading order pdfs~\cite{Ball:2012cx} and $\texttt{Madgraph 5}$ dynamical scale choice.  The events generated in the effective theory are then reweighted using the exact leading order matrix elements.  These were generated by implementing the full model in $\texttt{FeynArts}$~\cite{Hahn:2000kx} using the $\texttt{FeynRules}$ package~\cite{Christensen:2008py,Alloul:2013bka}.  Code for the exact matrix elements was then generated using $\texttt{FormCalc}$~\cite{Hahn:1998yk}.  These results were cross checked using $\texttt{HPAIR}$~\cite{Dawson:1998py}.  The reweighted events are then fed into \texttt{Pythia 6}~\cite{Sjostrand:2006za} for parton showering and hadronization and then to \texttt{Delphes 3}~\cite{deFavereau:2013fsa} for detector simulation.
\subsection{Backgrounds}

The dominant backgrounds for the trilepton signature can be classified into two categories: those involving fake leptons and those with three prompt leptons.

\subsubsection{Backgrounds with Fakes}

Of all relevant backgrounds, by far the largest in LHC trilepton searches for final states without an opposite-charge same-flavor pair and with non-negligible MET (see e.g.~Refs.~\cite{Khachatryan:2014qwa, Chatrchyan:2014aea, Aad:2014nua}) is that arising from $t\bar{t}$, where both tops decay leptonically, no $b$-jets are tagged, and with one additional non-prompt (``fake'') lepton. The fake can arise, for example, from a heavy flavor meson decay or from the mis-reconstruction of light hadrons as leptons. Due to the very large $t\bar{t}$ cross-section at 14 and 100 TeV, this background will be the largest for our trilepton search, despite the typically small fake rates.

Modeling the fake lepton background with Monte Carlo is challenging because fakes are rare and the fake rate depends on complicated detector effects. To estimate this background, we use the \textit{FakeSim} method proposed in Ref.~\cite{Curtin:2013zua}. In particular, we generate a $t\bar{t}$ sample in \texttt{MadGraph 5}~\cite{Alwall:2014hca} (matched up to one additional parton), showering/hadronizing in \texttt{Pythia 6}~\cite{Sjostrand:2006za} and utilizing \texttt{Delphes 3}~\cite{deFavereau:2013fsa} for fast detector simulation.  We take the output and manually convert one jet to a lepton in each event, rescaling the event weights by an efficiency, $\epsilon_{j\rightarrow \ell}$. This efficiency has to be normalized to existing experimental studies. Additionally, fake events typically retain only a portion of the parent object's momentum, with the rest contributing to the total missing energy in the event. For each event in our $t\bar{t}$ sample, we choose $\alpha \equiv 1- p_T^{\rm fake}/p_T^j$ out of a truncated Gaussian distribution with mean $\mu = 0.5$ and variance $\sigma = 0.3$. We take $\epsilon_{j\rightarrow \ell}$ to be independent of $p_T$ ($r_{10}=1$ in the parametrization of Ref.~\cite{Curtin:2013zua}), and so the jet to convert is chosen randomly out of the event. 

To fix the expected value of $\epsilon_{j\rightarrow\ell}$ at the LHC, we match onto the results of the 13 TeV CMS trilepton search in Ref.~\cite{CMS:2017fdz}, which targets event topologies somewhat similar to ours. In particular, we normalize to the expected number of non-prompt background events in the 0-OSSF channel, which requires exactly three leptons with no opposite-sign same-flavor pair. Matching our Monte Carlo onto the SRB01 bin yields $\epsilon_{j\rightarrow\ell} \simeq 1 \times 10^{-3}$, which we will use for our study. This value closely reproduces the expected non-prompt background in the other bins. Comparing to the early ATLAS estimate in Ref.~\cite{Blondel:2002nta} for the trilepton signature very similar to our search, this choice for $\epsilon_{j\rightarrow\ell}$ predicts a $t\bar{t}$ background roughly a factor of 3 larger than reported. However the estimates of Ref.~\cite{Blondel:2002nta} were not based on data and utilized different isolation criteria. In any case, since our choice predicts a somewhat larger background than that of Ref.~\cite{Blondel:2002nta}, our results should yield a conservative estimate for the reach in the trilepton channel.

The efficiency and transfer function parameters at a future 100 TeV collider are of course unknown and depend on background modeling and the ability to discriminate prompt from non-prompt leptons at a future detector. To obtain an estimate of the expected $t\bar{t}$ background, we again take $\epsilon_{j\rightarrow\ell} =  10^{-3}$ as a representative value, with the same transfer function parameters as in our LHC analysis. Our overall conclusions do not significantly change in varying $\epsilon_{j\rightarrow\ell}$ by $\mathcal{O}(1)$ factors, and this estimate could be improved on with future dedicated study.

Note also that there can be other backgrounds involving fakes. In particular, $Z/\gamma^* (\rightarrow \tau^+ \tau^-)+$ jets where the taus decay leptonically can be an important background in trilepton searches. However, we find this contribution to be significantly suppressed in our case, due to our requirement (discussed below) of at least two additional hard jets reconstructing to the $W$ mass, significant MET, and our cuts on the variable $m_T^{\rm min}$. This is consistent with the discussions found in Refs.~\cite{Khachatryan:2014qwa, Chatrchyan:2014aea, Khachatryan:2016kod, CMS:2016gvu, CMS:2017fdz} for the LHC. 

\subsubsection{Processes with three prompt leptons}

There are several sources of prompt leptons predicted in the SM that contribute to the trilepton background. The most important are 
\begin{itemize}
\item $WZ/\gamma^*$ where the $Z/\gamma^*$ decays to taus which both decay leptonically, as does the $W$
\end{itemize}
and rare Standard Model processes involving three particles comprising:
\begin{itemize}
\item $WWW$ where all three gauge bosons decay leptonically
\item $t\bar{t} W$ where both $b$-jets are untagged and the tops and additional gauge boson decay leptonically
\item $t \bar{t} Z/\gamma^*$ where both $b$-jets are untagged, the tops and additional boson decay leptonically and one of the leptons from the $Z/\gamma ^*$ is missed, or where $Z/\gamma^*\rightarrow \tau^+ \tau-$, the taus decay leptonically, one of the tops decays leptonically, and the other hadronically, again with both $b$-jets missed
\item $t \bar{t} h_1$ where $h_1$ decays to $2\ell 2\nu$, one top decays leptonically, the other hadronically, and the $b$-jets are untagged.
\end{itemize}
Although subdominant to $t\bar{t}$, we will see that there can be choices of cuts for which these processes comprise a non-negligible background to our search. There can also be a contribution from $ZZ/\gamma^*/h_1$ where one boson decays to taus, the other to light leptons and one lepton is missed, however in existing LHC trilepton searches this background is typically smaller than or at most comparable to the rare SM backgrounds (which are already quite negligible compared to $t\bar{t}$) in signal regions without an opposite-charge same flavor lepton pair. We have verified that this is the case at both 14 and 100 TeV. The same is true of other backgrounds from photon conversions and lepton charge misidentification (see e.g.~\cite{Khachatryan:2014qwa, CMS:2017fdz}) and so we do not consider them further here.

We simulate the above backgrounds using $\texttt{Madgraph 5}$~\cite{Alwall:2014hca}. Events are then passed to \texttt{Pythia 6}~\cite{Sjostrand:2006za} for showering and hadronization and then to \texttt{Delphes 3}~\cite{deFavereau:2013fsa} for fast detector simulation. All backgrounds are generated at leading order. The $WWW$ background is matched to two additional partons using the MLM scheme~\cite{Mangano} while $WZ/\gamma$ is matched onto only one additional parton to speed up generation. 
 For all except $WZ/\gamma^*$ and $t\bar{t} Z/\gamma^*$, we exclude $\tau$ leptons from our parton-level background and signal event generation to improve the efficiency of the Monte Carlo given the large number of events required, especially for $t\bar{t}$. Their inclusion would affect both signal and background similarly and should not appreciably change our results.

We have cross-checked these backgrounds at parton level with those listed in Ref.~\cite{Baur:2002qd}, where applicable, and find reasonable agreement. We neglect the effect of pile-up throughout our analysis.

\subsection{Discriminating Signal from Background}

In order to reduce the large Standard Model backgrounds present in the trilepton channel, it is necessary to consider discriminating kinematic variables. In addition to our basic selection criteria (outlined below), we find that the quantities $m_{T2}$, $m_T^{\min}$, and $m_{\rm vis}$ can be useful in distinguishing signal from background.

The $m_{T2}$ variable we utilize is a simple generalization of the usual definition used in analyzing decays of heavy resonances with missing energy in the final state \cite{Lester:1999tx,Barr:2003rg}. For our signal we know that two opposite sign leptons will come from the decays of one $h_2$, with the other lepton and two jets (at parton level) from the second $h_2$. We can therefore form two $m_{T2}$ variables, $m_{T2}^{1,2}$, corresponding to the two possible ways of grouping the two jets with highest $p_T$ with one of the same-sign leptons (we only include the two highest-$p_T$ jets in our $m_{T2}$ variable).  We then define $m_{T2}\equiv \operatorname{Min}(m_{T2}^{1,2})$. We expect the corresponding differential distribution to peak around the $h_2$ mass for the signal and decrease significantly for larger values. 

Following Ref.~\cite{Khachatryan:2016kod}, we define $m_{T}^{\rm min}$ as 
\begin{equation}
m_T^{\rm min} \equiv \operatorname{Min}\left\{m_T(\ell_1, \slashed{E}_T), m_T(\ell_2, \slashed{E}_T), m_T(\ell_3, \slashed{E}_T) \right\}.
\end{equation}
This quantity can be useful in rejecting backgrounds with non-prompt leptons, since leptons not produced from $W$ decays will have a low kinematic endpoint for their $m_T$ distribution. 
We also consider the quantity $m_{\rm vis}$, defined via
\begin{equation}
m_{\rm vis}^2 \equiv \left| \sum_{\rm i} p^{\rm vis}_i \right|^2,
\end{equation}
which is simply the total visible invariant mass in the event (here $p^{\rm vis}_i$ are 4-momenta). The usefulness of this variable in non-resonant scalar pair production was pointed out in Ref.~\cite{Baur:2002qd}.

There are of course other kinematic quantities one can consider, however we find that these above, in addition to our basic selection criteria, can already provide reasonably good discrimination between signal and background.

\section{The LHC}\label{sec:LHC}

Let us first consider the prospects for observing non-resonant $h_2 h_2$ production at the LHC. To do so, we simulate the signal and backgrounds as described above, utilizing the default CMS detector card included in the \texttt{Delphes 3} distribution. For the signal, we once again consider two particular values of $m_2$: $m_2=170$, 240 GeV, both with $\sin\theta = 0.05$. All other parameters are scanned over, as described in Section~\ref{sec:param_space}.

Our basic selection and isolation criteria are similar to those of Ref.~\cite{Baur:2002qd} and are as follows: we require events to have at least 2 jets and exactly three identified leptons, with two leptons of the same charge and same flavor, with the other of opposite charge and flavor. Jets are defined as  having  $p_T>20$ GeV and  $|\eta|<5.0$, while identified leptons must have $p_T>10$ GeV and $|\eta|<2.5$. Furthermore, we require the two leading jets to have $p_T>30$ GeV, $|\eta|<3.0$, that all jet pairs satisfy $\Delta R(j_m,j_ n) > 0.6$, and that at least one jet pair reconstructs to the $W$ mass, with 50 GeV $< m_{jj}<$ 110 GeV.

 \begin{table}[!t]
\centering
 \begin{tabular}{c | c c }

 Process & Basic Selection & All cuts \\
 \hline
 \hline
  $t\bar{t}$ & 8144 & 159 \\
  $WZ/\gamma^*$ &147 & 8 \\
  Rare SM  & 141 & 9 \\
  Signal ($m_2 = 170$ GeV) & 250 & 96
\end{tabular}
\caption{\label{tab:LHC} Expected number of events at the 14 TeV LHC given 3000 fb$^{-1}$ for the dominant backgrounds and signal optimization point ($\sin \theta = 0.05$, $a_2 = 8.5$, $b_3=0$) after applying our basic selection criteria and after applying all cuts. }
\end{table}

In addition to these criteria, we optimize cuts for each value of $m_2$ by choosing a particular point in the $a_2 - b_3$ plane and scanning over the boundaries of the cuts for the $m_{T2}$, $m_T^{\rm min}$,  $m_{\rm vis}$ variables, along with the $p_T$ requirements for the two leading jets and the leptons, selecting the cuts that maximize $S/\sqrt{S+B}$ while maintaining $>10$ events at 3000 fb$^{-1}$. For $m_2=170$ GeV, we optimize our cuts for $a_2=8.5$, $b_3 = 0$, which corresponds to the point with the largest value of $\lambda_{221}$ consistent with a strong first-order PT (the $h_2 h_2$ production cross-section is independent of $b_4$). Note that these couplings are large, and so caution should be applied in interpreting our perturbative phase transition and stability analysis for this particular point. The resulting requirements we obtain for $m_2=170$ GeV are:
\begin{equation}
p_T^{j_1,j_2}>30 \, {\rm GeV}, \quad p_T^{\ell_1,\ell_2}>25 \, {\rm GeV}, \quad m_{T2}<150 \,{\rm GeV}, 
\end{equation}
\begin{equation*}
m_{\rm vis} < 700 \, {\rm GeV}, \quad m_{T}^{\rm min}>20 \,{\rm GeV}.
\end{equation*}
We also require  $\slashed{E}_T>30$ GeV, since we expect missing energy from neutrinos in the final state, and the third lepton to satisfy $p_T^{\ell_3}>20$ GeV. Specific values for the expected number of signal and background events after basic selection and once all cuts are applied are provided in Table~\ref{tab:LHC}, and some details of the relevant kinematic distributions are provided in Appendix~\ref{sec:app_kin}.

\begin{figure}[!t]
\centering
\includegraphics[width=.45\textwidth]{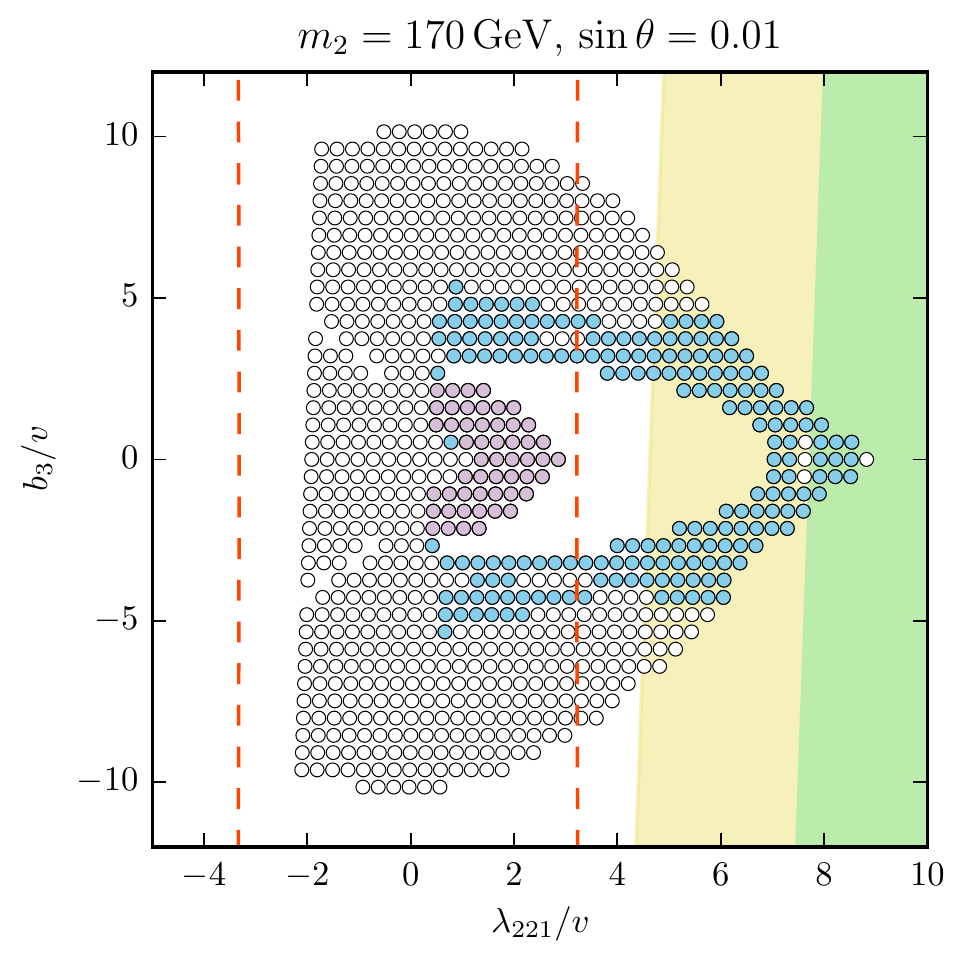}\,\includegraphics[width=.45\textwidth]{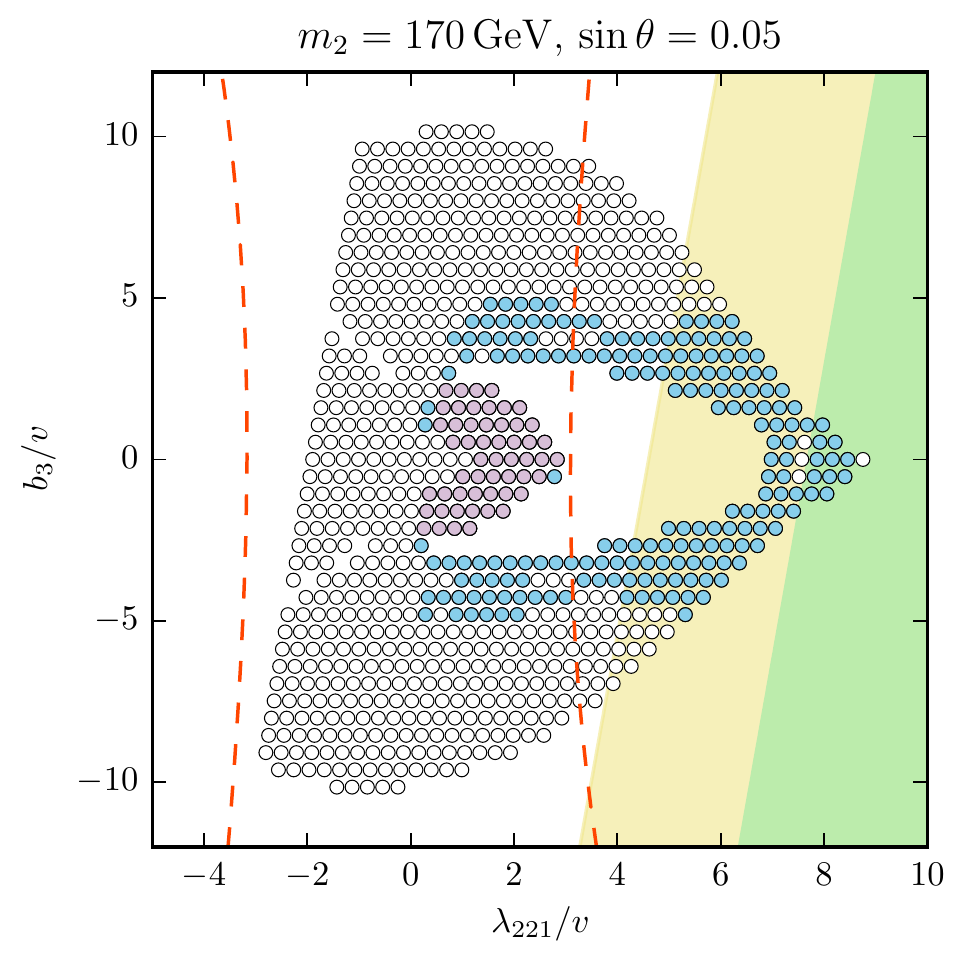}\\
\includegraphics[width=.45\textwidth]{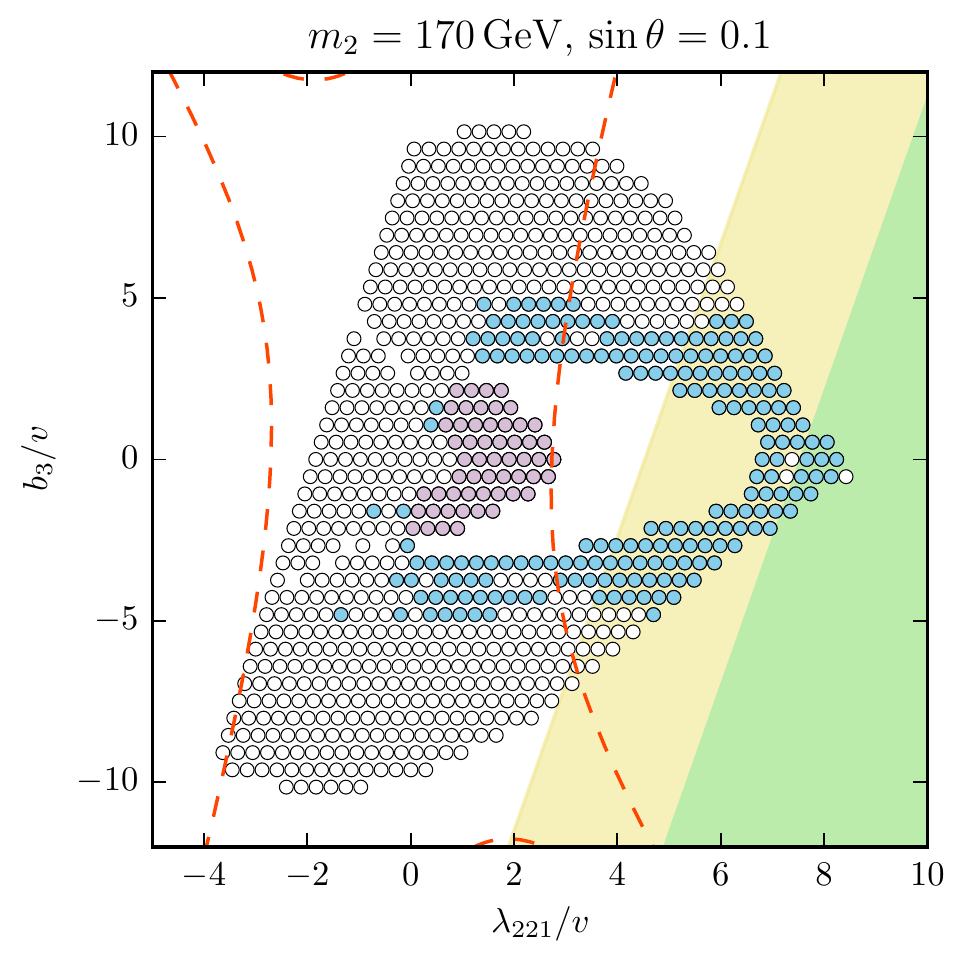}\,\includegraphics[width=.45\textwidth]{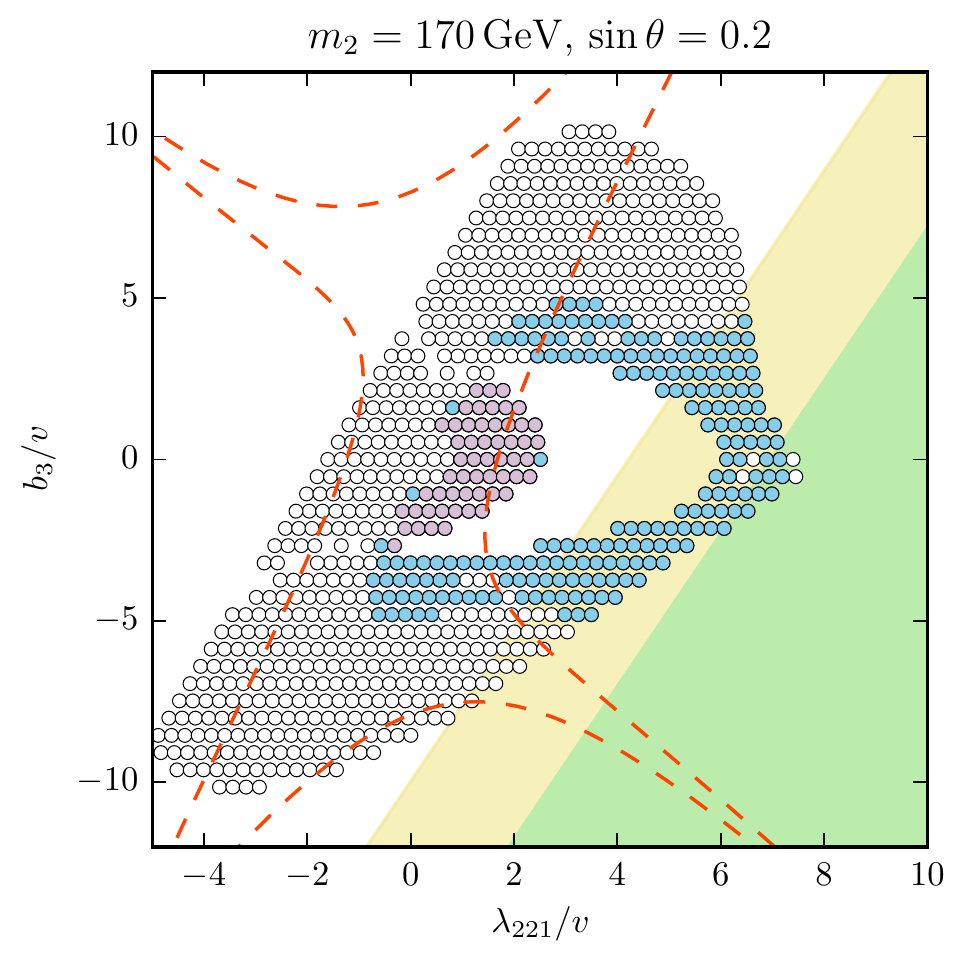}
\caption{\label{fig:LHC} Exclusion (yellow) and discovery (green) reach in the $h_2 h_2\rightarrow 2j 2\ell ^{\pm} \ell ^{\prime \mp} 3\nu$ channel at the high-luminosity LHC with 3000 fb$^{-1}$ for $m_2=170$ GeV. The coloring of the points corresponds to that used in Fig.~\ref{fig:PT}. Also shown is the approximate expected sensitivity of $h_1$ self-coupling measurements at the HL-LHC to the parameter space, given  3000 fb$^{-1}$ integrated luminosity: regions between the dashed contours would \emph{not} be probed by $h_1 h_1$ production at the LHC. Here, $\lambda_{111}$ is computed using Eq.~\ref{eq:L111_loop}, which includes the leading higher-order corrections. The purple region, for which our phase transition results are the most robust, will likely not be accessible to the channels we have considered at the LHC.}
\end{figure}

The impact of the high-luminosity LHC at 14 TeV given 3000 fb$^{-1}$ on the parameter space of interest is shown in Fig.~\ref{fig:LHC}. The yellow region  features $S/\sqrt{S+B}\geq 2$ and would roughly correspond to a 2$\sigma$ exclusion limit. The green region  features $S/\sqrt{S+B} \geq 5$ and would roughly correspond to a 5$\sigma$ discovery. In all of the shaded regions, $S/B>0.1$, and the reach is statistics limited\footnote{There are more sophisticated methods to calculate exclusion and discovery in the presence of systematic uncertainties~\cite{Cowan,Cowan:2010js,Kumar:2015tna}.  However, for for simplicity we adopt the prescription that the systematic uncertainties are subdominant for $S/B\gtrsim 0.1$.}. In terms of the total $h_2 h_2$ production cross-section,  the high-luminosity LHC should have sensitivity to regions of parameter space with 
\begin{equation}
\begin{aligned}
\sigma_{h_2 h_2} &\gtrsim 53 \, {\rm fb} \quad (2\sigma), \quad 147 \, {\rm fb} \quad (5\sigma)
\end{aligned}
\end{equation}
in the trilepton channel for $m_2 = 170$ GeV. From Fig.~\ref{fig:LHC} it is clear that, although the LHC can be sensitive to some points with large $\lambda_{221}$, much of the parameter space consistent with a strong EWPT would remain inaccessible by this channel even at 3000 fb$^{-1}$. The value of the excludable cross-section quoted above is considerably larger than the dilepton sensitivity estimate in Ref.~\cite{Huang:2016cjm}. We believe that this is due to the smaller trilepton branching ratio and the considerably lower signal efficiencies we have found in our collider study than those assumed in the estimate of Ref.~\cite{Huang:2016cjm}. It is also possible that a more sophisticated multivariate analysis performed by the LHC experimental collaborations could significantly improve our projected sensitivity. Note also that our fake rate estimate may be overly pessimistic. We therefore expect the results shown to represent a conservative estimate of the reach.

For $m_2 = 240$ GeV, virtually all of the viable parameter space consistent with a strong first-order PT features less than $\sim 10$ $h_2 h_2 \rightarrow 2j 2\ell ^{\pm} \ell ^{\prime \mp} 3\nu$ events at 3000 fb$^{-1}$ after applying our basic selection criteria. We thus conclude that non-resonant scalar pair production in the trilepton channel will be unable to probe $m_2 \gtrsim 240$ GeV at the high luminosity LHC.

\subsection{Additional Probes} \label{sec:indirect_LHC}

As mentioned above, there are additional measurements that can provide experimental sensitivity to the parameter space of the model consistent with a strong first-order electroweak phase transition~\cite{Curtin:2014jma}. The most important are measurements of the pair production cross-section for two Standard Model-like Higgses ($h_1 h_1$) at hadron colliders (discussed in Sec.~\ref{sec:nonres}), as well as measurements of the $Zh_1$ production cross-section at future lepton colliders~\cite{Craig:2013xia, Curtin:2014jma}.

In scenarios where the only new contribution to double-Higgs production arises from modifications to the Higgs self-coupling, the high-luminosity LHC is expected to be able to constrain $\sim 30-50$\% modifications of the SM triple Higgs coupling at 3000 fb$^{-1}$~\cite{Curtin:2014jma}, as discussed in Sec.~\ref{sec:nonres}. For small values of $|\sin \theta|$, radiative corrections to $\lambda_{111}$ dominate the corrections to the SM $h_1h_1$ production cross-section, as opposed to the leading order mixing angle effects reflected in Figs.~\ref{fig:b3vslam221prod}-\ref{fig:sthvsm2prod} (which are important for larger $|\sin\theta|$). The 1-loop correction to the $h_1$ trilinear self-coupling can be written, to $\mathcal{O}(\sin \theta)$, as
\begin{equation} \label{eq:L111_loop}
\Delta \lambda_{111}^{\rm 1-loop} =  \frac{1}{16\pi^2}\left(\frac{1}{2 m_2^2} a_2^3 v^3 + 27 \frac{m_1^4}{v^3} + \frac{3}{ m_2^2} a_2^2 b_3 v^2 \sin\theta + \mathcal{O}(\sin^2\theta) \right)
\end{equation} 
We approximate the regions accessible to $h_1 h_1$ production cross-section measurements as those for which $|(\lambda_{111}+\Delta \lambda^{\rm 1-loop}_{111})-\lambda_{111}^{\rm SM}|/\lambda_{111}^{\rm SM} > 30$\%, with $\Delta \lambda_{111}^{\rm 1-loop}$ computed to $\mathcal{O}(\sin \theta)$, as in Eq.~\ref{eq:L111_loop}. The corresponding regions for $m_2=170$ GeV lie outside of the dashed red contours in Fig.~\ref{fig:LHC}. Note that, as discussed previously for the leading order result, this is likely an optimistic estimate of the reach, given the additional diagram and Yukawa suppression that contribute to the $h_1 h_1$ amplitude relative to models in which new physics only alters the SM Higgs self-coupling.

The presence of an additional singlet scalar also alters the $Zh_1$ production cross-section relative to its SM expectation, by virtue of new contributions to the wave-function renormalization of the Standard Model-like Higgs~\cite{Craig:2013xia}, as well as mixing angle suppression of the $h_1$ couplings to gauge bosons. The corresponding fractional shift of the $Zh_1$ production cross-section is approximately
\begin{equation} \label{eq:Zh}
\delta_{Zh} \approx - \sin^2\theta + \frac{\lambda_{221}^2}{32 \pi^2 m_1^2}\left[1+F(\tau_s) \right],
\end{equation}
where we have dropped higher-order terms in $\sin \theta$ and the couplings,
\begin{equation}
F(\tau) = - \frac{\sin^{-1}(\sqrt{\tau})}{\sqrt{\tau (1-\tau)}},\label{eq:F}
\end{equation}
and $\tau_s \equiv m_1^2/(4 m_2^2)$. A derivation of this result is presented in Appendix~\ref{sec:app_Zh}.  Refs.~\cite{Dawson:2013bba, Curtin:2014jma, Fujii:2015jha, dEnterria:2016fpc} suggest that deviations from the SM $Zh_1$ production section, $\delta_{Zh}$, of order $\sim 7\%$ should be accessible to the high-luminosity LHC. In all of the parameter space shown in Fig.~\ref{fig:LHC}, $|\delta_{Zh}|<5\%$, and so we conclude that the HL-LHC will not be able to probe any of the EWPT-compatible regions considered through this channel. Future lepton colliders will likely be able to measure the $Zh_1$ production cross-section much more accurately. We consider the corresponding prospects when discussing future colliders below. 

\subsection{Summary of LHC results}

Fig.~\ref{fig:LHC} demonstrates that, while non-resonant $h_2 h_2$ production at the high luminosity LHC will likely be sensitive to some of the parameter space with large $\lambda_{221}$ and low $m_2$, measurements of the $h_1 h_1$ production cross-section are expected to provide better sensitivity to regions with a strong first-order EWPT. Neither channel is likely to probe the region in which the gauge-invariant high-$T$ approximation predicts a strong phase transition. However, we stress that considering additional $h_2h_2$ final states and/or achieving better discrimination between prompt and non-prompt leptons than we have assumed will likely improve the prospects for non-resonant singlet-like scalar pair production. 

While we have shown results down to $\sin \theta = 0.01$, our conclusions remain essentially unchanged for smaller mixing angles (provided $h_2$ decays promptly). As we increase $m_2$, the $h_2 h_2$ reach quickly decreases, with the high-luminosity LHC unlikely to probe any of the parameter space with $m_2\gtrsim 240$ GeV in the trilepton final state. When $m_2>250$ GeV, additional sensitivity will be provided by resonant $h_1 h_1$ production, however this channel will also become ineffective at small $\sin \theta$.

\section{100 TeV Collider}\label{sec:100TeV}

As we have seen, observing non-resonant scalar pair production LHC can be difficult, even at high luminosity. We now move on to consider the situation at a future 100 TeV collider.  The prospects at 100 TeV are more encouraging, in part due to the larger center-of-mass energy, as well as to likely improvements in tracking and detector technology that would be incorporated into such a machine.

 \begin{table}[!t]
\centering
 \begin{tabular}{c | c c c}

 Process & Basic Selection & All cuts: $m_2=170$ GeV & $m_2=240$ GeV \\
 \hline
 \hline
  $t\bar{t}$ & 498049 & 6486 & 1869 \\
  $WZ/\gamma^*$ & 40134 & 284 & 142 \\
  Rare SM & 34288 & 332 & 306 \\
  $m_2 = 170$ GeV & 7411 & 1311 & - \\
  $m_2 = 240$ GeV & 1024 & - & 112 
\end{tabular}
\caption{\label{tab:FCC} Expected number of events at a future 100 TeV collider given 30 ab$^{-1}$ for the dominant backgrounds and signal optimization points ($a_2 = 2$, $b_3=0$ for $m_2= 170$ GeV; $a_2=3.5$, $b_3=0$ for $m_2=240$ GeV, both with $\sin\theta = 0.05$) after applying our basic selection criteria and after all cuts.}
\end{table}

Once again, we simulate signal and background events as discussed above in Sec.~\ref{sec:nonres}. Although there is no detector design in place for a 100 TeV collider, detector effects can be important in realistically assessing the viability of our proposed searches. We thus pass showered and hadronized events to  \texttt{Delphes 3} for a fast detector simulation. We use the default FCC-hh detector card included in the \texttt{Delphes 3} distribution. This detector configuration features tracking out to $\eta=4$ as well as higher assumed tracking efficiencies, resulting in an improved signal efficiency as compared to the LHC.

The basic selection and isolation criteria we use are the same as those listed in Sec.~\ref{sec:LHC}, except that the lepton identification criterion is extended to the forward region with $|\eta|<4.0$. To further reduce the backgrounds, we proceed as before and perform a simple scan over the cut boundaries for specific parameter space points. We optimize the cuts taking $a_2 = 2$, $b_3 = 0$, $\sin\theta = 0.05$ for $m_2 = 170$ GeV and maximizing $S/\sqrt{S+B}$. For $m_2=240$ GeV with $\sin\theta=0.05$, we take $a_2 = 3.5$, $b_3=0$ and optimize cuts by maximizing $S/B$, since the cross-section is considerably smaller. Following this procedure, we arrive at the following set of requirements for each mass:
\begin{itemize}
\item $m_2 = 170$ GeV:
\begin{equation}
p_T^{j_1,j_2}>30 \, {\rm GeV}, \quad p_T^{\ell_1,\ell_2}>25 \, {\rm GeV}, \quad m_{T2}<150 \,{\rm GeV}, 
\end{equation}
\begin{equation*}
m_{\rm vis} < 600 \, {\rm GeV}, \quad m_T^{\rm min} > 60 \, {\rm GeV}
\end{equation*}

\item $m_2 = 240$ GeV:
\begin{equation}
p_T^{j_1,j_2}>50 \, {\rm GeV}, \quad p_T^{\ell_1,\ell_2}>40 \, {\rm GeV}, \quad m_{T2}<200 \,{\rm GeV}, 
\end{equation}
\begin{equation*}
m_{\rm vis} < 400 \, {\rm GeV},\quad m_T^{\rm min} > 80 \, {\rm GeV}.
\end{equation*}
\end{itemize}
In all cases, we also require $p_T^{\ell_3} > 20$ GeV and $\slashed{E}_T > 50$ GeV. The expected number of background and signal events given our basic selection criteria and the cuts listed above\footnote{In some instances for the rare SM backgrounds we consider, our cut-and-count analysis yields no expected events. In these cases, we conservatively estimate the impact of the corresponding background by including the background's unit weight multiplied by a factor of 3. This would correspond to the largest cross-section expected to yield 0 events with $\gtrsim$ 5\% probability in a counting experiment, assuming Poisson statistics. } are given in Table~\ref{tab:FCC}.

\begin{figure}[!t]
\centering
\includegraphics[width=.45\textwidth]{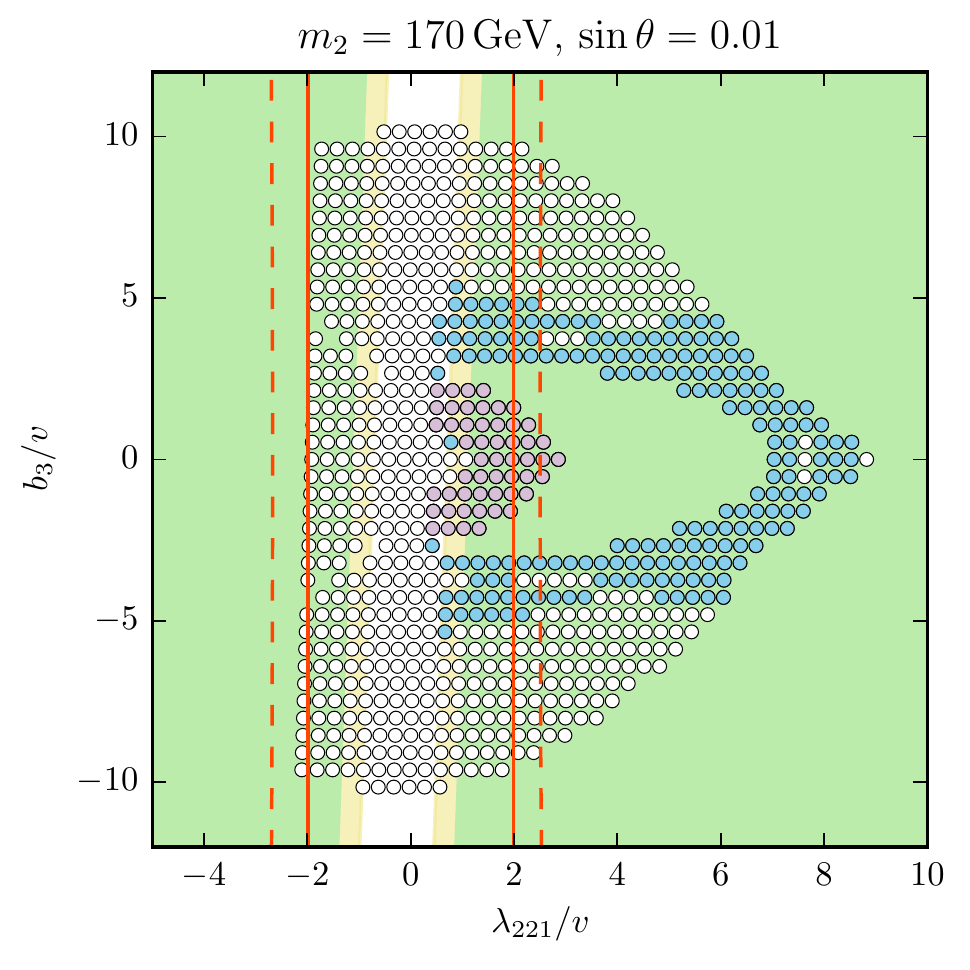}\,\includegraphics[width=.45\textwidth]{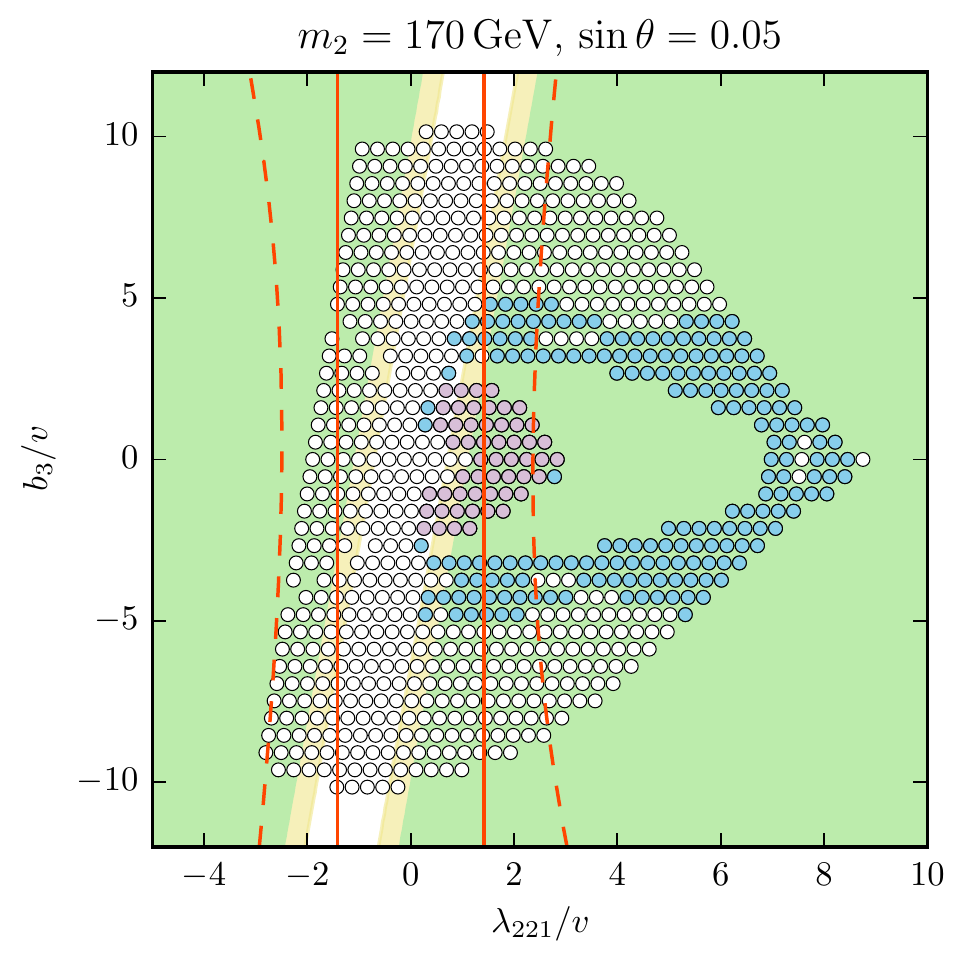}\\
\includegraphics[width=.45\textwidth]{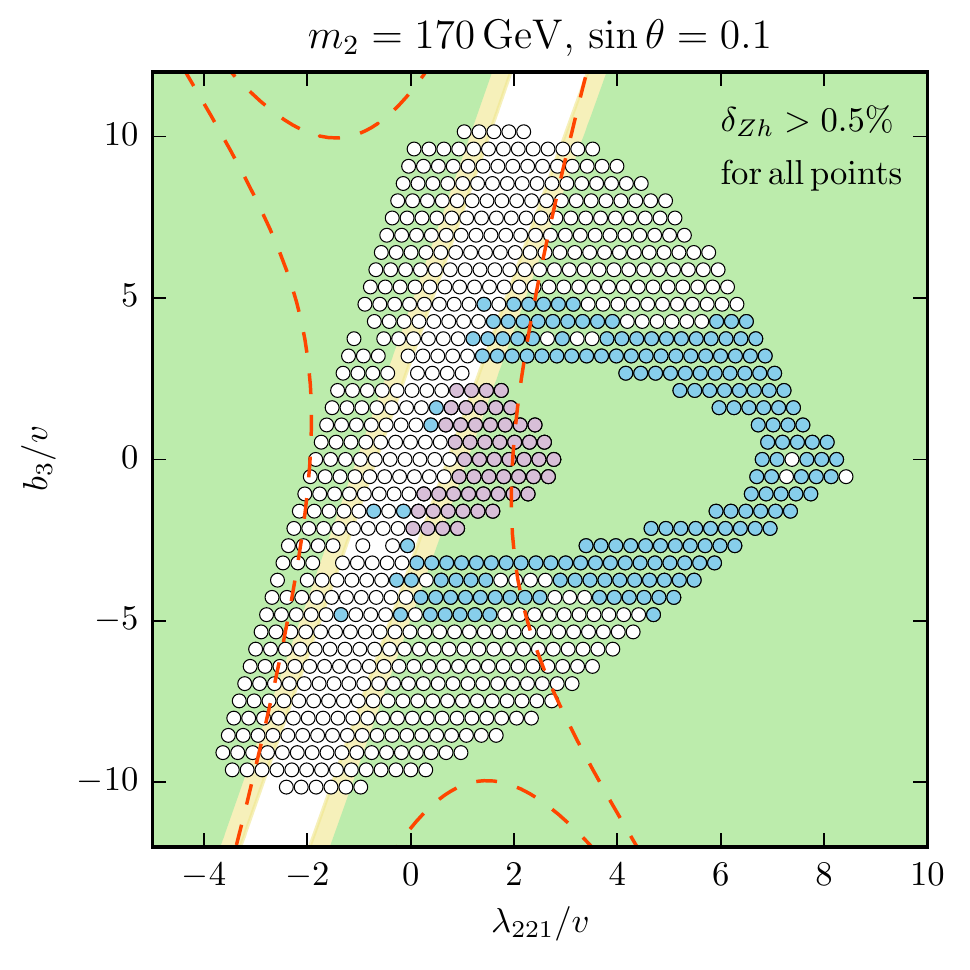}\,\includegraphics[width=.45\textwidth]{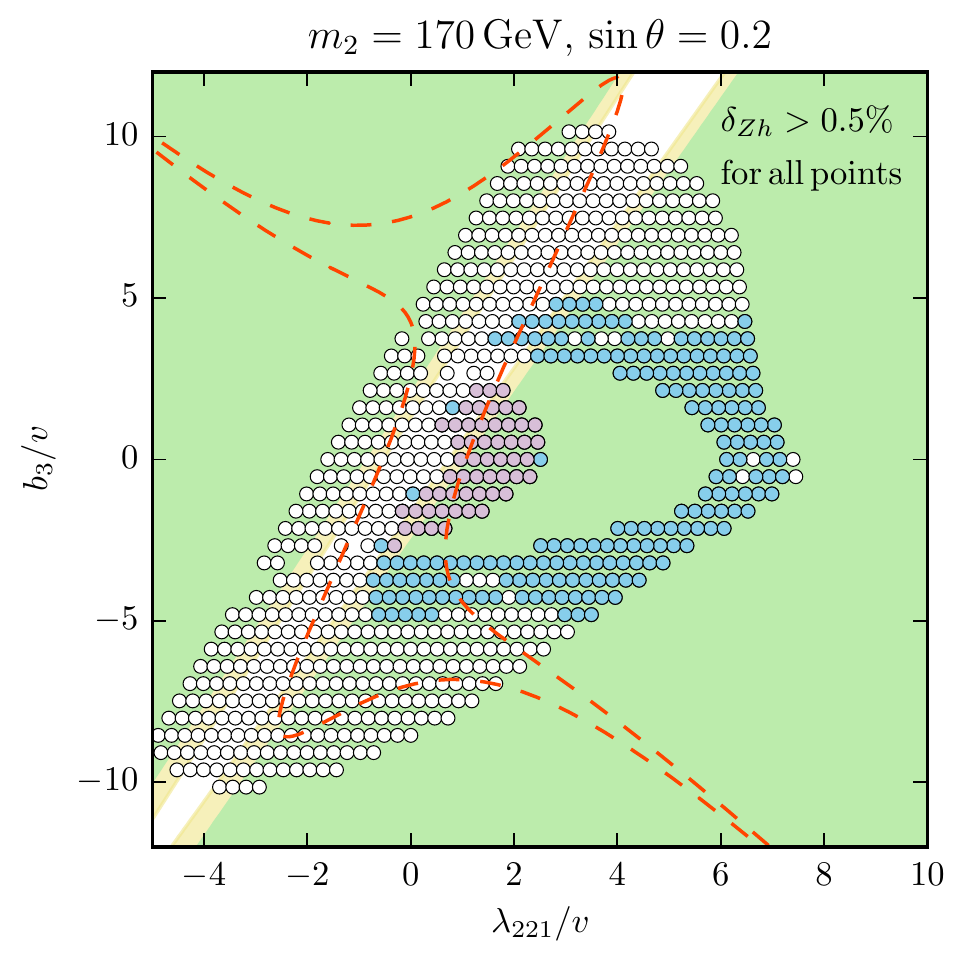}
\caption{\label{fig:100TeV} Discovery (green) and exclusion (yellow) reach in the $h_2 h_2\rightarrow 2j 2\ell ^{\pm} \ell ^{\prime \mp} 3\nu$ channel at a future 100 TeV collider with 30 ab$^{-1}$ for $m_2=170$ GeV and various values of $\sin\theta$. Also shown is the approximate corresponding reach of $h_1$ self-coupling measurements at a future 100 TeV collider with 30 ab$^{-1}$ (dashed contours) and of measurements of $\delta_{Zh}$ at a future lepton collider, such as the CEPC, FCC-ee or ILC (solid contours); points lying within the regions bounded by these contours would \emph{not} be probed by the corresponding measurement. Note that for $\sin\theta = 0.1$, 0.2, the entire region shown features $\delta_{Zh}>0.5\%$ and would likely be accessible to $Zh_1$ cross-section measurements at future lepton colliders. $h_2 h_2$ production is expected to be the best probe of the EWPT-compatible regions for small mixing angles. At larger mixing angles, the $h_2 h_2$ sensitivity to the EWPT-compatible regions is expected to be comparable to that of $Zh_1$ measurements. }
\end{figure}

\begin{figure}[!t]
\centering
\includegraphics[width=.45\textwidth]{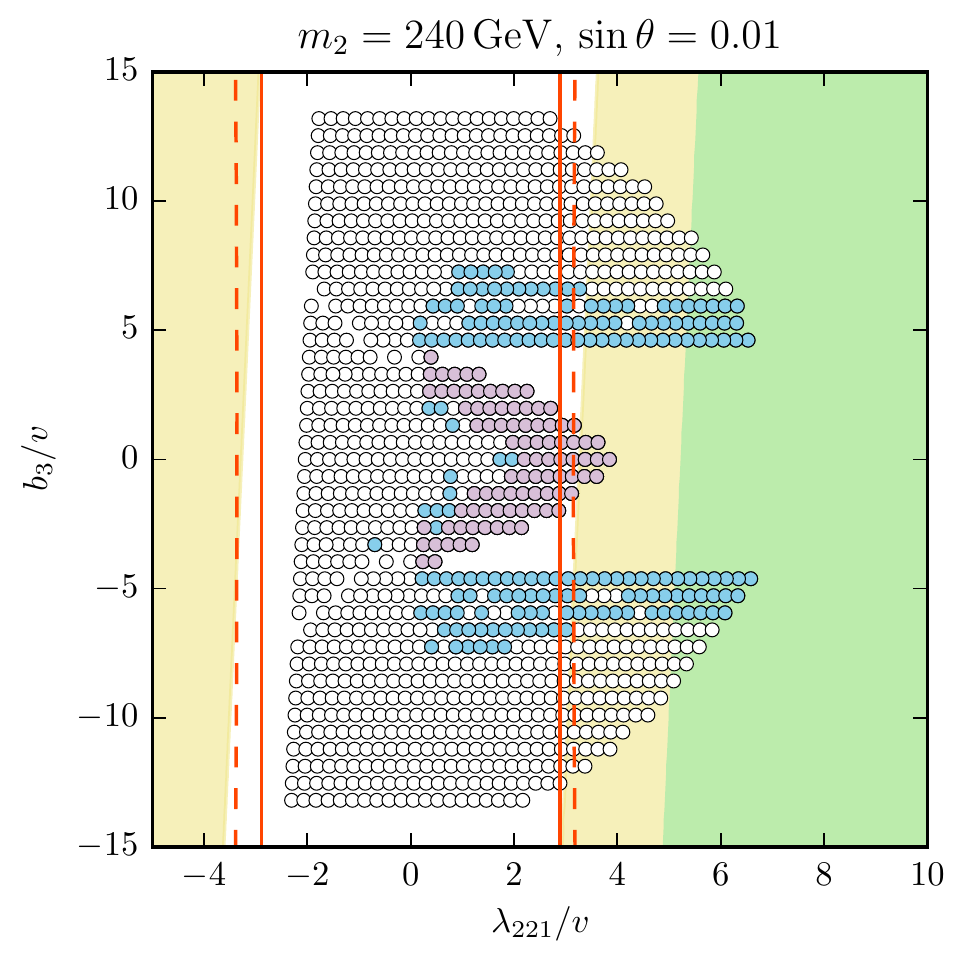}\,\includegraphics[width=.45\textwidth]{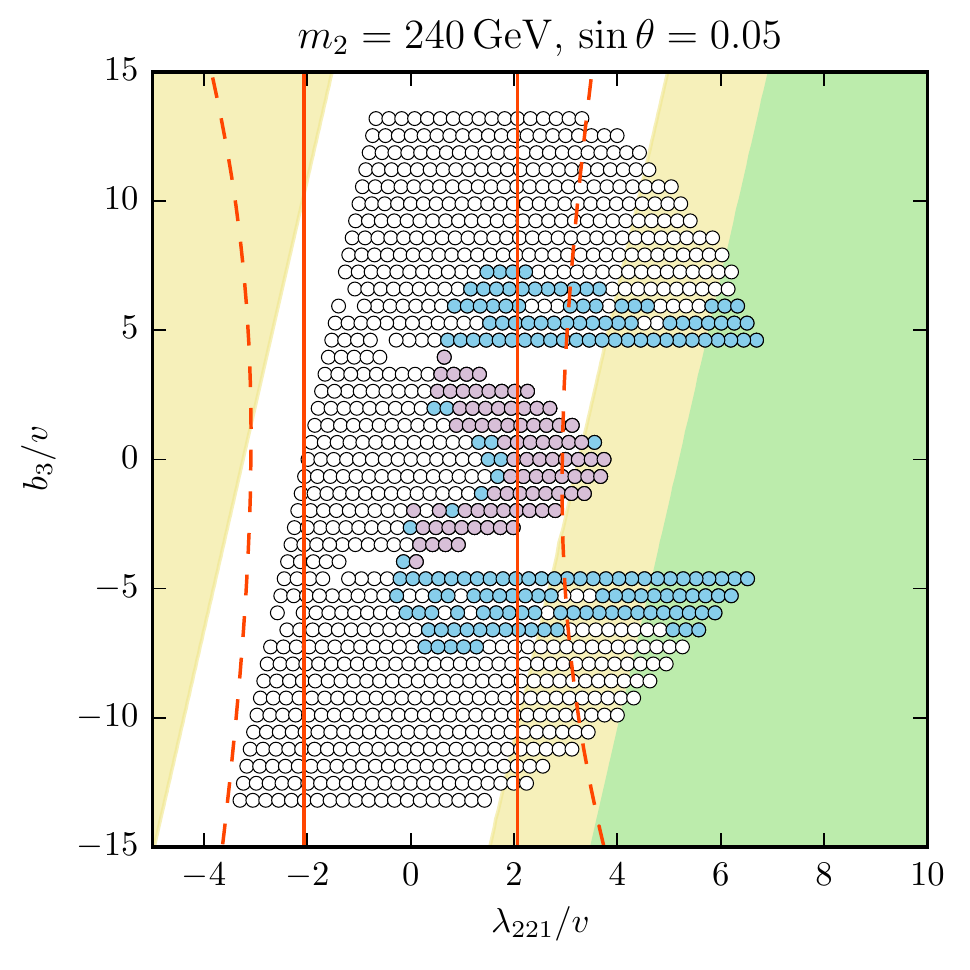}\\
\includegraphics[width=.45\textwidth]{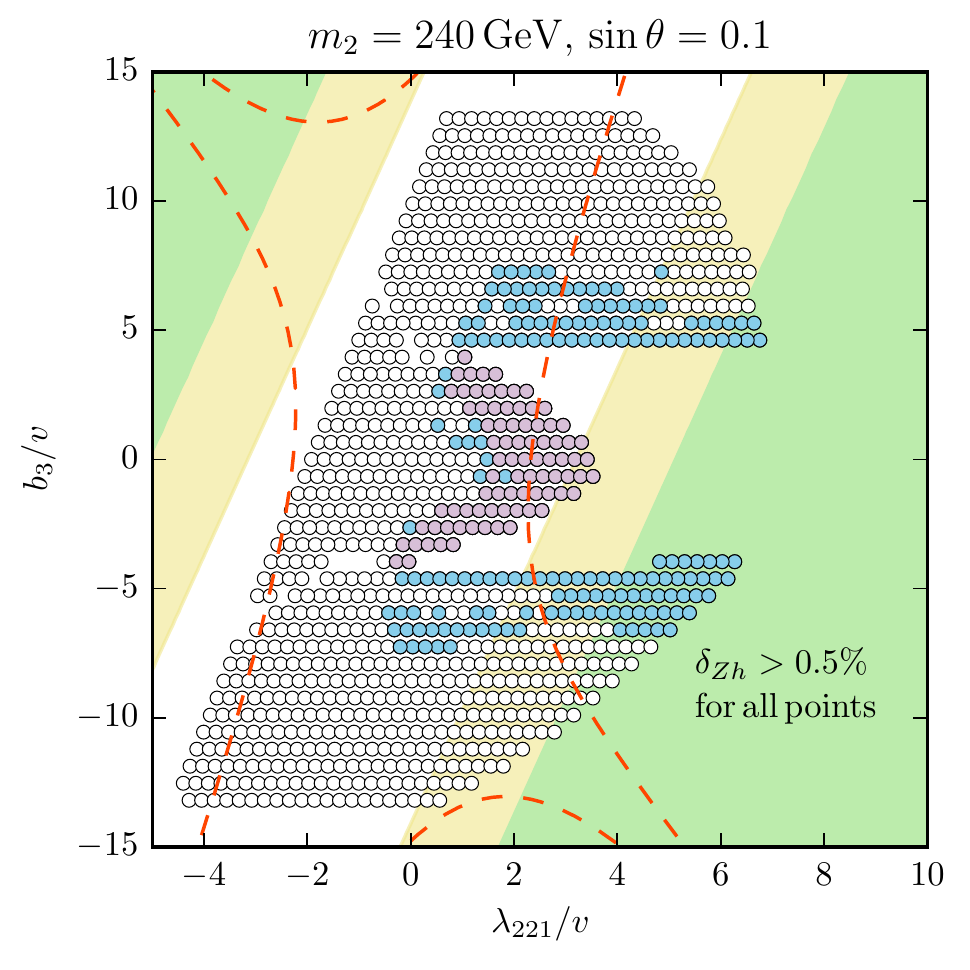}\,\includegraphics[width=.45\textwidth]{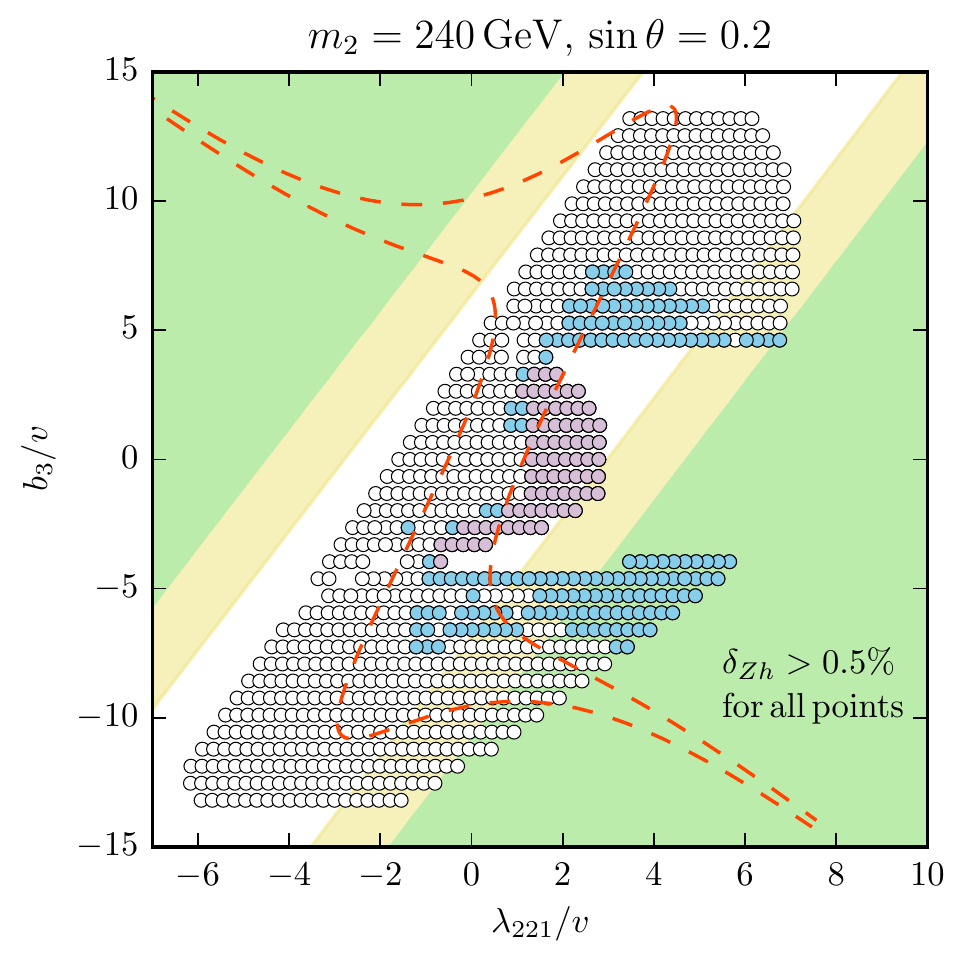}
\caption{\label{fig:100TeV_2} As in Fig.~\ref{fig:100TeV} but for $m_2 = 240$ GeV. $h_2 h_2$ production is expected to provide sensitivity to the EWPT-compatible regions comparable to that of $Zh_1$ and $h_1 h_1$ measurements for small mixing angles. }
\end{figure}

The impact of a future 100 TeV collider on the parameter space of interest, given 30 ab$^{-1}$ of integrated luminosity, is shown in Figs.~\ref{fig:100TeV} and~\ref{fig:100TeV_2}. The green shaded regions feature $S/\sqrt{S+B}\geq 5$ and would roughly correspond to a 5$\sigma$ discovery. The yellow regions feature $S/\sqrt{S+B}\geq 2$ and would roughly correspond to a 2$\sigma$ exclusion limit. For $m_2 = 170$ GeV, we find that $S/B \gtrsim 0.1$ over most of the shaded regions. For $m_2=240$ GeV, values of $S/B$ are lower, but are $\gtrsim 0.04$ over all of the parameter space shaded green. Some portions of this parameter space with smaller $\lambda_{221}$ may thus require reducing the corresponding systematic uncertainties in the $t\bar{t}$ background prediction to conclusively probe.

 In terms of the total $h_2 h_2$ production cross-section, our results suggest that a future 100 TeV collider can have sensitivities to 
\begin{equation}
\begin{aligned}
m_2 = 170 \, {\rm GeV:} \quad \sigma_{h_2 h_2} &\gtrsim 56 \, {\rm fb} \quad (2\sigma), \quad 142 \, {\rm fb} \quad (5\sigma)\\
m_2 = 240 \, {\rm GeV:} \quad \sigma_{h_2 h_2} &\gtrsim 202 \, {\rm fb} \quad (2\sigma),  \quad 519 \, {\rm fb} \quad (5\sigma)
\end{aligned}
\end{equation}
in the trilepton channel and for the choices of cuts we considered. Note that the observable cross-sections for $m_2 = 240$ GeV are significantly higher than those for $m_2=170$ GeV due to the decreased $WW$ branching ratio and the fact that we maximized $S/B$ (and not $S/\sqrt{S+B}$) for the signal to be at a roughly observable level once systematic uncertainties are taken into account. 
For $m_2=170$ GeV, most of the parameter space featuring a strong first-order electroweak phase transition can be probed at 2$\sigma$ by the trilepton channel given 30 ab$^{-1}$. This conclusion holds for all values of $0<|\sin \theta| \lesssim 0.2$ considered, provided that $h_2$ decays promptly. As one increases $m_2$ to 240 GeV, the cross-section and corresponding sensitivity is reduced, but a 100 TeV collider can still cover a significant portion of the parameter space consistent with a strong first-order electroweak phase transition.

\subsection{Additional Probes}

We can once again compare our results to the expected sensitivities of future experiments to deviations in the $h_1h_1$ and $Zh_1$ production cross-sections from their predicted Standard Model values. We proceed in the same way as discussed in Sec.~\ref{sec:indirect_LHC}. 

Whereas the high luminosity LHC is expected to probe $|\lambda_{111}-\lambda_{111}^{\rm SM}| /\lambda_{111}^{\rm SM}\sim 30$\%, a future 100 TeV $pp$ collider is expected to extend this sensitivity to  $|\lambda_{111}-\lambda_{111}^{\rm SM}| /\lambda_{111}^{\rm SM}\sim 15$\% with 30 ab$^{-1}$~\cite{Barr:2014sga}. The corresponding parameter space lies outside of the dashed red contours in Figs.~\ref{fig:100TeV}-\ref{fig:100TeV_2}. 

As discussed in Refs.~\cite{Dawson:2013bba, Curtin:2014jma, Fujii:2015jha, dEnterria:2016fpc,Durieux:2017rsg}, future lepton colliders are expected to be sensitive to $\delta_{Zh}\sim 0.5$\%  or better. For $\sin \theta = 0.01$, 0.05, the corresponding parameter space lies outside of the solid red contours in Figs.~\ref{fig:100TeV}-\ref{fig:100TeV_2}, using Eq.~\ref{eq:Zh} to compute $\delta_{Zh}$. For $\sin \theta \gtrsim 0.07$, a $\sim 0.5 \%$ precision in the $Zh_1$ production cross-section measurement would probe all of the parameter space shown.

\subsection{Summary of 100 TeV Results}

Non-resonant $h_2 h_2$ production should be observable at a future 100 TeV collider across much of the parameter space with a strong first-order electroweak phase transition and $m_2 \lesssim 2 m_1$. The sensitivity to this process in the trilepton channel is likely to exceed or be comparable to that of non-resonant $h_1 h_1$ production and measurements of the $Zh_1$ production cross-section at lepton colliders, especially for small mixing angles. We expect masses below $m_2=170$ GeV to also be probed via $h_2 h_2$ production, however for $m_2<140$ GeV the $h_2$ would decay primarily to $b$ quarks, necessitating a separate study to properly address. For $m_2\gtrsim 2m_1$, a combination of $h_1 h_1$ and $Z h_1$ measurements will likely be required to probe the regions with a strong first-order EWPT. It is possible that non-resonant $h_2 h_2$ production could yield some additional sensitivity for small $|\sin \theta|$ in this region via other final states with a larger branching ratio, such as $h_2 h_2 \rightarrow 4h_1$, as pointed out in Ref.~\cite{Arkani-Hamed:2015vfh}. This would be worthwhile to investigate in the future.

\section{Outlook and Conclusions}\label{sec:concl}

In this study, we have analyzed the parameter space of the general real singlet extension of the Standard Model compatible with a strong first-order electroweak phase transition and its corresponding collider signatures. Regions supporting a strong first order electroweak phase transition typically feature significant couplings of the SM-like Higgs to pairs of singlet-like scalars.  This led us to consider non-resonant scalar pair production at both the LHC and a future 100 TeV collider as a direct probe of the electroweak phase transition in this setup. We compared the various non-resonant production cross-sections across the parameter space, and focused particularly on pair production of the singlet-like state with masses $140$ GeV$<m_2<2 m_1$ such that it decays predominantly to gauge bosons. 

The high luminosity LHC should have some sensitivity to $h_2 h_2$ production in the trilepton channel we considered. Measurements of the $h_1$ self-coupling (via non-resonant $h_1 h_1$ production) may provide better coverage of the regions with a strong first-order EWPT. However, as shown in Sec.~\ref{sec:nonres}, current estimates on the sensitivity of the LHC to the Higgs self-coupling may be overly optimistic when applied to the singlet model, and a more thorough study will be needed to draw any firm conclusions.  Also, our projected sensitivities to $h_2 h_2$ production were based on a conservative estimate of the $t\bar{t}$ lepton fake rate and considering only the trilepton channel (to enhance $S/B$). Including more $h_2h_2$ decay modes with higher backgrounds, such as $h_2h_2 \rightarrow 4W \rightarrow 4j 2\ell^\pm 2\nu$, will likely improve the LHC sensitivity to $h_2h_2$ production relative to our estimates.  

The situation is improved at future colliders. Much of the available parameter space with a strong first-order phase transition with $m_2 \lesssim 2 m_1$ can be probed by $h_2 h_2$ production in the trilepton channel given 30 ab$^{-1}$ at a 100 TeV collider. Future lepton colliders alone will likely be able to access some, but not all, of the parameter space accessible to a 100 TeV $pp$ collider in this mass range. In this sense these machines would be highly complementary, with lepton colliders probing larger mixing angles and direct $h_2 h_2$ production at 100 TeV closing the gap for small $\sin \theta$, provided that $m_2$ is not too large. For singlet-like masses heavier than $\sim 300$ GeV, it is unlikely that $h_2 h_2$ production will provide much sensitivity to regions with a strong first-order EWPT in the trilepton channel. Here instead one should look to di-Higgs ($h_1 h_1$) production, provided the mixing angle is not too small, or to measurements of the $Zh_1$ production cross-section at lepton colliders to probe the EWB-compatible regions. These conclusions reaffirm the point made in Refs.~\cite{Curtin:2014jma, Arkani-Hamed:2015vfh, Assamagan:2016azc}: \emph{a multi-faceted approach, including a future lepton collider and high-energy hadron collider, is required to conclusively probe the nature of electroweak symmetry breaking in the early Universe}. 

Our study opens up several new directions for future collider studies:

\begin{itemize}

\item While we considered one particular final state ($2j 2\ell ^{\pm} \ell ^{\prime \mp} 3\nu$), it would be interesting to investigate different channels as well, for example involving $b$ quarks to gain sensitivity to lower masses, or other gauge boson final states with an increased branching ratio to reach larger $m_2$. 
\item It will also be important to consider the sensitivity to $h_1 h_2$ production at the LHC and a 100 TeV collider. This process could potentially provide additional coverage to regions with a strong first-order electroweak phase transition, given its complementarity to $h_2 h_2$ production as discussed in Sec.~\ref{sec:nonres}. 
\item In the singlet model where the Higgs and singlet can mix, there is no longer a one-to-one correspondence between the non-resonant $h_1h_1$ production rate and $h_1$ trilinear coupling (see Sec.~\ref{sec:nonres}).  A more thorough study will be needed to determine sensitivity of $h_1h_1$ production measurements to the various trilinear scalar couplings.  In particular, kinematic distributions have been shown to increase the LHC sensitivity to the SM Higgs trilinear couplings~\cite{Huang:2015tdv,Kling:2016lay}.  These distributions may be even more useful in the singlet model since the additional diagrams with new particles contribute to $h_1h_1$ production.
\item For light singlet-like states, exotic Higgs decays would likely also provide coverage to the EWB-compatible regions.
\end{itemize}

One could also narrow down the available parameter space by imposing additional requirements, allowing for sharper predictions. Requiring $\phi_h/T_c \gtrsim 1$ is not a sufficient condition for electroweak baryogenesis. One must also ensure, for example, that the phase transition in fact completes by computing the tunneling rate and comparing to the expansion rate of the Universe. It is possible that in some of the parameter space we considered with large $T=0$ barriers the tunneling does not occur quickly enough to allow for a graceful exit from the false vacuum. Also, singlet models are known to suffer from fast bubble walls~\cite{Bodeker:2009qy, Kozaczuk:2015owa}: if the bubble wall velocity in the plasma frame is larger than the sound speed, the conventional picture of non-local electroweak baryogenesis cannot occur\footnote{There are, however, other mechanisms that can account for baryogenesis in the fast-wall regime; see e.g.~Refs.~\cite{No:2011fi, Caprini:2011uz, Katz:2016adq}.}, which would again likely exclude some regions with large barriers (although these points are often interesting from the standpoint of observable gravitational radiation~\cite{Caprini:2015zlo}). Requiring the non-existence of low-lying Landau poles and/or perturbativity significantly above the electroweak scale would also likely narrow down the available parameter space. It would be worthwhile to investigate the impact of these considerations in the future~\footnote{Very recently, Ref.~\cite{Kurup:2017dzf} appeared, which applies some of these constraints to the $\mathbb{Z}_2$-symmetric case. See also Ref.~\cite{Chala:2016ykx}.}.

To obtain more reliable predictions, our perturbative treatment of the phase transition should be compared to other methods (e.g. utilizing the gauge-invariant $\hbar$-expansion~\cite{Patel:2011th}, or the ``optimized partial dressing'' technique discussed in Ref.~\cite{Curtin:2016urg}), and ultimately to a full non-perturbative study (e.g.~utilizing the dimensional reduction outlined in Ref.~\cite{Brauner:2016fla}). Progress on this front is essential for conclusively testing the nature of the EWPT, and thereby the viability of electroweak baryogenesis or gravitational wave generation in these models, experimentally.

Ultimately, it would be illuminating to combine our results with those of the complementary search strategies described above to comprehensively map out discovery strategies for the full parameter space consistent with a strong first-order electroweak phase transition. Maximizing experimental coverage to this scenario may allow next generation experiments to unearth evidence for a strong first-order electroweak phase transition in the early Universe.

\section*{Acknowledgements}
We thank David Curtin, Peisi Huang, Zhen Liu, Andrew Long, Stephen Martin, Jose Miguel No, Hiren Patel, Michael Ramsey-Musolf, and Graham Wilson for many illuminating discussions, suggestions and comments throughout the course of this work.  The work of C.-Y.C is supported by NSERC, Canada. Research at the Perimeter Institute is supported in part by the Government of Canada through NSERC and by the Province of Ontario through MEDT. While at TRIUMF, the work of JK  was supported by the Natural Sciences and Engineering Research Council of Canada (NSERC). TRIUMF receives federal funding via a contribution agreement with the National Research Council of Canada. IML was supported in part by the U.S. Department of Energy under contract DE-AC02-76SF0051  and the University of Kansas General Research Fund allocation 2302091.

\appendix
\section{Appendix A: The One-loop Effective Potential} \label{sec:app_renorm}

To compute the one-loop effective potential, one requires the field-dependent masses of the various particles. For the singlet model under consideration, these are given by~\cite{Espinosa:2011ax}:
\begin{equation}
\begin{aligned}
m_t^2(\phi_h) = \frac{1}{2} &y_t^2 \phi_h^2, \qquad m_W^2(\phi_h) = \frac{1}{4}g^2 \phi_h^2, \qquad m_Z^2(\phi_h) = \frac{g^2 + g^{\prime 2}}{4}\phi_h^2, \\
& m_G^2(\phi_h, \phi_s) = \lambda \phi_h^2 - \mu^2 + \frac{1}{2} \left(a_1+a_2 \phi_s\right) \phi_s.
\end{aligned}
\end{equation}
The field-dependent masses of the two real CP-even scalars are given by the eigenvalues of the matrix
\begin{equation} \label{eq:massmatrix}
\mathcal{M}^2(\phi_h,\phi_s) = \left(
\begin{array}{c c c}
-\mu^2+3 \lambda \phi_h^2 + \frac{1}{2} \left(a_1+a_2 \phi_s\right) \phi_s & \frac{1}{2} \left(a_1+2 a_2 \phi_s\right) \phi_h \\
\frac{1}{2} \left(a_1+2 a_2 \phi_s\right) \phi_h & b_2+ 3b_4 \phi_s^2 + 2 b_3 \phi_s + \frac{1}{2}a_2 \phi_h^2 
\end{array}
\right).
\end{equation}

To compute the daisy contributions at finite temperature we also require the finite-$T$ self-energies, which yield thermal masses for the various particles 
\begin{equation}
m^2_i(\phi_h,\phi_s,T) \equiv m^2_i(\phi_h, \phi_s) + \Pi_i(\phi_h, \phi_s, T).
\end{equation}
These thermal masses are then fed into Eq.~\ref{eq:daisy} for the ring contribution. To compute the self-energies, we employ the high-$T$ approximation which renders the thermal self-energies functions of the temperature alone. For the Goldstone bosons, we find
\begin{equation}
\Pi_G(T) = \left(\frac{3}{16} g^2 + \frac{1}{16}g^{\prime 2} + \frac{1}{4} y_t^2 + \frac{1}{2}\lambda + \frac{1}{24} a_2 \right) T ^2.
\end{equation}
For the real neutral scalars, the thermal contribution to the self-energies amounts to adding the following 2$\times$2 matrix to that in Eq.~\ref{eq:massmatrix}:
\begin{equation}
\Delta \mathcal{M}^2(\phi_h,\phi_s,T) = \left(
\begin{array}{c c c}
\frac{3}{16} g^2 + \frac{1}{16}g^{\prime 2} + \frac{1}{4} y_t^2 + \frac{1}{2}\lambda + \frac{1}{24} a_2  & 0 \\
0 & \frac{1}{6} a_2 + \frac{1}{4} b_4
\end{array}
\right) T^2.
\end{equation}
The thermal masses $m_{1,2}^2(\phi_h, \phi_s, T)$ appearing in Eq.~\ref{eq:daisy} are then the eigenvalues of $\mathcal{M}^2+\Delta \mathcal{M}^2$. The corresponding expressions for the gauge bosons can be found in the literature (see e.g.~Ref.~\cite{Quiros:1999jp}). 

The effective potential also depends on the choice of renormalization scheme. We choose our renormalization scheme to minimize the one-loop contributions to the various scalar trilinear and quartic couplings. Although the physical couplings should be defined at finite external momenta, the resulting corrections to the zero-external momenta couplings derived from the effective potential are numerically small and neglected throughout our study.
 
Decomposing the zero-temperature one-loop correction to $V_{\rm eff}$ in terms of the (unrenormalized) loop and counterterm pieces ( $\Delta V_1$ and $\Delta V_{\rm ct}$, respectively), we impose the following eight independent renormalization conditions:
\begin{equation}
\begin{aligned}
\left. \frac{\partial\left(\Delta V_1 +\Delta V_{\rm ct} \right)}{\partial \phi_h}\right|_{\phi_h = v, \phi_s = 0} =& 0, \qquad \left. \frac{\partial\left(\Delta V_1 +\Delta V_{\rm ct} \right)}{\partial \phi_s}\right|_{\phi_h = v, \phi_s = 0} = 0, \\
 \left. \frac{\partial^2\left(\Delta V_1 +\Delta V_{\rm ct} \right)}{\partial \phi_i \partial \phi_j}\right|_{\phi_h = v, \phi_s = 0} =& 0, \qquad \left. \frac{\partial^3\left(\Delta V_1 +\Delta V_{\rm ct} \right)}{\partial \phi_s^3}\right|_{\phi_h = v, \phi_s = 0} = 0, \\
 \left. \frac{\partial^3\left(\Delta V_1 +\Delta V_{\rm ct} \right)}{\partial \phi_s^2 \partial \phi_h}\right|_{\phi_h = v, \phi_s = 0} =& 0, \qquad \left. \frac{\partial^4\left(\Delta V_1 +\Delta V_{\rm ct} \right)}{\partial \phi_s^4}\right|_{\phi_h = v, \phi_s = 0} = 0.
\end{aligned}
\end{equation}
Explicitly writing out $\Delta V_{\rm ct}$ as 
\begin{equation}
\begin{aligned}
\Delta V_{\rm ct} = & \delta \mu^2 \left|H\right|^2 + \delta \lambda \left|H\right|^4 + \frac{\delta a_1}{2} \left|H\right|^2 S  + \frac{\delta a_2}{2}\left|H\right|^2S^2 \\
& + \delta b_1 S + \frac{\delta b_2}{2}S^2 +\frac{\delta b_3}{3}S^3 + \frac{\delta b_4}{4} S^4,
\end{aligned}
\end{equation}
we arrive at the following expressions for the various counterterms:
\begin{equation}
\begin{aligned}
\delta a_1 = &\left. \frac{-2}{v}\frac{\partial^2  \Delta V_1}{\partial \phi_h \partial \phi_s}\right|_{\phi_h = v, \phi_s =0}, \qquad \delta b_1 = \left.\left(-\frac{\partial \Delta V_1}{\partial \phi_s}+\frac{v}{2}\frac{\partial^2 \Delta V_1}{\partial\phi_h \partial\phi_s}\right)\right|_{\phi_h = v, \phi_s =0},\\
\delta a_2 = &\left. -\frac{1}{v}\frac{\partial^3 \Delta V_1}{\partial \phi_s^2 \partial \phi_h}\right|_{\phi_h=v, \phi_s=0}, \qquad \delta b_2 = \left.\left(-\frac{\partial^2 \Delta V_1}{\partial \phi_s^2}+\frac{v}{2}\frac{\partial^3 \Delta V_1}{\partial\phi_s^2 \partial\phi_h}\right)\right|_{\phi_h = v, \phi_s =0},\\
\delta b_3 = &\left. -\frac{1}{2}\frac{\partial^3 \Delta V_1}{\partial \phi_s^3}\right|_{\phi_h=v, \phi_s=0}, \qquad \delta\mu^2 = \left.\left(\frac{1}{2}\frac{\partial^2\Delta V_1}{\partial \phi_h^2}-\frac{3}{2v}\frac{\partial \Delta V_1}{\partial\phi_h}\right)\right|_{\phi_h = v, \phi_s =0},\\
\delta b_4 = &\left. -\frac{1}{6}\frac{\partial^4 \Delta V_1}{\partial \phi_s^4}\right|_{\phi_h=v, \phi_s=0},  \qquad \delta\lambda = \left.\frac{1}{2 v^3}\left(\frac{\partial\Delta V_1}{\partial \phi_h}-v\frac{\partial^2 \Delta V_1}{\partial\phi_h^2}\right)\right|_{\phi_h = v, \phi_s =0}.
\end{aligned}
\end{equation}
We evaluate these expressions numerically for a given choice of $\Lambda$ in Eq.~\ref{eq:V1loopT0_2} for $\Delta V_1$. The total resulting one-loop contribution is independent of the cutoff.

\section{Appendix B: Scalar Pair Production Cross-sections}\label{sec:app_signal}

Here we discuss the various scalar pair production cross-sections at hadron colliders. The leading order amplitude for gluon production of two scalar bosons, $g^{A,\mu}(p_1)+g^{B,\nu}(p_2)\rightarrow h_2(k_1) h_2(k_2)$, is
\begin{eqnarray}
\mathcal{A}^{\mu\nu}_{AB}=\frac{\alpha_s}{8\pi v^2}\delta_{AB}\left(F_1(s,t,u,m_t^2)P_1^{\mu\nu}(p_1,p_2)+F_2(s,t,u,m_t^2)P_2^{\mu\nu}\right),
\end{eqnarray}
where $P_1^{\mu\nu}$ and $P_2^{\mu\nu}$ are the spin-0 and spin-2 projections operators, respectively:
\begin{eqnarray}
P_1^{\mu\nu}&=&g^{\mu\nu}-\frac{p_1^\nu}{p_2^\mu}{p_1\cdot p_2}\\
P_2^{\mu\nu}&=&g^{\mu\nu}+\frac{1}{p_T^2 (p_1\cdot p_2)}\left[k_1^2 p_1^\nu p_2^\mu-2(p_2\cdot k_1) p_1^\nu k_1^\mu-2 (p_1\cdot k_1) k_1^\nu p_2^\mu+2 (p_1\cdot p_2) k_1^\nu k_1^\mu\right]\nonumber,
\end{eqnarray}
and the transverse momentum and Mandelstam variables are
\begin{eqnarray}
p_T^2 &=& \frac{2(p_1\cdot k_1)(p_2\cdot k_1)}{p_1\cdot p_2}-k_1^2\\
s &=& (p_1+p_2)^2\nonumber\\
t &=& (p_1-k_1)^2\nonumber\\
u &=& (p_1-k_2)^2.\nonumber
\end{eqnarray}
The partonic cross section is then
\begin{eqnarray}
\frac{d\hat \sigma}{dt}=\frac{\alpha_s^2}{2^{15}\pi^3 v^4 s^2}\left(|F_1(s,t,u,m_t^2)|^2+|F_2(s,t,u,m_t^2)|^2\right).
\end{eqnarray}
Both the triangle and box diagrams contribute to the spin-$0$ function $F_1$, while only the box contributes to the spin-2 function $F_2$:
\begin{eqnarray}
F_1&=&-s\left(\frac{\cos\theta \lambda_{221}v}{s-m_1^2+i\,m_1\,\Gamma_1}-\frac{\sin\theta \lambda_{222} v}{s-m_2^2+i\,m_2\,\Gamma_2}\right)F_{\Delta}(s,m_t^2)+s\sin^2\theta F_{\square}(s,t,u,m_t^2)\nonumber\\
F_2&=& s\,\sin^2\theta G_{\square}(s,t,u,m_t^2).\label{F1F2}
\end{eqnarray}
The triangle and box loop function $F_{\Delta},\,F_{\square}$ and $G_{\square}$ are known analytically~\cite{Plehn:1996wb,Glover:1987nx}.  The forms of $F_1$ and $F_2$ for different double scalar production modes $h_1h_2$ and $h_2h_2$ can be found by inserting the correct trilinear coupling for the triangle contributions, combinations of $\sin\theta$ and $\cos\theta$ suppressions for the top Yukawa coupling contributions to the box diagrams, and symmetry factors.

For non-resonant production, the hadronic level cross section can be given numerically.  Here we give results for the masses we studied.  As mentioned in Sec.~\ref{sec:nonres},   we generate the cross sections by implementing our model into $\texttt{FeynArts}$~\cite{Hahn:2000kx} via $\texttt{FeynRules}$ package~\cite{Christensen:2008py,Alloul:2013bka} and using $\texttt{FormCalc}$~\cite{Hahn:1998yk}.  We use the NNPDF2.3QED leading order~\cite{Ball:2013hta} parton distribution functions (pdf) with $\alpha_s(M_Z)=0.119$.  These are implemented via LHAPDF~\cite{Buckley:2014ana}.  The factorization and renormalization scales, $\mu_f,\mu_r$, are both set to be the diboson invariant mass.

At the 14 TeV LHC we have:
\begin{eqnarray}
m_2&=&170~{\rm GeV}:\\
\sigma(h_1h_1)&=&16\left|\lambda_{111}\cos\theta\frac{m_t}{v^2}-(1.2-0.082\,i)\lambda_{211}\sin\theta\frac{m_t}{v^2}-(2.7+1.4\,i)\cos^2\theta\frac{m_t^2}{v^2}\right|^2~{\rm fb}\nonumber\\
\sigma(h_1h_2)&=&17\left|\lambda_{211}\cos\theta\frac{m_t}{v^2}-\left(1.1-0.048\,i\right)\lambda_{221}\sin\theta\frac{m_t}{v^2}+\left(3.6+1.4\,i\right)\sin\theta\cos\theta\frac{m_t^2}{v^2}\right|^2~{\rm fb}\nonumber\\
\sigma(h_2h_2)&=&4.9\left|\lambda_{221}\cos\theta\frac{m_t}{v^2}-\left(1.1-0.029\,i\right)\lambda_{222}\sin\theta\frac{m_t}{v^2}-\left(4.4+1.7\,i\right)\sin^2\theta\frac{m_t^2}{v^2}\right|^2~{\rm fb}\nonumber\\
m_2&=&240~{\rm GeV}:\\
\sigma(h_1h_1)&=&16\left|\lambda_{111}\cos\theta\frac{m_t}{v^2}-(2.3-1.3\,i)\lambda_{211}\sin\theta\frac{m_t}{v^2}-(2.7+1.4\,i)\cos^2\theta\frac{m_t^2}{v^2}\right|^2~{\rm fb}\nonumber\\
\sigma(h_1h_2)&=&6.7\left|\lambda_{211}\cos\theta\frac{m_t}{v^2}-\left(1.3-0.11\,i\right)\lambda_{221}\sin\theta\frac{m_t}{v^2}+\left(5.1+1.9\,i\right)\sin\theta\cos\theta\frac{m_t^2}{v^2}\right|^2~{\rm fb}\nonumber\\
\sigma(h_2h_2)&=&0.65\left|\lambda_{221}\cos\theta\frac{m_t}{v^2}-\left(1.2-0.065\,i\right)\lambda_{222}\sin\theta\frac{m_t}{v^2}-\left(6.3+3.7\,i\right)\sin^2\theta\frac{m_t^2}{v^2}\right|^2~{\rm fb}\nonumber
\end{eqnarray}

At the 100 TeV $pp$ collider we have:
\begin{eqnarray}
m_2&=&170~{\rm GeV}:\\
\sigma(h_1h_1)&=&550\left|\lambda_{111}\cos\theta\frac{m_t}{v^2}-(1.2-0.078\,i)\lambda_{211}\sin\theta\frac{m_t}{v^2}-(2.9+1.5\,i)\cos^2\theta\frac{m_t^2}{v^2}\right|^2~{\rm fb}\nonumber\\
\sigma(h_1h_2)&=&650\left|\lambda_{211}\cos\theta\frac{m_t}{v^2}-\left(1.1-0.048\,i\right)\lambda_{221}\sin\theta\frac{m_t}{v^2}+\left(3.8+1.6\,i\right)\sin\theta\cos\theta\frac{m_t^2}{v^2}\right|^2~{\rm fb}\nonumber\\
\sigma(h_2h_2)&=&200\left|\lambda_{221}\cos\theta\frac{m_t}{v^2}-\left(1.1-0.0041\,i\right)\lambda_{222}\sin\theta\frac{m_t}{v^2}-\left(4.5+2.0\,\right)\sin^2\theta\frac{m_t^2}{v^2}\right|^2~{\rm fb}\nonumber\\
m_2&=&240~{\rm GeV}:\\
\sigma(h_1h_1)&=&550\left|\lambda_{111}\cos\theta\frac{m_t}{v^2}-(2.1-1.2\,i)\lambda_{211}\sin\theta\frac{m_t}{v^2}-(2.9+1.5\,i)\cos^2\theta\frac{m_t^2}{v^2}\right|^2~{\rm fb}\nonumber\\
\sigma(h_1h_2)&=&280\left|\lambda_{211}\cos\theta\frac{m_t}{v^2}-\left(1.3-0.12\,i\right)\lambda_{221}\sin\theta\frac{m_t}{v^2}+\left(5.1+1.9\,i\right)\sin\theta\cos\theta\frac{m_t^2}{v^2}\right|^2~{\rm fb}\nonumber\\
\sigma(h_2h_2)&=&34\left|\lambda_{221}\cos\theta\frac{m_t}{v^2}-\left(1.1-0.067\,i\right)\lambda_{222}\sin\theta\frac{m_t}{v^2}-\left(6.4+3.9\,i\right)\sin^2\theta\frac{m_t^2}{v^2}\right|^2~{\rm fb}\nonumber
\end{eqnarray}

\section{Appendix C: Trilepton Search Kinematics} \label{sec:app_kin}

\begin{figure}[!t]
\centering
\includegraphics[width=.45\textwidth]{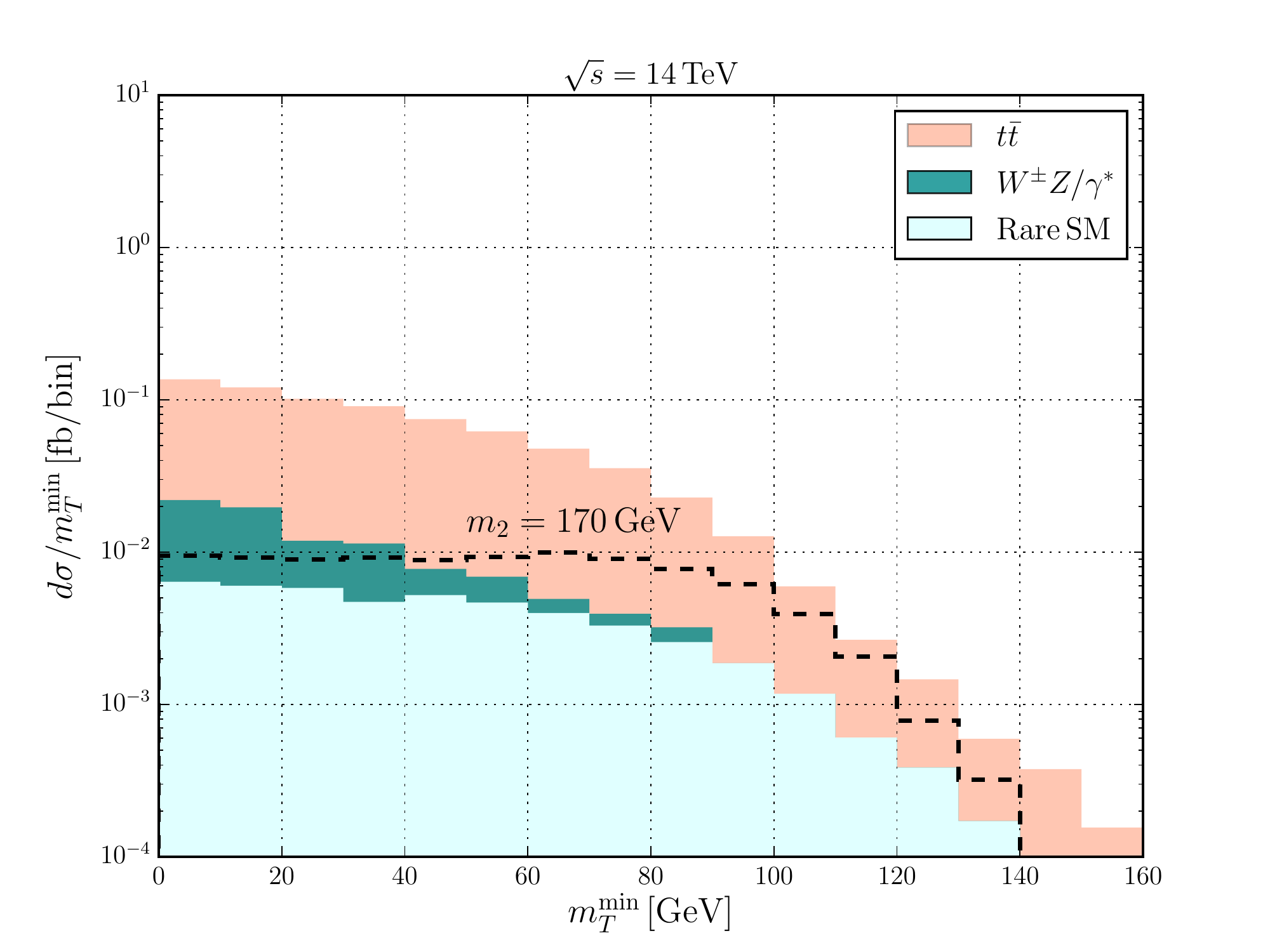}\,\includegraphics[width=.45\textwidth]{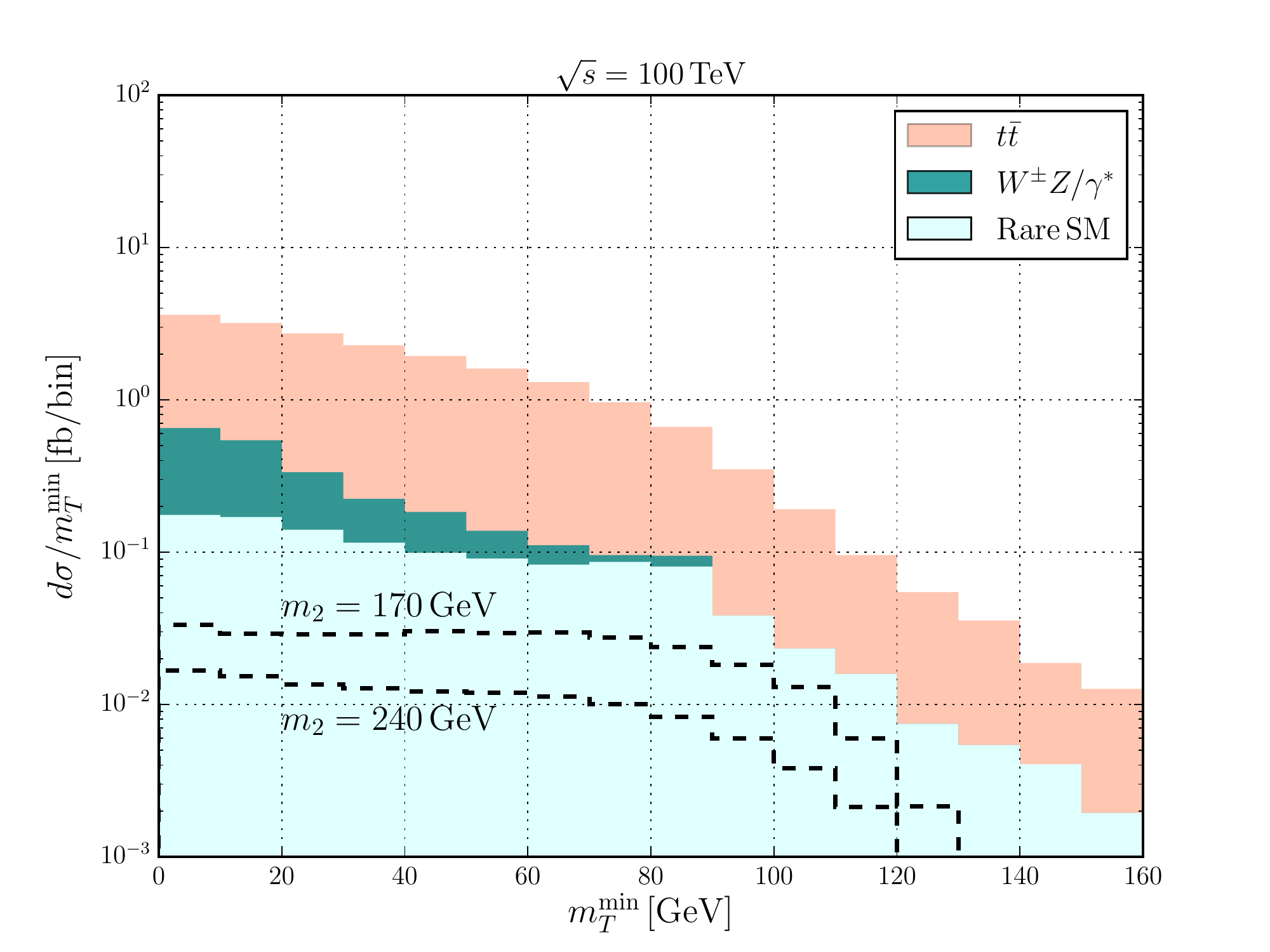}\\
\includegraphics[width=.45\textwidth]{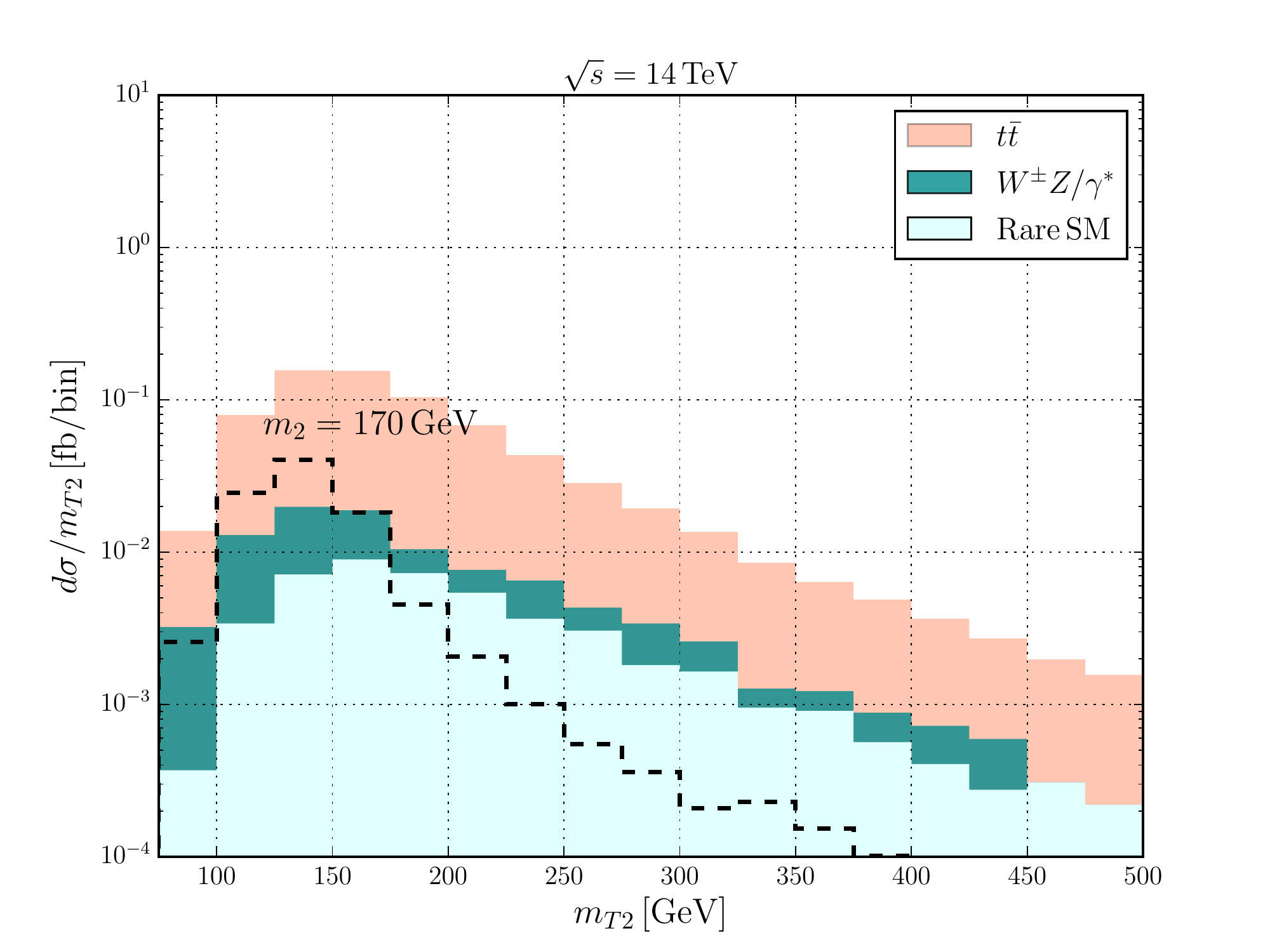}\,\includegraphics[width=.45\textwidth]{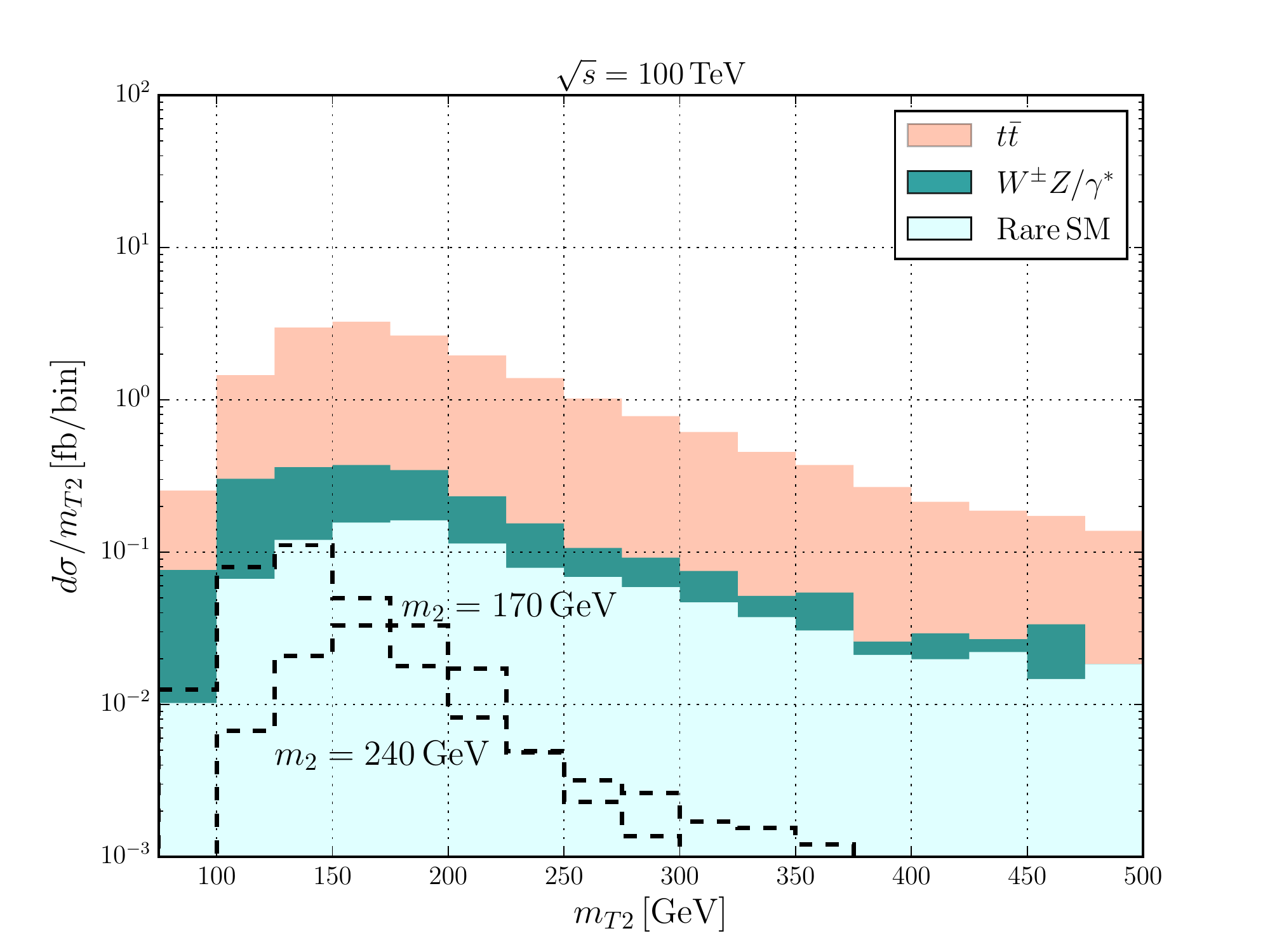}\\
\includegraphics[width=.45\textwidth]{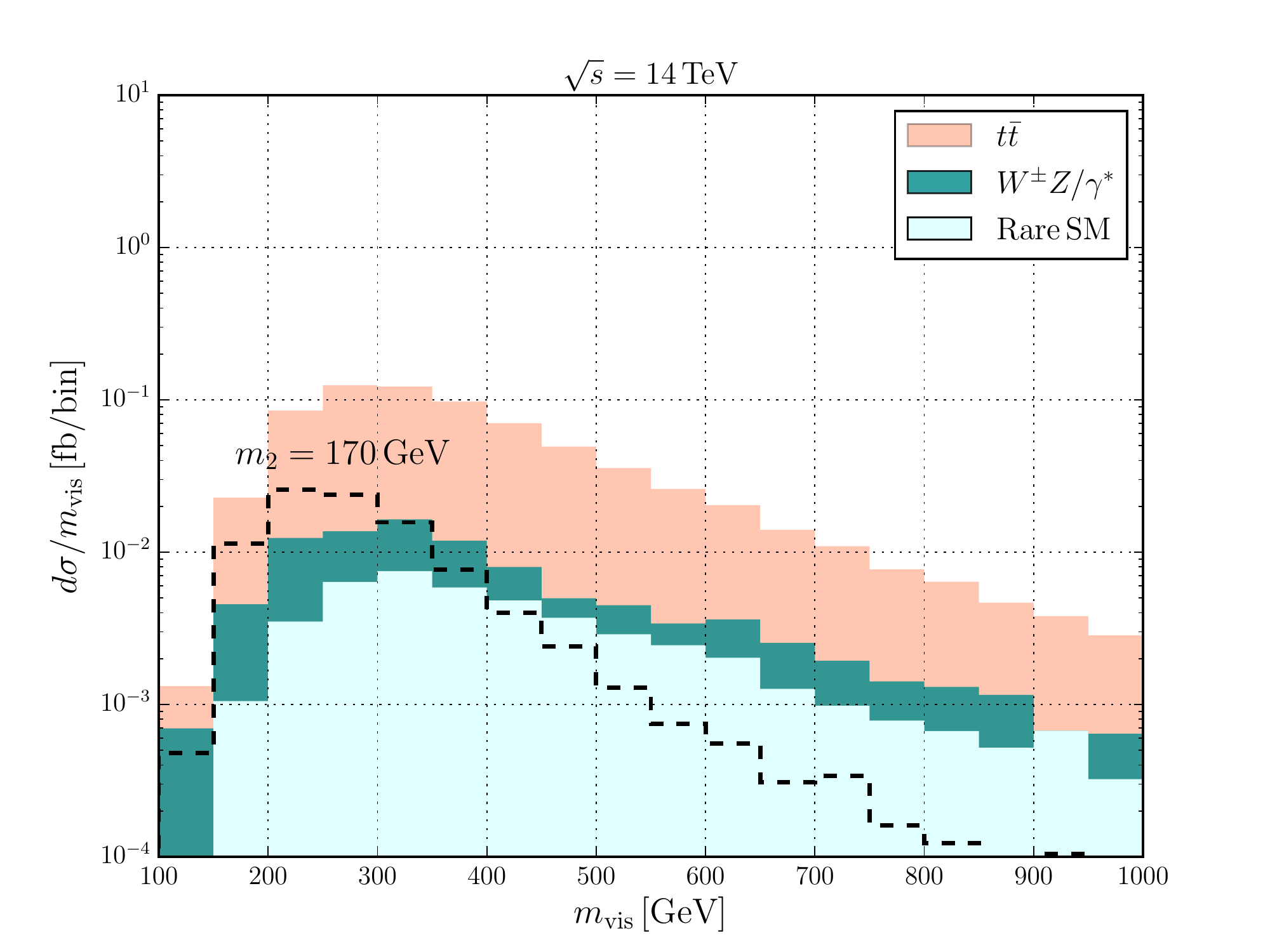}\,\includegraphics[width=.45\textwidth]{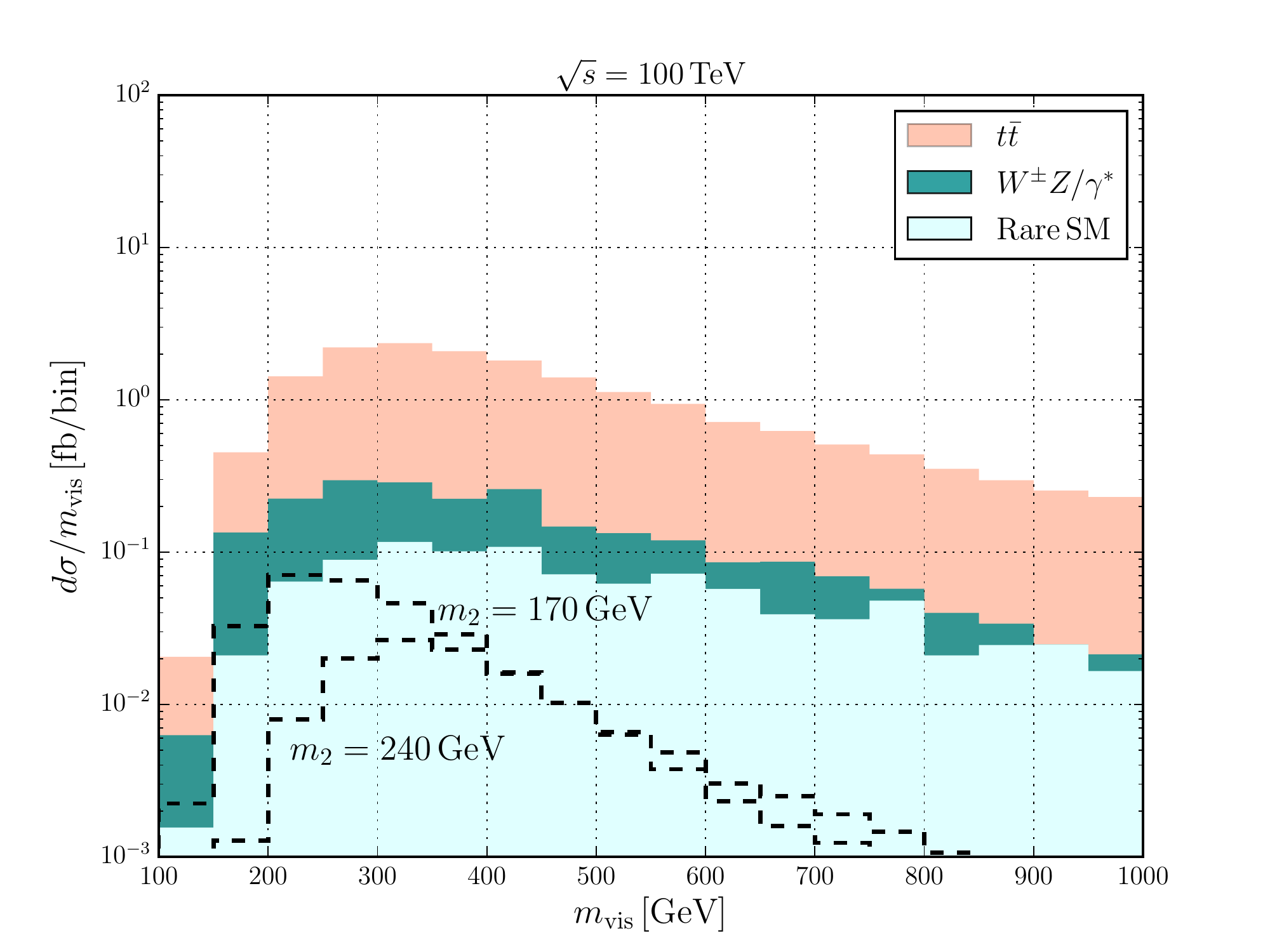}
\caption{\label{fig:kin} Distributions of $m_T^{\rm min}$, $m_{T2}$, and $m_{\rm vis}$ at 14 and 100 TeV for the various background processes and signal points after basic acceptance. The various signal points shown correspond to those used in optimizing cuts for the 14 and 100 TeV analysis, as described in Secs.~\ref{sec:LHC} and~\ref{sec:100TeV}.}
\end{figure}

In this appendix we provide some details of the kinematic distributions relevant for our analysis of $h_2 h_2$ production in the trilepton final state. Distributions of $m_T^{\rm min}$, $m_{T2}$, and $m_{\rm vis}$ at 14 and 100 TeV are shown in Fig.~\ref{fig:kin} for the various backgrounds and for the signal points used in optimizing our cuts. For 14 TeV, the signal point corresponds to $m_2=170$ GeV, $\sin \theta = 0.05$, $a_2=8.5$, $b_3=0$. For our 100 TeV analysis, the signal points shown are $m_2=170$ GeV, $\sin \theta = 0.05$, $a_2=2$, $b_3=0$ and $m_2=240$ GeV, $\sin \theta = 0.05$, $a_2=3.5$, $b_3=0$.

From Fig.~\ref{fig:kin}, we can see that there are notable features distinguishing the signal from backgrounds. For example, the signal features a peak in the $m_{T2}$ and $m_{\rm vis}$ distributions, and falls off less quickly with increasing $m_T^{\rm min}$ than the backgrounds. In our study, we performed a cut-and-count analysis with simple rectangular cuts on these quantities. A more sophisticated multivariate analysis may significantly improve the sensitivity to the trilepton final state, given the distinct kinematic features of the signal.

\section{Appendix D: Calculation of $\delta_{Zh}$} \label{sec:app_Zh}
In this appendix we give an overview of the calculation of the shift in the $Z-h_1$ production cross-section relative to the SM expectation.  There are contributions from the Higgs wavefunction renormalization~\cite{Craig:2013xia, McCullough:2013rea} as well as shifts in the tree-level $Z-Z-h_1$ coupling and loops that directly effect this vertex.  In the limit in which the scalar potential exhibits an unbroken $\mathbb{Z}_2$ symmetry $S\rightarrow -S$, only the wave-function counterterms contribute and the results can be found in Refs.~\cite{Craig:2013xia, McCullough:2013rea}.  In the model considered here, all contributions, including the effect of mixing between the two scalar bosons, must be considered. Electroweak corrections to the singlet extended SM have been studied before~\cite{Bojarski:2015kra,Costa:2017gky}, including full one-loop corrections to the $Z-h_1$ couplings~\cite{Kanemura:2015fra}.  In our case, we are interested in the small mixing limit, $\sin\theta\ll1$.  Accordingly, we adopt the parametric counting
\begin{eqnarray}
\sin\theta\sim \frac{\lambda}{4\pi},\label{eq:sth}
\end{eqnarray}
where $\lambda$ is an order one parameter.  From the appendix of Ref.~\cite{Chen:2014ask}, we count the scalar couplings as
\begin{eqnarray}
&\lambda_{111}^{\rm SM}\sim \lambda_{221}\sim \lambda_{222}\sim \lambda,\, \lambda_{111}=\lambda_{111}^{\rm SM}+\mathcal{O}(\sin^2\theta),\, \lambda_{211}\sim \mathcal{O}(\sin\theta)\label{eq:TriScal}\\
&\lambda_{1111}^{\rm SM}\sim \lambda_{2211}\sim \lambda_{2222}\sim \lambda,\, \lambda_{1111}=\lambda_{1111}^{\rm SM}+\mathcal{O}(\sin^2\theta),\, \lambda_{2111}\sim\lambda_{2221}\sim \mathcal{O}(\sin\theta)\label{eq:QuadScal}
\end{eqnarray}
where the superscript ${\rm SM}$ indicate the  the SM values.   We will perform our calculation to ``one-loop'' order in scalar couplings.  This amounts to $\mathcal{O}(\lambda^2/(4\pi)^2)=\mathcal{O}(1/(4\pi)^2)$ in the self-energy diagrams and $\mathcal{O}(g_W^2\lambda/(4\pi)^2)=\mathcal{O}(g_W^2/(4\pi)^2)$ where $g_W$ and $\lambda$ are generic weak and order one couplings, respectively.  Using the parameter counting in Eqs.~\ref{eq:sth}, \ref{eq:TriScal}, and \ref{eq:QuadScal} together with working to ``one-loop'' order, we effectively ignore loop contributions that are additionally suppressed by the scalar mixing angle. This is reasonable in the small mixing limit, although a more complete calculation is needed away from this limit.

First, we renormalize the scalar Lagrangian:
\begin{eqnarray}
\mathcal{L}=\left| D_\mu H^{0}\right|^2+\frac{1}{2}(\partial_\mu S^{0})^2-V(H^{0},S^{0}),
\end{eqnarray}
where we add the superscript $0$ to indicate bare fields. After mixing in Eq.~\ref{eq:mix1}, the scalar Lagrangian is
\begin{eqnarray}
\mathcal{L}_{\rm Scalars}=\frac{1}{2}(\partial_\mu h_1^{0})^2+\frac{1}{2}(\partial_\mu h_2^{0})^2-V(h_1^{0},h_2^{0}),
\end{eqnarray}
where the scalar potential is
\begin{eqnarray}
V(h_1^{0},h_2^{0})&=&\frac{1}{2}(m_1^{0})^2\,h_1^{0}\,h_1^{0}+\frac{1}{2}(m_2^{0})^2\,h_2^{0}\,h_2^{0}+\frac{\lambda^{0}_{111}}{6}(h_1^{0})^3+\frac{\lambda^{0}_{211}}{2}(h_1^{0})^2h_2^{0}+\frac{\lambda_{221}^{0}}{2}h_1^{0}(h_2^{0})^2\nonumber\\
&+&\frac{\lambda_{222}^{0}}{6}(h_2^{0})^2+\frac{\lambda_{1111}^{0}}{4!}(h_1^{0})^4+\frac{\lambda_{2111}^{0}}{3!}(h_1^{0})^3 h_2^{0}+\frac{\lambda_{2211}^{0}}{4}(h_1^{0})^2 (h_2^{0})^2+\frac{\lambda_{2221}^{0}}{3!}h_1^{0} (h_2^{0})^3\nonumber\\
&+&\frac{\lambda_{2222}^{0}}{4!}(h_2^0)^4
\end{eqnarray}
and the $Z$ boson couplings to scalars are
\begin{eqnarray}
\mathcal{L}_{Z}=\frac{M_Z^2}{v}\cos\theta\, h_1^0\,Z_\mu Z^\mu-\frac{M_Z^2}{v}\sin\theta\, h_2^0\, Z_\mu Z^\mu.\label{eq:LZh}
\end{eqnarray}

In order to calculate the $Z-Z-h_1$ coupling at one-loop order, the scalar fields and masses must be renormalized.  Since the scalars have the same quantum numbers, both off-diagonal wave-function and mass renormalization constants appear.  To one-loop order, the wavefunction renormalization is given by
\begin{eqnarray}
\begin{pmatrix} h_1^{0} \\ h_2^{0} \end{pmatrix} = \begin{pmatrix} 1+\frac{1}{2}\delta Z_{11} & \frac{1}{2}\delta Z_{12} \\ \frac{1}{2}\delta Z_{21} & 1+\frac{1}{2}\delta Z_{22}\end{pmatrix}\begin{pmatrix}h_1 \\ h_2 \end{pmatrix},
\end{eqnarray}
where $\delta Z_{ij}$ are wavefunction counterterms.  The relationship between the bare and renormalized masses is
\begin{eqnarray}
(m_i^0)^2=m_i^2+\delta m_i^2,
\end{eqnarray}
where $\delta m_i^2$ are the diagonal mass counterterms and $i=1,2$.  In principle there is also an off-diagonal mass term $\delta m_{12}^2$.  Finally, note that we are calculating the shift of the $Z-Z-h_1$ coupling away from the SM value.  The sources of the shifted $Z-Z-h_1$ coupling comprise the scalar mixing as well as the trilinear and quartic scalar couplings.  Since the scalar trilinear and quartic couplings only appear at one-loop order, any higher order corrections to these terms will be at least two-loop in the $Z-Z-h_1$ coupling.  Hence, we can just replace the bare scalar trilinear and quartic couplings with the renormalized couplings.   The relevant renormalized scalar Lagrangian is
\begin{eqnarray}
\mathcal{L}_{\rm Scalars}&=&\frac{1}{2}h_1\left[-(1+\delta Z_{11})(\partial^2 +m_1^2)-\delta m_1^2\right]h_1+\frac{1}{2}h_2\left[-(1+\delta Z_{22})(\partial^2+m_2^2)-\delta m_2^2\right]h_2\nonumber\\
&+&h_1\left[-\frac{\delta Z_{12}}{2}(\partial^2+m_1^2)-\frac{\delta Z_{21}}{2}(\partial^2+m_2^2)-\delta m_{12}^2\right]h_2-\frac{\lambda_{111}}{6}(h_1)^3-\frac{\lambda_{211}}{2}(h_1)^2h_2\nonumber\\
&-&\frac{\lambda_{221}}{2}h_1(h_2)^2-\frac{\lambda_{222}}{6}(h_2)^2-\frac{\lambda_{1111}}{4!}(h_1)^4-\frac{\lambda_{2111}}{3!}(h_1)^3 h_2-\frac{\lambda_{2211}}{4}(h_1)^2 (h_2)^2\nonumber\\
&-&\frac{\lambda_{2221}}{3!}h_1 (h_2)^3-\frac{\lambda_{2222}}{4!} (h_2)^4.\label{eq:LScal}
\end{eqnarray}
and the renormalized $Z-Z-h_1$ coupling is 
\begin{eqnarray}
\mathcal{L}_{Zh_1}=\frac{M_Z^2}{v}Z_\mu Z^\mu h_1\left(\cos\theta+\frac{1}{2}\cos\theta\,\delta Z_{11}-\frac{1}{2}\sin\theta\, \delta Z_{21}\right).\label{eq:Zren}
\end{eqnarray}
Hence, we require the wavefunction counterterms $\delta Z_{11}$ and $\delta Z_{21}$.

\begin{figure}[t]
\begin{center}
\includegraphics[width=0.32\textwidth]{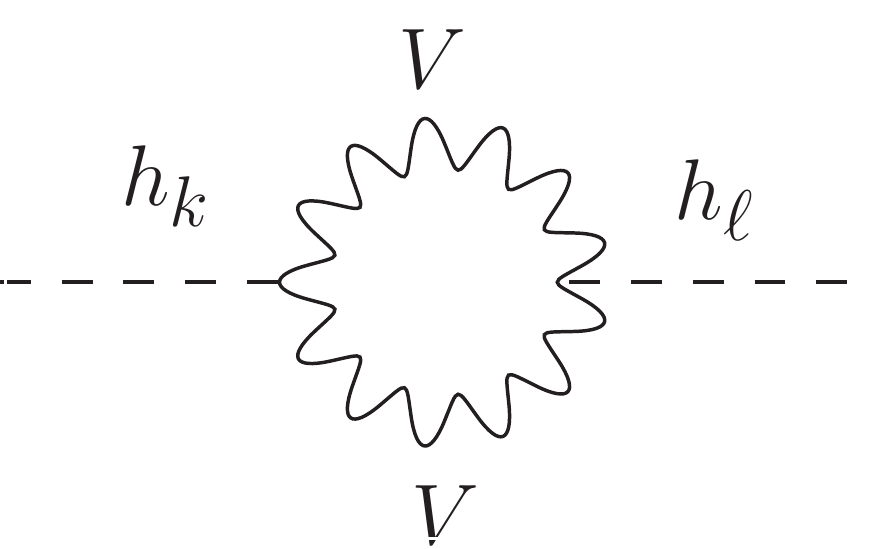}
\includegraphics[width=0.3\textwidth]{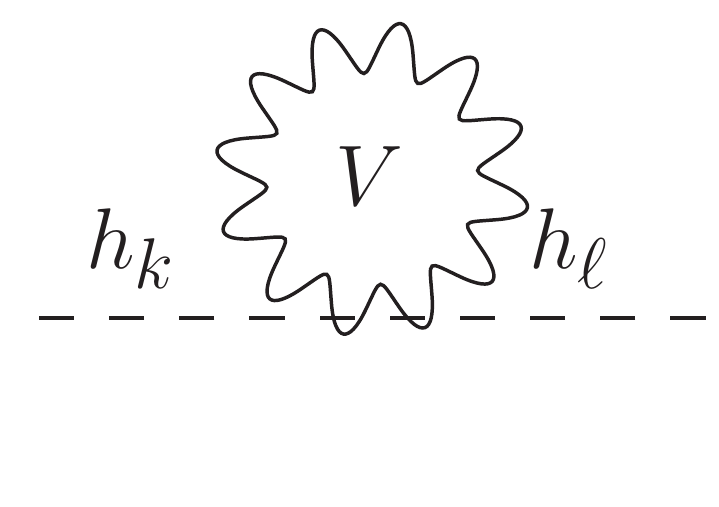}
\includegraphics[width=0.33\textwidth]{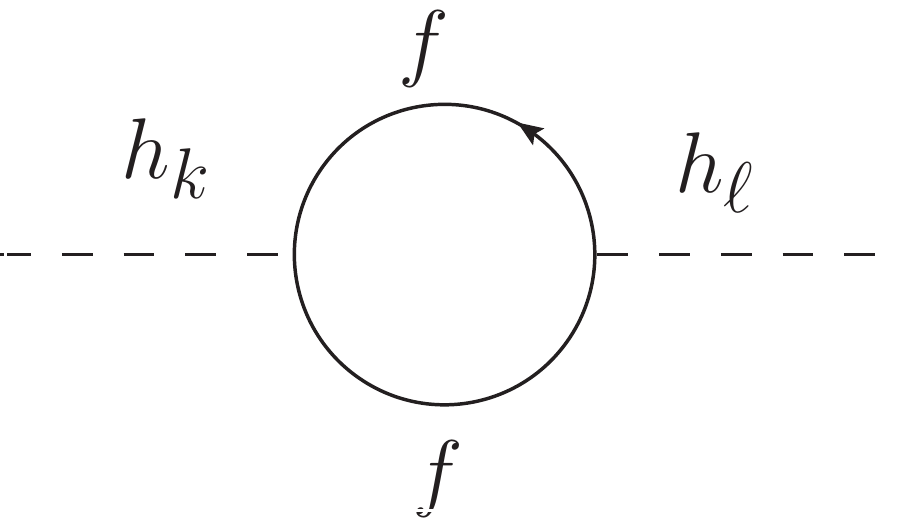}\\
\includegraphics[width=0.33\textwidth]{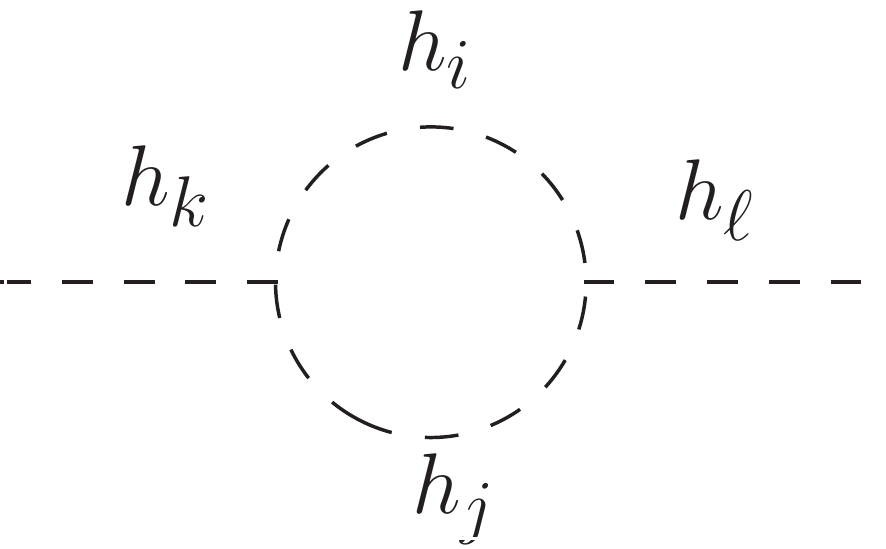}
\includegraphics[width=0.3\textwidth]{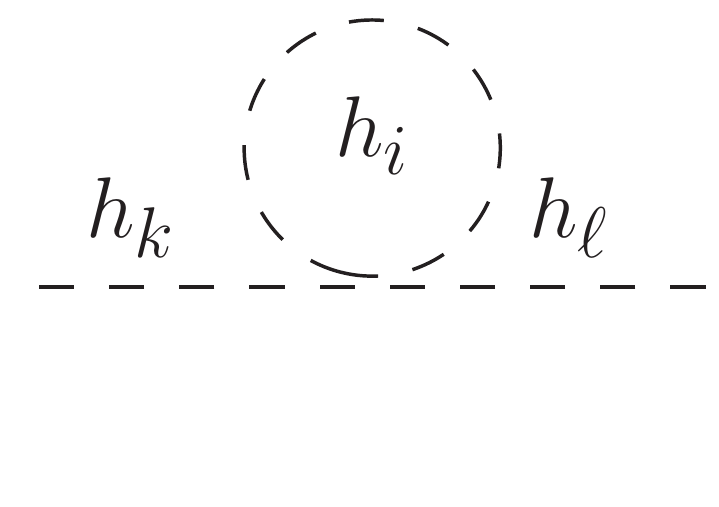}
\end{center}
\caption{\label{fig:wvfcn} One-loop corrections to scalar wavefunction and mass renormalization from (top-left, top-middle) gauge bosons, (top-right) fermions, (bottom) scalars.}
\end{figure}

The contributions to scalar wavefunction and mass renormalization are shown in Fig.~\ref{fig:wvfcn}.  We first consider the diagonal wavefunction renormalization contribution to $Z-Z-h_1$: $\cos\theta\,\delta Z_{11}$.  Let $-i\hat{M}_{11}(p^2)$ denote the one-particle irreducible two-point functions.  Then, using the counterterms in Eq.~\ref{eq:LScal}, the inverse propagator for $h_1$ is
\begin{eqnarray}
iS_{11}^{-1}(p^2)=(1+\delta Z_{11})(p^2-m_1^2)-\delta m_1^2-\hat{M}_{11}(p^2).
\end{eqnarray}
We choose the on-shell renormalization conditions such that the renormalized mass is the pole mass and that the residue of the propagator is $i$.  The pole mass condition gives
\begin{eqnarray}
0=iS_{11}^{-1}(m_1^2)\quad\Rightarrow\quad\delta m_1^2 = \hat{M}_{11}(m_1^2).
\end{eqnarray}
For the residue condition, we follow the normal procedure and Taylor expand $\hat{M}_{11}(p^2)$ about $p^2=m_1^2$ and find the propagator
\begin{eqnarray}
S_{11}(p^2)=\frac{i(1-\delta Z_{11}+\hat{M}'_{11}(m_1^2))}{p^2-m_1^2},
\end{eqnarray}
where 
\begin{eqnarray}
\hat{M}'_{11}(m_1^2)=\frac{d \hat{M}_{11}(p^2)}{dp^2}|_{p^2=m_1^2}.
\end{eqnarray}
Hence, we have
\begin{eqnarray}
\delta Z_{11}=\hat{M}'_{11}(m_1^2) .\label{eq:Z11}
\end{eqnarray}
It has been pointed out that the on-shell renormalization scheme may not be gauge independent~\cite{Bojarski:2015kra}.  As we will show, the only contributions relevant for our calculation are scalar loops, which are manifestly gauge independent.  Hence, there is no ambiguity in our choice of renormalization scheme.

In the following, we break down the contribution to $\cos\theta\,\delta Z_{11}$ according to Fig.~\ref{fig:wvfcn}.  Since we are looking for deviations from the SM prediction, we do not explicitly calculate the SM-like contributions and generically label them as $SM$.  For the gauge boson and fermion contributions in the first row of Fig.~\ref{fig:wvfcn} we have
\begin{eqnarray}
\cos\theta\, \delta Z_{11}^{V,f}=\cos^3\theta\times SM=SM+\mathcal{O}\left(\frac{1}{(4\pi)^4}\right)\approx SM,
\end{eqnarray}
which is just the SM contribution.  The scalar loop contribution to $\delta Z_{11}$ is
\begin{eqnarray}
\cos\theta\, \delta Z_{11}=\frac{\cos\theta}{1+\delta_{ij}}\frac{\lambda_{1ij}^2}{m_1^2(4\pi)^2}(1+F(\tau_s)),
\end{eqnarray}
where~$\tau_s = m_1^2/4m_2^2$ and $F(\tau)$ is defined in Eq.~\ref{eq:F}.
Keeping terms up to $\mathcal{O}(\lambda^2/(4\pi))^2$ we find
\begin{eqnarray}
\cos\theta\, \delta Z_{11}=SM +\frac{1}{2m_1^2}\left(\frac{\lambda_{221}}{4\pi}\right)^2(1+F(\tau_s))+\mathcal{O}\left(\frac{\lambda^4}{(4\pi)^4}\right),
\end{eqnarray}
where $SM$ comes from the $h_1h_1$ internal state and the new physics contribution comes from the $h_2h_2$ contribution.

The gauge boson and fermion contributions to the off-diagonal wavefunction counterterm contribution to $Z-Z-h_1$ scale as 
\begin{eqnarray}
\sin\theta\,\delta Z_{21}^{V,f}\sim \frac{\sin^2\theta}{(4\pi)^2}\sim \frac{1}{(4\pi)^4}.
\end{eqnarray}
For the scalar contributions, different internal states must be considered separately.  The parameter counting of Eq.~\ref{eq:TriScal} is used and we only keep the leading terms.  
\begin{itemize}
\item First, consider diagrams with the topology of the bottom left Feynman diagram of Fig.~\ref{fig:wvfcn}.  The contributions to $\sin\theta\, \delta Z_{21}$ are then
\begin{eqnarray}
\sin\theta \,\delta Z_{21}^{h_ih_j}\sim \frac{\sin\theta \lambda_{1ij}\lambda_{2ij}}{(4\pi)^2}\sim \frac{\lambda\, \lambda_{1ij}\lambda_{2ij}}{(4\pi)^3},
\end{eqnarray}
where $\lambda$ is a generic $\mathcal{O}(1)$ coupling.
\item Second, consider diagrams with the topology of the bottom right Feynman diagram of Fig.~\ref{fig:wvfcn}:
\begin{eqnarray}
\sin\theta\, \delta Z_{21}^{h_i}\sim \frac{\sin\theta \lambda_{21ii}}{(4\pi)^2}\sim \frac{\lambda\, \lambda_{21ii}}{(4\pi)^3},
\end{eqnarray}
where $\lambda$ is a generic $\mathcal{O}(1)$ coupling.
\end{itemize}
Clearly, all contributions from $\sin\theta\,\delta Z_{21}$ are higher order than $\cos\theta\,\delta Z_{11}$ and can be neglected.  

Although there are one-loop diagrams that appear directly in $Z-Z-h_1$ couplings, it can be shown that these reduce to the SM contribution plus higher order corrections.  Hence, to order $\mathcal{O}(\lambda^2/(4\pi)^2)$ we find the fractional shift in the $Z-Z-h_1$ coupling from the SM value is
\begin{eqnarray}
\delta g_{Zh}=\frac{g_{Zh_1}}{g^{\rm SM}_{Zh_1}}-1\approx-\frac{1}{2}\sin^2\theta+\frac{1}{4\,m_1^2}\left(\frac{\lambda_{221}}{4\pi}\right)^2(1+F(\tau_s)),
\end{eqnarray}
where we have expanded the tree-level modification $\cos\theta$ to $\mathcal{O}(\lambda^2/(4\pi))^2$, $g_{Zh_1}$ is the coupling constant in Eq.~\ref{eq:LZh}, and $g_{Zh_1}^{\rm SM}$ is the SM limit.  The fractional shift in the $Z-h_1$ cross section is a factor of two larger:
\begin{eqnarray}
\delta_{Zh}=\frac{\sigma_{Zh_1}}{\sigma^{\rm SM}_{Zh_1}}-1\approx -\sin^2\theta+\frac{\lambda^2_{221}}{32\pi^2m_1^2}(1+F(\tau_s)).
\end{eqnarray}
 This expression is compatible with the result in the $\mathbb{Z}_2$-symmetric limit~\cite{Craig:2013xia, McCullough:2013rea} with the addition of the tree-level shift.  That is, in the small mixing limit, the effect of the scalar mixing only appears at tree level.  Ref.~\cite{Huang:2016cjm} contains additional terms in the shift of the $Z-Z-h_1$.  However, we find that those terms are higher order and may be of similar size to off-diagonal wavefunction renormalization effects.

\end{document}